\documentclass[12pt,openany]{mythesis}
\usepackage{a4,epsf,fancyhdr}
\begin{document}
\oddsidemargin  0in    
\evensidemargin 0in    
\title{{\Huge \bf \sf Light-front Hamiltonian field theory}\\[1cm]
\em towards a relativistic description of bound states}
\vspace{2cm}
\author{Nico Schoonderwoerd}
\date{}
\clearpage{\pagestyle{empty}\cleardoublepage}
\maketitle
\renewcommand{\thepage}{\roman{page}}
\pagestyle {empty}

\hbox{}
\vfill
\begin{tabbing}
\= {\sf Lichtfront Hamiltoniaanse veldentheorie}\\
\> {\em Naar een relativistische beschrijving van gebonden  
                      toestanden}\\
\end{tabbing}
\noindent
{\tt http://xxx.lanl.gov/abs/hep-ph/9811xxx}\\[.5cm]
\noindent
\begin{tabbing}
Leescommissie: \= prof.dr. J.F.J. van den Brand\\
               \> dr. A.E.L. Dieperink\\
               \> prof.dr. G. McCartor\\
               \> prof.dr. P.J.G.~Mulders\\
               \> prof.dr. H.-C. Pauli\\
\end{tabbing}

\begin{tabbing}
Grafisch ontwerp: \= Alex Henneman\\ 
                  \> Nico Schoonderwoerd\\
\end{tabbing}

\begin{figure}[h]
\epsfxsize=2.3cm
  \epsffile{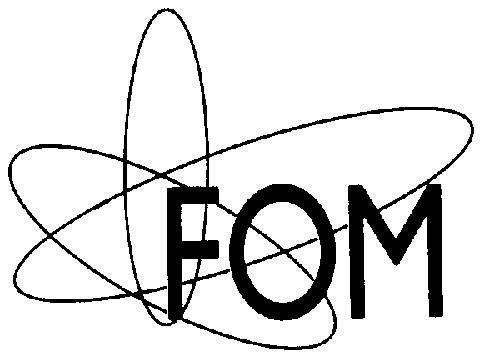}
\end{figure}
\noindent
Dit werk maakt deel uit van het onderzoekprogramma van de Stichting
voor Fundamenteel Onderzoek der Materie (FOM) die financieel wordt
gesteund door de Nederlandse Organisatie voor Wetenschappelijk
Onderzoek (NWO). \\[1cm]
\mbox{\small \copyright 1998, Nico Schoonderwoerd}

\newpage

\begin{center}
VRIJE UNIVERSITEIT \\
\vspace*{2.0cm}
{\Large\bf \sf Light-front Hamiltonian field theory} \\[.5\baselineskip]
{\large\em  towards a relativistic description of bound states} \\

\vspace*{2.0cm}
ACADEMISCH PROEFSCHRIFT \\
\vspace*{1.0cm}
ter verkrijging van de graad van doctor aan 		\\[.4\baselineskip]
de Vrije Universiteit te Amsterdam,			\\[.4\baselineskip]
op gezag van de rector magnificus			\\[.4\baselineskip]
prof.dr. T. Sminia,					\\[.4\baselineskip]
in het openbaar te verdedigen				\\[.4\baselineskip]
ten overstaan van de promotiecommissie			\\[.4\baselineskip]
natuurkunde en sterrenkunde 				\\[.4\baselineskip]
van de faculteit der exacte wetenschappen	        \\[.4\baselineskip]
op donderdag 14 januari 1999 om 13.45 uur 		\\[.4\baselineskip]
in het hoofdgebouw van de universiteit,			\\[.4\baselineskip]
De Boelelaan 1105 					\\[.4\baselineskip]
\vspace*{1.5cm}

door \\

\vspace*{0.8cm}

{\bf Nicolaas Cornelis Johannes Schoonderwoerd} \\
\vspace*{0.3cm}
geboren te Breukelen \\

\end{center}

\newpage

\noindent
\begin{tabular}{ll}
Promotor:    & prof.dr. P.J.G.~Mulders \\
Copromotor:  & dr. B.L.G.~Bakker\\
\end{tabular}
\vfill
\newpage
{\hfill\em Aan Joop, Bea en Cosander\hspace{2cm}}
\hbox{} \vfill

\pagestyle{fancy}
\fancyhf{}
\fancyhead[LE,RO]{\bfseries\thepage}
\fancyhead[LO]{\bfseries\rightmark}
\fancyhead[RE]{\bfseries\leftmark}
\renewcommand{\chaptername}{}
\renewcommand{\chaptermark}[1]{\markboth{\footnotesize \sf Chapter \thechapter\ \hspace{.3cm} #1}{\footnotesize \sf #1}\markright{\footnotesize \sf Chapter \thechapter\ \hspace{.3cm} #1}}
\renewcommand{\sectionmark}[1]{\markright{\footnotesize \sf 
              \thesection\ \hspace{.3cm} #1}}
\renewcommand{\thechapter}{\Roman{chapter}}
\renewcommand{\thesection}{\S{\hspace{.010 cm}}\arabic{section}}
\renewcommand{\theequation}{\mbox{\thechapter-\arabic{equation}}}
\renewcommand{\thefigure}{\mbox{\thechapter-\arabic{figure}}}
\renewcommand{\thetable}{\mbox{\thechapter-\arabic{table}}}
\renewcommand{\theenumi}{{\sf \arabic{enumi}}}
\def \slash#1{\not\! #1}
\def \r#1{(\ref{#1})}
\def \inp#1#2{#1\cdot\!#2}
\def \inperp#1#2{#1^\perp\!\!\cdot\!#2^\perp}
\def \sec{}
\def \sex{}

\tableofcontents
\chapter{\label{chap1}Introduction to light-front Hamiltonian dynamics}
\renewcommand{\thepage}{\arabic{page}}
\setcounter{page}{1}

\begin{quote}
{\em Einstein's great achievements, the principle of relativity, imposes
conditions which all physical laws have to satisfy. It profoundly
influences the whole of physical science, from cosmology, which deals
with the very large, to the study of the atom, which deals with the
very small.}

\raggedleft Paul~Dirac
\end{quote}
This quote reflects the work that I have done in
physics so far, which began with an investigation of new black hole solutions
to the Einstein equation~\cite{S94} when I was a student in Groningen,
and which now ends with the research presented in this Ph.D.-thesis on
models to describe bound states of elementary
particles~\cite{SB96,SB98a,SB98b,SBK99,BS99,thesis}.

The above quote contains the first two lines of an important article
that Dirac~\cite{Dir49} wrote in 1949, in the middle of a century that
has produced enormous progress in the understanding of the properties
of matter. Not only the development of relativity, but also the rise of
Quantum Mechanics was instrumental for this progress.  At the end of
this century, these and many other advances have resulted in a model
that ambitiously is called the Standard Model. It describes all
elementary particles that have been discovered until now; the leptons
and the quarks, and their interactions.

However, this does not imply we have to call it the end of the day
for high-energy physics. A number of problems of a fundamental nature
remain in the Standard Model. As an example we mention the question of
the neutrino mass, that may or may not point to physics not included
in the Standard Model.

There are many practical
problems when one wants to calculate a physical amplitude.
If the interactions are sufficiently weak, perturbation theory is
usually applied and gives in many cases extremely accurate results.
However, in the case of strong interactions, or when bound states
are considered, nonperturbative methods must be developed.

We have to ensure that in such methods covariance is maintained. 
We mean by this that measured quantities, like cross sections and
masses, are relativistic invariants.
When the equations are written down in covariant form it is clear
that the outcome will satisfy relativistic invariance and we refer
to such methods as manifestly covariant.

For example, the Bethe-Salpeter equation is manifestly covariant, but
suffers from numerical intractability beyond the ladder approximation.
Great progress is made by Lattice Field Theory, which is now
able to give quantitative predictions. However, it depends on the
choice of a specific frame of reference and the advances in its
application  rely strongly on a continued increase of the speed of
computers.

A very intuitive picture of a bound state is provided by Hamiltonian 
methods. However, the ``classical'' method of setting up a
relativistic Hamiltonian theory by quantization on the equal-time
plane, so-called instant-form (IF) quantization, suffers from 
problems such as a square root in the energy operator, which results in
the existence of both positive and negative energy eigenstates, and the
complexity of the boost operators, which keeps us away from determining
the wavefunctions in an arbitrary reference frame.
Weinberg~\cite{Wei66} proposed to use the Infinite Momentum Frame
(IMF), because in this limit time-ordered diagrams containing vacuum
creation or annihilation vertices vanish, and therefore the total
number of contributing diagrams is significantly reduced. It is found
that it provides a picture which connects to the one of the constituent
quark model. However, its big disadvantage is that the IMF is connected
to the rest frame by a boost for which one takes the limit of the boost
parameter to infinity. It is dubitable whether this limit commutes
with others that are taken in field theory.

It was only in the seventies that one began to realize that a theory
with the same advantages as the IMF, but without the disadvantages, had
already been suggested by Dirac some decades before: light-front (LF)
quantization, i.e., quantization on a plane tangent to the light-cone.
Of the ten Poincar\'{e} generators, seven are kinematic, i.e., can be
formulated in a simple way and correspond to conserved quantities in
perturbation theory. Most important is that these seven operators
include one of the boost operators, allowing us to determine the
wavefunction in a boosted frame if it is known in the rest frame. This
property is not found in IF quantization. As a drawback one finds that
not all rotations are kinematic, and therefore rotational invariance is
not manifest in LF quantization, a problem which is discussed
frequently in the literature.  In particular, our interest was
triggered by an article by Burkardt and Langnau~\cite{BL91} who claimed
that rotational invariance is broken for $S$-matrix elements in the
Yukawa model. Instead of a lack of manifest rotational invariance, we
prefer to talk about lack of manifest covariance, as this is a property
that all Hamiltonian theories share.  Because in each form of
quantization dynamical operators that involve creation or annihilation
of particles are present, in any relativistic
Hamiltonian theory particle number is not conserved, implying that each
eigenstate has to be represented as a sum over Fock states of arbitrary
particle number.  However, light-front dynamics (LFD) is the only
Hamiltonian dynamical theory which has the property that the
perturbative vacuum is an eigenstate of the (light-front) Hamiltonian,
provided that zero-modes are neglected (in this thesis zero-modes will
not explicitly be discussed). Bound states are also eigenstates and are
distinct from the LF vacuum, which simplifies their analysis.

In this thesis we shall not solve the eigenvalue problem, an interacting
Hamiltonian will not even be written down! The goal of this thesis will
not be to calculate a spectrum, but to illuminate two important
properties of LF Hamiltonian dynamics. The first is:

\begin{quote}
\samepage{{\sf 1.} \em Light-front dynamics provides a covariant framework for the
      treatment of bound states. }
\end{quote}

Although the calculation of bound states requires  nonperturbative methods,
these usually involve ingredients encountered in perturbation theory,
e.g., the driving term in a Lippmann-Schwinger or Bethe-Salpeter approach. 
We prove that LF perturbation theory is equivalent to covariant
perturbation theory. By equivalent we mean that physical observables
in LF perturbation theory are the same as those obtained in
covariant perturbation theory.  This can be done by showing that the
rules for constructing LF time-ordered diagrams can be obtained
algebraically from covariant diagrams by integration over the LF energy
$k^-$. Two technical difficulties, namely that the integration over
$k^-$ can be ill-defined, and that divergences in the transverse
directions may remain, are solved in Chapters~\ref{chap3} and
\ref{chap4} respectively, for the Yukawa model, which is
introduced in Chapter~\ref{chap2}.

In Chapter~\ref{chap5} we discuss the entanglement of covariance
and the Fock-space expansion, and show another important property
of LF Hamiltonian dynamics:

\begin{quote}
\samepage{{\sf 2.} \em Higher Fock state contributions in LF Hamiltonian 
      field theory are typically small, 
      in particular much smaller than in IF Hamiltonian field
      theory, and therefore the ladder approximation gives accurate results
      for  the spectrum. }
\end{quote} 
It has been known for a long time that on the light-front one has to
take into account fewer diagrams than in the instant-form of Hamiltonian
dynamics.  On top of this,  diagrams involving higher Fock states are
numerically smaller, as we will show.  We look at two nucleons
interacting via boson exchange, and we compare the contributions of the
diagrams with one boson in the air, to diagrams where two bosons are
simultaneously exchanged. The latter are ignored if we use the ladder
approximation.  We show in numerical calculations involving scalar
particles that this approximation is viable for both scattering
amplitudes and off energy-shell states, if masses and momenta are
chosen in such a way that they are relevant for the deuteron.

\section{\label{sec11}Forms of relativistic dynamics}
An important first step on the path to a Hamiltonian description
of a dynamical system was taken by Dirac in 1949, in his famous
article 'Forms of Relativistic Dynamics' \cite{Dir49}. One foot of this work
 is in special relativity, when Dirac writes:
\begin{quote}
{\em \dots physical laws shall be invariant under transformations
from one such coordinate system to another.}
\end{quote}

The other foot of his method is in Quantum Mechanics because Dirac writes:
\begin{quote}
{\em \dots the equations of motion shall be expressible in the
Hamiltonian form.}
\end{quote}
In more technical terms, this condition tells us that any two dynamical
variables have a Poisson bracket, later to be associated with
(anti-)commutation relations. We restrict the transformations further
to continuous ones, therefore excluding space inversion and time
reversal.
  In the forthcoming subsections we are going to work out these two
principles and construct the generators of the Poincar\'{e}  group.

\subsection{The Poincar\'{e} group}
The transformations mentioned in the first quote are the four
translations $P^\mu$, the three rotations $J^i = \frac{1}{2}
\epsilon^{ijk}M_{jk}$, and the three boosts $K^i = M^{0i}$, where $M$
is an anti-symmetric tensor. These transformations should satisfy

\begin{eqnarray}
\label{pb1}
[P^\mu,P^\nu] &=& 0,\\
\label{pb2}
\left[ M^{\mu\nu}, P^\rho \right] &=& - g^{\mu\rho} P^\nu + g^{\nu\rho} P^\mu,\\
\label{pb3}
\left[ M^{\mu\nu}, M^{\rho\sigma} \right] &=& 
- g^{\mu\rho} M^{\nu\sigma} + g^{\nu\rho} M^{\mu\sigma}
- g^{\mu\sigma} M^{\rho\nu} + g^{\nu\sigma} M^{\rho\mu}.
\end{eqnarray}
Setting up a dynamical system is equivalent to finding a solution to
these equations. The solution of the ten generators is generally such
that some of them are simple, and correspond to conserved quantities.
These are labeled as {\em kinematical}, indicating that they do not
contain any interaction. Others are more complicated and describe the
dynamical evolution of the system as the Hamiltonian does in
nonrelativistic dynamics. Therefore these are called {\em dynamical},
which means that they do contain interaction. It seems obvious that one
should want to setup the framework in such a way that the number of dynamical
operators is small. A simple solution of the Eqs.~\r{pb1}-\r{pb3} can
be found if we define a point in space-time to be given by the
dynamical variable $x^\mu$ and its conjugate momentum by $p^\nu$. Using

\begin{equation} 
[x^\mu,x^\nu] = 0, \quad [p^\mu,p^\nu] = 0, \quad [x^\mu,p^\nu] = i g^{\mu\nu},
\end{equation}
a solution is now given by
\begin{equation}
\label{basic}
P^\mu = p^\mu, \quad M^{\mu\nu} = x^\mu p^\nu- x^\mu p^\nu .
\end{equation}
As already mentioned by Dirac, this solution may not be of practical
importance, however, it can serve as a building block for future
solutions.

Another important ingredient for the dynamical theory is that we have
to specify boundary conditions. We do this by taking a
three-dimensional surface $\Sigma$ in space-time not containing
time-like directions, at which we specify the initial conditions of the
dynamical system.  The ten generators then split into two groups, namely those
that leave $\Sigma$ invariant and those that do not. The first group is
called the stability group. The larger the stability group of $\Sigma$,
the smaller the dynamical part of the problem. We can ensure a large
stability group by demanding that it acts transitively
on $\Sigma$: every point on $\Sigma$ may be mapped on any other point
of $\Sigma$ by applying a suitable element of the stability group. This
ensures that all points on the initial surface are equivalent.

\begin{figure}
\begin{center}
\begin{tabular}{ccc}
(a) the instant-form && (b) the light-front\\
&&\\
\epsfxsize=6cm \epsffile{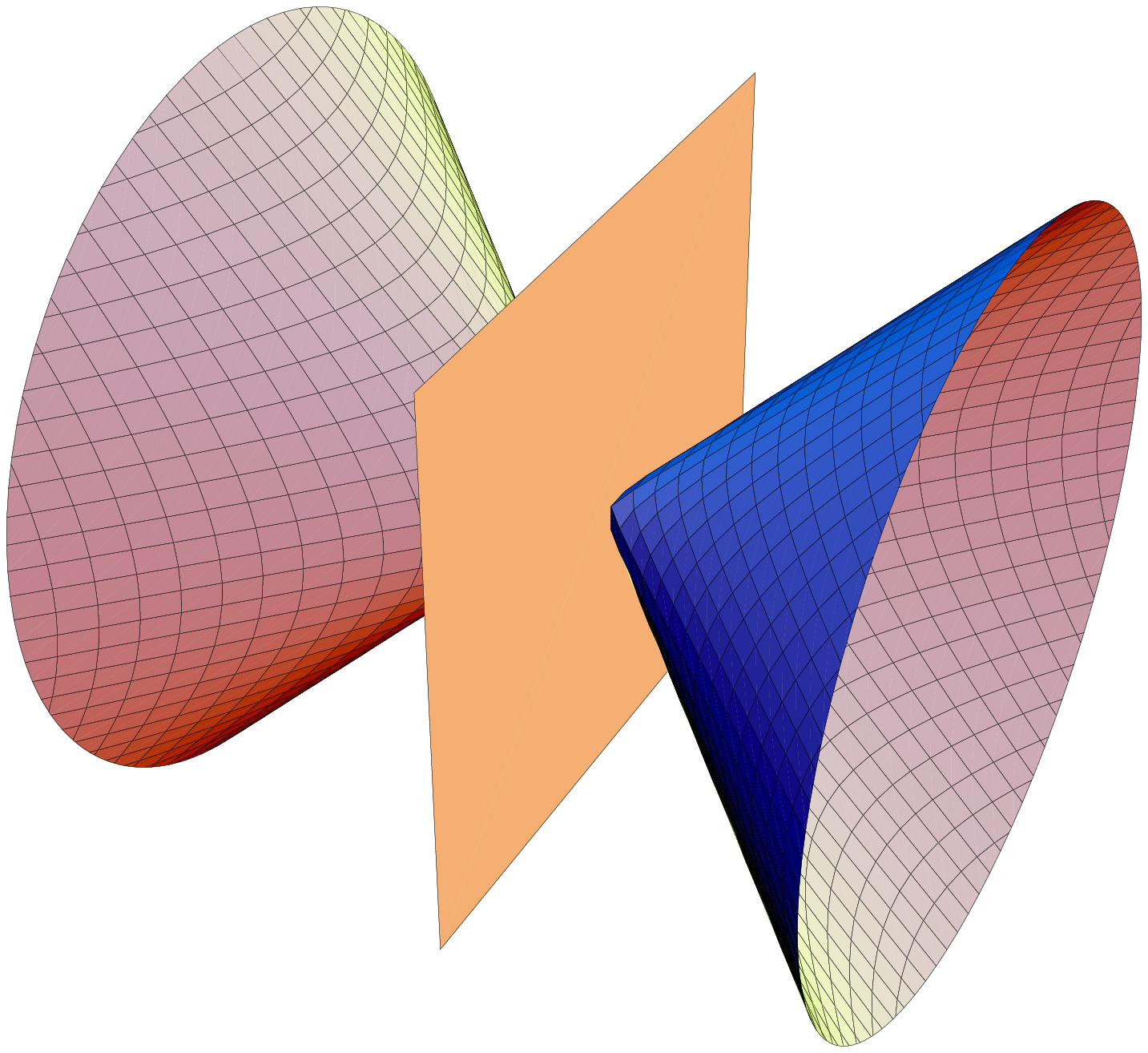} &
\epsfxsize=3cm \raisebox{1.3cm}{\epsffile{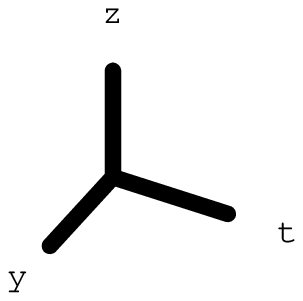}} &
\epsfxsize=6cm \epsffile{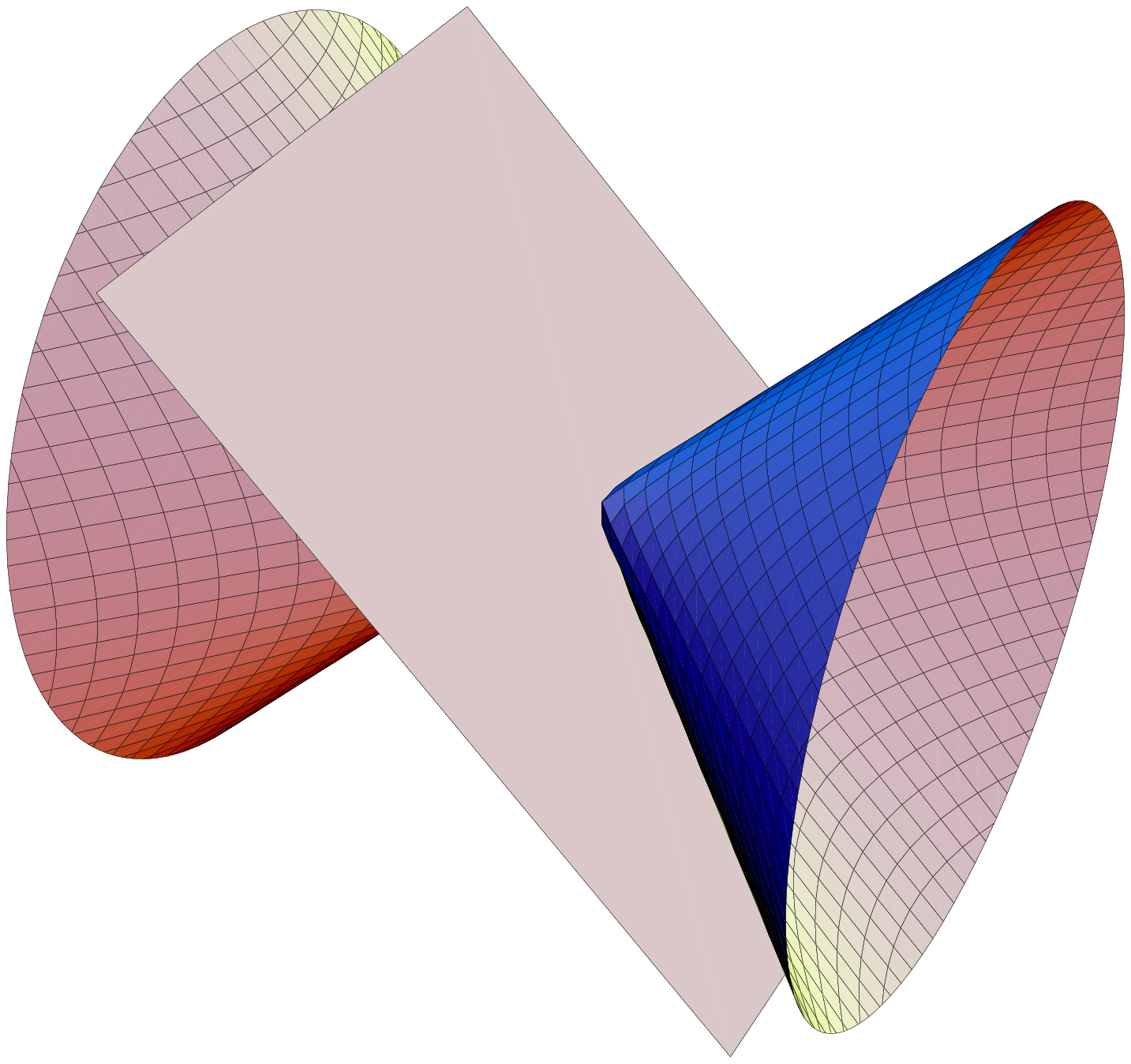} \\
\end{tabular}
\end{center}
\caption{Projection of two initial surfaces onto $y$-$z$-$t$-space.
\label{figforms}
In (a) the initial surface is at $t=0$, and in (b) it is tangent to
the light-cone.}
\end{figure}
The restriction of relativistic causality reduces the number of world lines,
and therefore increases the number of surfaces that one can choose for $\Sigma$.
Dirac found three independent choices for the initial
surface that fulfill these conditions. In total there are five,
as was pointed out by Leutwyler and Stern~\cite{LS78}. 
We, however, only discuss the two most important
ones. They are listed in Fig.~\ref{figforms}. In 
IF Hamiltonian dynamics one quantizes on the equal-time plane, given by
\begin{equation}
\label{ifplane}
x^0 = 0.
\end{equation}
This is the form of dynamics closest to nonrelativistic Quantum Mechanics.
Another important possibility for quantization is offered by a plane tangent
to the light-cone.  The light-front is given by the equation
\begin{equation}
\label{lfplane}
c x^0 + x^3 = 0.
\end{equation}
Notice that this plane contains light-like directions.
It is common to use the $z$-direction to define the light-front. The
different status of the other space-like directions $x$ and $y$ leads
to the fact that the symmetry of  rotational invariance becomes
nonmanifest on the light-front. In explicitly covariant
LFD~\cite{CDKM98}, one defines the light-front by its normal vector
$\vec{\omega}$, which is not fixed. This method will be encountered in
Chapter~\ref{chap5}.

Note that in the nonrelativistic limit ($c \rightarrow \infty$) the
planes \r{ifplane} and \r{lfplane} coincide. This degeneracy is a
feature of Hamiltonian relativistic dynamics. In this limit only one
operator remains dynamical, namely the Hamiltonian, the operator that
generates time translations.

\subsection{Light-front coordinates}
In the remainder of this chapter, we show some of the advantages of
LF quantization. We define so-called longitudinal coordinates
\begin{eqnarray}
A^- =  \frac{c A^0 - A^3}{\sqrt{2}},\\
A^+ =  \frac{c A^0 + A^3}{\sqrt{2}},
\end{eqnarray}
and transverse coordinates
\begin{eqnarray}
A^\perp = (A^1,A^2),   
\end{eqnarray}
such that the spatial coordinates $x^\perp$ and $x^-$ define a
coordinate system on the light-front, and $x^+$ plays the role of time.
From now on we will put the velocity of light equal to unity: $c=1$.
The indices of the four-vectors 
can be lowered and raised using the following LF
metric $g^{\mu\nu}$:
\begin{equation}
g^{\mu\nu} = 
\left( \begin{array}{p{0.7cm}p{0.7cm}p{0.7cm}p{0.7cm}}
       0 & 1 & 0 & 0 \\
       1 & 0 & 0 & 0 \\
       0 & 0 & -1& 0 \\
       0 & 0 & 0 & -1\\
       \end{array}
\right),
\end{equation}
where the first and second row/column refer to the longitudinal components,
and the third and fourth to the transverse components.
Unfortunately, a number of conventions are frequently used for LF
coordinates. We will stick to the one given above, commonly referred to as the
Kogut-Soper convention \cite{KS70}. The stability
group on the light-front has seven elements, as can be verified by writing out
the commutation relations between these operators and $x^+$:
\begin{eqnarray}
\label{eq1816}
&&[ x^+, P^\perp ] = [ x^+, P^+  ] = [ x^+, M^{12}  ] = [ x^+, M^{+\perp} ] = 0,
\nonumber\\
&&[ x^+, M^{+-} ] = - 2 i x^+ = 0.
\end{eqnarray} 
The other three operators are dynamical, as can be seen by the fact that they do not
commute with $x^+$,
\begin{eqnarray}
&&[ x^+, P^- ] = - i \nonumber\\
&& [ x^+, M^{-\perp} ] = - i x^\perp.
\end{eqnarray}
If we look at Fig.~\ref{figforms}b, we see that we can describe the
operation of $P^-$ as a translation perpendicular to the light-front.
The operators $M^{-\perp}$ correspond to rotations of the light-front
about the light-cone. Using these two words in one sentence clearly
indicates why the common expression ``light-cone quantization'' is badly
chosen. We prefer to use the phrase ``light-front quantization''.

\subsection{The initial surface}
In LF quantization, we first solve the Poincar\'{e} algebra on the
surface $x^+ = 0$.  The stability group is the group generated by
transformations of this surface into itself. We already met these
operators in Eq.~\r{eq1816}. As $x^+$
is fixed, the dynamical variable $p^-$ has lost its meaning and,
according to Dirac, it should be eliminated. We can add to the
generators in Eq.~\r{basic} multiples of $(p_\sigma p^\sigma - m^2)$:

\begin{eqnarray}
\label{nobasic}
P^\mu &=& p^\mu + \lambda^\mu (p_\sigma p^\sigma - m^2), \\ 
M^{\mu\nu} &=& x^\mu p^\nu- x^\mu p^\nu + \beta^{\mu\nu} (p_\sigma p^\sigma - m^2).\nonumber
\end{eqnarray}
We then construct the $\lambda^\mu$ and $\beta^{\mu\nu}$ in such a way that on the 
light-front
the $p^-$-dependence drops from these equations. For the elements of the
stability group we find:
\begin{eqnarray}
\label{eq1654ii}
P^\perp = p^\perp,&&\quad P^+ = p^+, \\
M^{+\perp} = - x^\perp p^+, \quad M^{+-} \!\!&=&\!\!
 - x^- p^+, \quad M^{12} = x^1 p^2 - x^2 p^1,
\nonumber
\end{eqnarray}
and for the three dynamical operators we find:
\begin{eqnarray}
\label{eq1655ii}
P^- &=& \frac{{p^\perp}^2 + m^2}{2 p^+}, \\
M^{-1}= x^- p^1 + x^1 \frac{{p^\perp}^2 + m^2}{2 p^+},&& \quad
M^{-2}= x^- p^2 + x^2 \frac{{p^\perp}^2 + m^2}{2 p^+}.
\label{eq1655iii}
\end{eqnarray}
When one quantizes in the instant-form, one finds four operators to be dynamical, which is one more than in LF quantization. 
However, more important is the form of the energy operator. In the instant-form, it is
\begin{equation}
\label{p0bla}
P^0 = \sqrt{\vec{p}^{\,2} + m^2}.
\end{equation}
The presence of the square root causes the degeneracy of positive and
negative energy solutions in IF dynamics, whereas on the light-front
they are kinematically separated, as can be seen from Eq.~\r{eq1655ii}:
positive longitudinal momentum $p^+$ corresponds to positive LF energy
$P^-$, and vice versa.  This effect leads to the spectrum condition,
which is explained in the next section.

The dynamical operators \r{eq1655ii} and \r{eq1655iii} reveal a little
of the problems encountered on the light-front:  the infrared problem
for $p^+ = 0$, which can be associated with the so-called zero-modes.
As the path of quantization on the light-front is beset by problems,
such as the nonuniqueness of the solution of the
Cauchy problem, attempts have been made to find another path.
Inspiration may be found in Quantum Field Theory, which leads
to expressions for propagators of particles, and finally for $S$-matrix
elements. They may serve as a starting point to derive rules for
time-ordered diagrams.

\section{\label{sec12}Light-front quantization}
The first to set foot on the new path towards a LF perturbation
theory were Chang and Ma~\cite{CM69}, and Kogut and Soper~\cite{KS70}.
Their work relies on the Feynman rules that are
constructed in Quantum Field Theory.
To determine the LF time-ordered propagator we take the Feynman
propagator and integrate out the energy component. 

For the types of
theories that are discussed in this thesis, two are of importance:
the scalar propagator and the fermion propagator.

\subsection{The scalar propagator}
The Klein-Gordon propagator for a particle of mass $m$ is well-known:
\begin{equation}
\label{KGprop}
\Delta_{\rm F}(x) = \frac{1}{(2 \pi)^4} 
\int_{\rm Min} {\rm d}^4k \frac{e^{- i k_\mu x^ \mu}}{k^2 - m^2 + i \epsilon},
\end{equation}
where the subscript ``Min'' denotes that the integral is over Minkowski
space. The inner products of the Lorentz vectors can be written in
LF coordinates:
\begin{eqnarray}
k^2 = k_\mu k^\mu = 2 k^+ k^- - {k^\perp}^2, \\
k_\mu x^\mu  = k^- x^+ + k^+ x^- - \inperp{k}{x}.
\end{eqnarray}
Following Kogut and Soper~\cite{KS70}, we separate the energy integral 
from the integral over the kinematical components of $k$, indicated
by $\vec{k}$:
\begin{equation}
\vec{k} = \left( k^+, k^\perp \right),
\end{equation}
We then find for the propagator of Eq.~\r{KGprop}:
\begin{equation}
\label{KGprop2}
\Delta_{\rm F}(x) = \frac{1}{(2 \pi)^4} 
\int \frac{{\rm d}^3 \vec{k} }{2 k^+} e^{- i \vec{k} \vec{x}}
\int {\rm d}k^- \frac{ e^{- i k^- x^+}}{k^- - k^-_{\rm on}},
\end{equation}
in which we use the definition
\begin{equation}
\vec{k} \vec{x} = k^+ x^- - \inperp{k}{x},
\end{equation}
and where $k^-_{\rm on}$ is the on mass-shell value, or, in other words,
the pole in the complex $k^-$-plane:
\begin{equation}
\label{kon}
k^-_{\rm on} = \frac{{k^\perp}^2 + m^2 - i \epsilon}{2 k^+}.
\end{equation}
Forward propagation in LF time requires $x^+ \geq 0$. Then, we
can only evaluate the integral over $k^-$ by closing the contour in the
lower complex half-plane, because of the presence of the factor
$e^{-ik^- x^+}$ in the integrand. For $k^+ > 0$ the pole is below the
real axis.  Therefore application of Cauchy's theorem gives a
nonvanishing result only in this region:

\begin{equation}
\label{KGprop3}
\Delta_{\rm F}(x) = \frac{- i}{(2 \pi)^3 }
\int \frac{{\rm d}^3 \vec{k} \; \theta(k^+) }{2 k^+}
e^{- i k^\mu_{\rm on} x_\mu},
\end{equation}
where the on mass-shell four-vector $k$ is given by
\begin{equation}
k^\mu_{\rm on} = \left(k^-_{\rm on},k^+,k^\perp \right).
\end{equation}

\subsection{The fermion propagator}
The well-known propagator for a spin-$1/2$ particle is
related to the Klein-Gordon propagator by the following relation:
\begin{equation}
\label{fermprop}
S_{\rm F}(x) = ( i \partial_\mu \gamma^\mu + m ) \Delta_{\rm F}(x),
\end{equation}
where $\partial_\mu$ is short for $\partial/(\partial x^\mu)$. 
We interchange differentiation and integration. Differentiation
of the integrand in Eq.~\r{KGprop} gives:
\begin{equation}
\label{fermprop2}
S_{\rm F}(x) = \frac{1}{(2 \pi)^4}
\int_{\rm Min} {\rm d}^4k \frac{\slash{k} + m}{k^2 - m^2 + i \epsilon}
e^{- i k_\mu x^ \mu},
\end{equation}
where the Feynman slash for an arbitrary four-vector $p$ is defined by 
\begin{equation}
\slash{p} = p_\mu \gamma^\mu.
\end{equation}
An important difference with Eq.~\r{KGprop2} is that the numerator contains
the LF energy $k^-$. We can remove it by rewriting the numerator,
\begin{equation}
\label{ksexp}
\slash{k} + m = (k^- - k^-_{\rm on}) \gamma^+ + (\slash{k}_{\rm on} + m ).
\end{equation}
Upon substitution of this expansion into Eq.~\r{fermprop2} we see that
the first term of Eq.~\r{ksexp} cancels against a similar factor in
the denominator. Integration over the LF energy gives the LF time-ordered
fermion propagator:
\begin{eqnarray}
S_{\rm F}(x) &=& \frac{- i}{(2 \pi)^3} 
\int \frac{{\rm d}^3 \vec{k} \; \theta(k^+) }{2 k^+} (\slash{k}_{\rm on} + m)
e^{- i k^\mu_{\rm on} x_\mu} \nonumber\\
\label{fermprop3}
             &+& \frac{- i}{(2 \pi)^3}
\int \frac{{\rm d}^3 \vec{k}}{2 k^+} \; \gamma^+
\; \delta(x^+).
\end{eqnarray}
The first term is the same as the result for the scalar propagator
\r{KGprop3}, except for the factor $(\slash{k}_{\rm on} + m)$. The
second term on the right-hand side of \r{fermprop3} has lost its
propagating part, resulting in the appearance of a $\delta$-function
in $x^+$.  This explains why it is called the instantaneous part. The
decomposition of the covariant propagator into the propagating and the
instantaneous fermion will occur frequently in this thesis and is an
important ingredient in establishing equivalence between LF and
covariant perturbation theory.

\subsection{\label{secspec}The spectrum condition} An important result
that we infer from the previous subsections is that the time-ordered
propagators \r{KGprop3} and \r{fermprop3} contain $\theta$-functions
restricting the longitudinal momentum.  This will severely reduce the
size of phase-space.

\begin{figure}
\[
\epsfxsize=4cm \epsffile{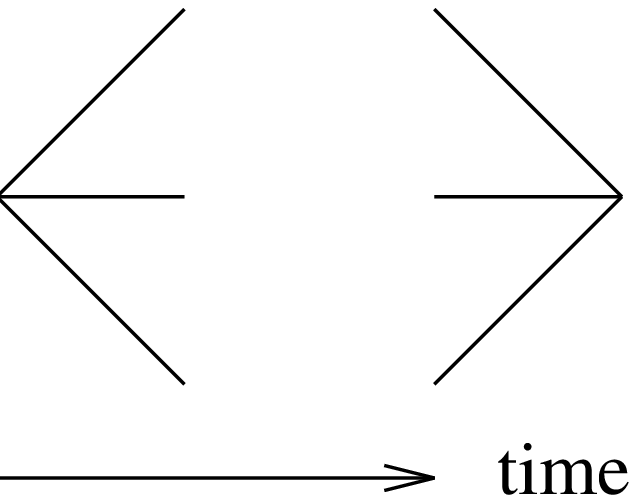}
\]
\caption{\label{figspec}Two vertices that cannot
occur in LF time-ordered diagrams with massive particles.  Light-front time goes from left to
right. The left-hand side shows a vacuum creation, the right-hand side
a vacuum annihilation.} 
\end{figure}

Moreover, the longitudinal momentum
is a conserved quantity and therefore all LF time-ordered diagrams
containing either vacuum creation or annihilation contributions will vanish,
as can be explained by looking at Fig.~\ref{figspec}.
According to Eq.~\r{eq1655ii}, every massive particle in
Fig.~\ref{figspec} should have positive $p^+$-momentum. As the
longitudinal momentum is a kinematical quantity, it should be conserved
at each vertex. However, the vacuum has $p^+=0$.  Therefore diagrams
containing vacuum creation or annihilation vertices are not allowed in
a series of LF time-ordered diagrams. In IF 
dynamics there is no such reduction of the number of diagrams, because
there is no restriction on the IF momentum $\vec{P}$.

\subsection{The energy denominator}
From now on, we shall write the Feynman diagrams in the momentum representation.
In this subsection we show where the energy denominators originate from.
Let us choose as a simple example Compton-like scattering in $\phi^3$ theory:
\begin{equation}
\label{jul32}
P \left\{ \; \epsfxsize=3cm \raisebox{-.7cm}{\epsffile{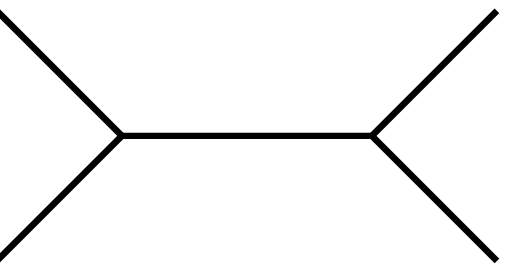}}
\begin{picture}(10,10)(0,0)
\put(0,0){$\phantom{i}$}
\put(-46,8){$p$}
\end{picture} 
\right.
= \frac{1}{p^2 - m^2 - i \epsilon}
= \frac{1}{2 p^+} \; \frac{1}{P^- - \frac{{p^\perp}^2 + m^2 + i \epsilon}{2 p^+}},
\end{equation}
where $P$ is the total momentum, and $p$ is the momentum of the 
intermediate particle. Because of momentum conservation, they are the same. 
However, we make this distinction, to be able to write it in the following form:
\begin{equation}
\epsfxsize=3cm \raisebox{-.7cm}{\epsffile{comcov.eps}}=
\frac{1}{2 p^+} \; \frac{1}{P^- - p^-_{\rm on}} = 
\label{eq1837}
\epsfxsize=3cm \raisebox{-.9cm}{\epsffile{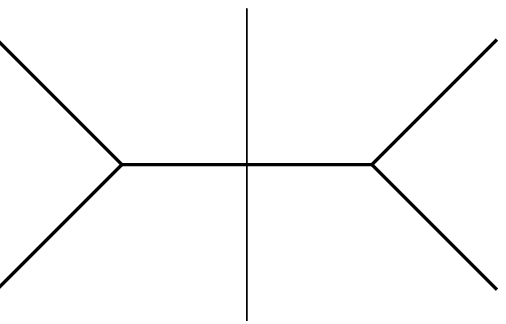}},
\end{equation}
where $p^-_{\rm on}$ is the on mass-shell value of $p^-$, conform
Eq.~\r{kon}. To stress that the diagram on the right-hand side is a
time-ordered diagram, we draw a vertical thin line, indicating an
energy denominator.  The first denominator in Eq.~\r{eq1837}
 is the phase-space factor,
constructed by taking for each intermediate particle the plus-momentum
and a factor 2 because of the Kogut-Soper
convention~\cite{KS70,BPP97}. The direction of the momenta should be
chosen forward in time, such that the plus-component, satisfying the
spectrum condition, is positive.  The energy denominator is constructed
by taking the total energy $P^-$ and subtracting from it the on
mass-shell values of the minus-momentum of the particles in the
corresponding intermediate state.  Thus, it is proportional to the
energy that is ``borrowed'' from the vacuum. This explains why highly off
energy-shell intermediate states are suppressed.  In the next
subsection we present examples where the energy denominators are more
complicated because different time-orderings of the vertices are
involved.

\subsection{Light-front time-ordering}
The most trivial example of time-ordering of vertices was already
discussed in the previous subsection. In Compton-like scattering there are
two time-orderings, however, one, the so-called Z-graph, is excluded
because of the spectrum condition.

\subsubsection{The one-boson exchange}
If we look at a similar amplitude as Eq.~\r{jul32}, now with the 
exchanged particle in the $t$-channel, both time-orderings can contribute.
\begin{equation}
\begin{array}{c}q\\ \\  p
\end{array}
\epsfxsize=2cm \raisebox{-.4cm}{\epsffile{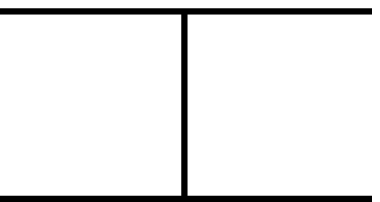}}
\begin{array}{c}q'\\ \\ p'
\end{array}
= \frac{1}{k^2  - m^2 - i \epsilon} =
\frac{1}{2 k^+ 
\left(p^- - p'^- - \frac{m^2 + {k^\perp}^2 + i \epsilon}{2k^+}\right)},
\end{equation}
where the momentum of the intermediate particle $k = p' - p = q - q'$.
The sign of $k^+$ determines the time-ordering of the vertices:
\begin{equation}
\epsfxsize=2cm \raisebox{-.35cm}{\epsffile{onebe.eps}} =
\epsfxsize=2cm \raisebox{-.47cm}{\epsffile{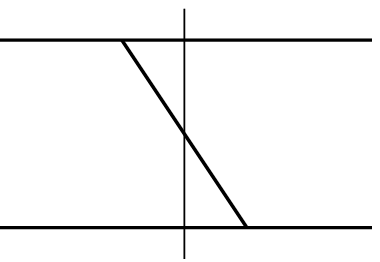}} +
\epsfxsize=2cm \raisebox{-.47cm}{\epsffile{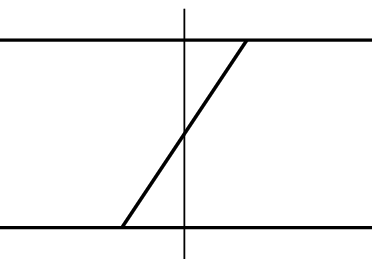}} ,
\end{equation}
with 
\begin{eqnarray}
\epsfxsize=2cm \raisebox{-.5cm}{\epsffile{onebea.eps}} &=&
\theta(k^+) \; \frac{1}{2 k^+} \; 
\frac{1}{P^-- p'^- - \frac{m^2 + {k^\perp}^2 + i \epsilon}{2k^+} - q^-},
\\
\label{jul31}
\epsfxsize=2cm \raisebox{-.5cm}{\epsffile{onebeb.eps}} &=&
\theta(-k^+) \; \frac{1}{2 (- k^+)} \; 
\frac{1}{P^-- p^- - \frac{m^2 + {k^\perp}^2 + i \epsilon}{2(-k^+)} - q'^-},
\end{eqnarray}
where $P^- = p^- +q^-$, which, if the external states are on energy-shell,
coincides with the energy of the system. Again we see that the 
energy denominators are constructed by subtracting from the total
energy $P^-$ the on mass-shell values of the minus-momentum of the particles in the intermediate
state. Because of our choice of momenta, for the diagram~\r{jul31} the
momentum flow of the intermediate particle is backward in time. 
If we substitute $k'^+= - k^+$, then the plus-momentum becomes positive, and
the particle can be reinterpreted as going forward in LF time.
In Chapter~\ref{chap5} we will again encounter these two time-orderings
when we describe the interaction of two nucleons by the exchange of
bosons.

\subsubsection{The scalar shower}
The next example is used to illustrate that some 
algebraic manipulations are needed to construct all LF time-ordered diagrams. 
We look at the decay of a particle into four scalars, again in $\phi^3$
theory.
\begin{equation}
P \; \epsfxsize=2cm \raisebox{-1.17cm}{\epsffile{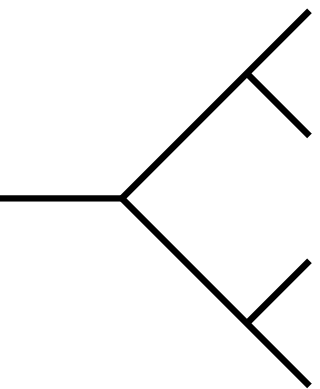}}
\label{eq1609} 
\begin{array}{c}p_4\\ \\p_3\\ p_2\\\\ p_1
\end{array} =
\frac{1}{(p^2_{12} - m^2 - i \epsilon)(p^2_{34} - m^2 - i \epsilon)}
= \frac{1}{p^+_{12} p^+_{34}} \; \frac{1}{(p^-_{12} - p^-_{12 {\rm on}})
(p^-_{34} - p^-_{34 {\rm on}})},
\end{equation}
where the two intermediate scalars have momentum $p_{12} = p_1 + p_2$ and
$p_{34} = p_3 + p_4$ respectively.  We now use the algebraic identity 
\begin{equation}
\label{splittric}
\frac{1}{(a-b)(c-d)}=\frac{1}{a+c-b-d} \; \left( \frac{1}{a-b} + \frac{1}{c-d} \right),
\end{equation}
This splitting can be used for the covariant amplitude in Eq.~\r{eq1609}.
We find: 

\begin{equation}
\epsfxsize=2cm \raisebox{-1.17cm}{\epsffile{treecov.eps}} \; = \;
\epsfxsize=2cm \raisebox{-1.17cm}{\epsffile{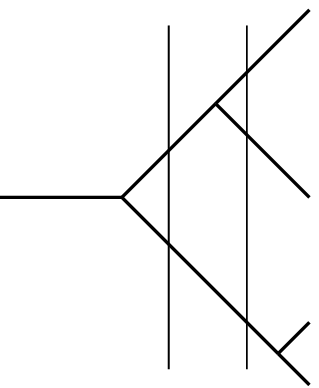}}  \;+ \;
\epsfxsize=2cm \raisebox{-1.17cm}{\epsffile{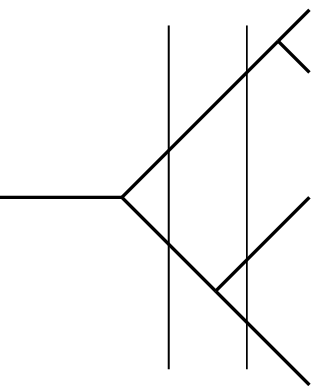}} \;\;,
\end{equation}
with 
\begin{eqnarray}
\epsfxsize=2cm \raisebox{-1.17cm}{\epsffile{treeto1.eps}} &=& 
\frac{1}{4p^+_{12} p^+_{34}} \; \frac{1}{(P^- - p^-_{12\;{\rm on}} - p^-_{34\;{\rm on}})
(P^- - p^-_{12\;{\rm on}} - p^-_3 - p^-_4)}, \\
\epsfxsize=2cm \raisebox{-1.17cm}{\epsffile{treeto2.eps}} &=&
\frac{1}{4p^+_{12} p^+_{34}} \; \frac{1}{(P^- - p^-_{12\;{\rm on}} - p^-_{34\;{\rm on}})
(P^- - p^-_1 - p^-_2 - p^-_{34\;{\rm on}})},
\end{eqnarray}
which again are energy denominators as defined above.
This way to split the covariant denominator using the trick of
\r{splittric} is the simplest example of a complicated
recombination scheme for denominators which can be found in the article
of Ligterink and Bakker~\cite{LB95b}.

\subsection{Loop diagrams}
\begin{figure}
\[
{\rm (a)} \;\;\epsfxsize=4.3cm \epsffile{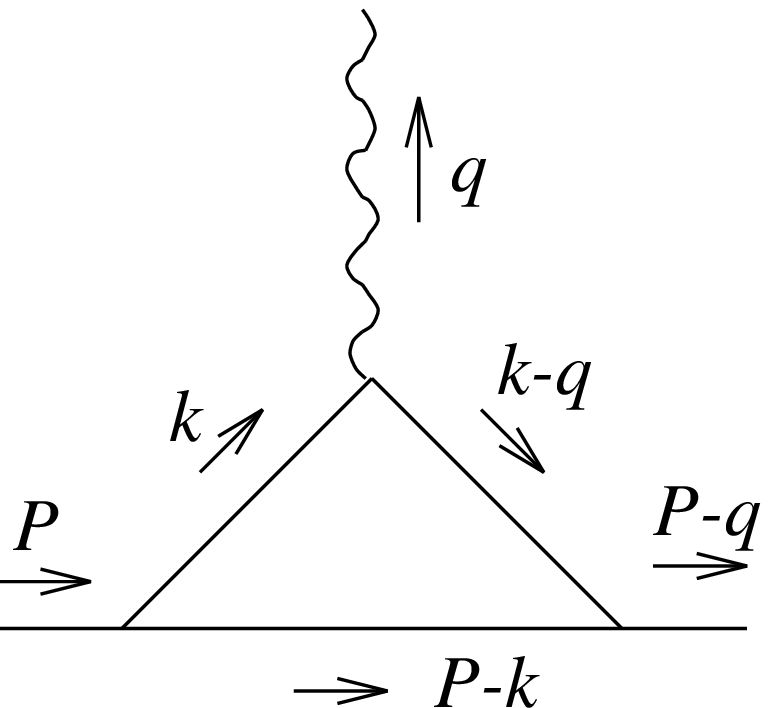} \;\;\;
{\rm (b)} \epsfxsize=9cm \raisebox{0cm}{\epsffile{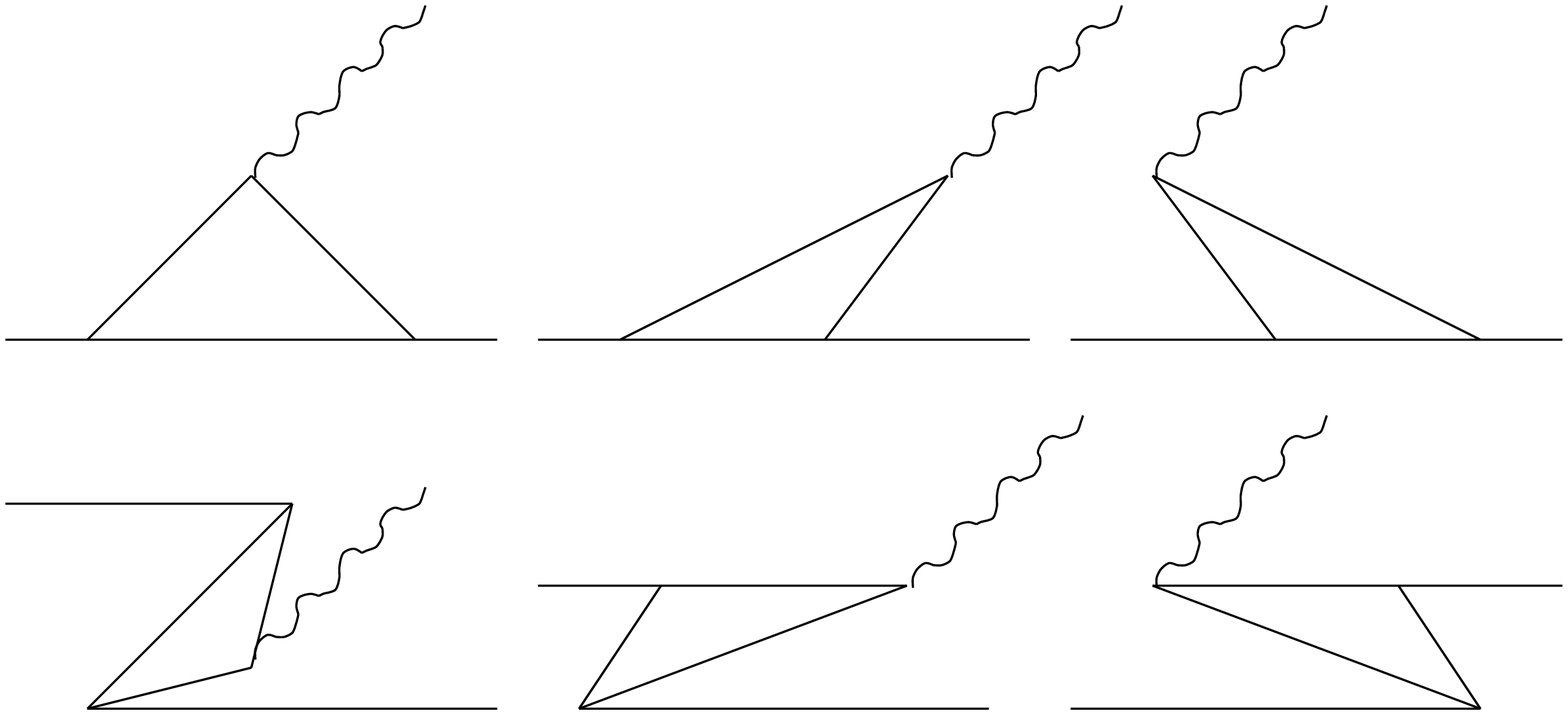}}
\]
\caption{(a) Kinematics for the current. (b) 
\label{figEMform}The six time-orderings contributing to the 
electro-magnetic form factor in $\phi^3$ theory. The vertical lines
denoting energy denominators have been omitted.}
\end{figure}

In the case of a loop diagram, covariant Feynman rules require an
integration over the internal four-momentum $k$. Time-ordered diagrams
have an integration over the three kinematical components. As was
found by Kogut and Soper~\cite{KS70}, a relation between these types
of diagrams can be established if we integrate out the energy component
$k^-$ from the covariant diagram. Upon doing this integration one
finds all LF time-ordered diagrams with the vertices time-ordered in
all possible ways, however, respecting the spectrum condition. For
an arbitrary number of particles in a loop, the proof was only
recently given by Ligterink and Bakker~\cite{LB95b}. As an example, which
also shows some of the problems encountered in LF Hamiltonian dynamics,
we discuss the electromagnetic form factor in $\phi^3$ theory, 
given earlier by Sawicki~\cite{Saw91}:

\begin{equation}
\label{EMformcov}
J^\mu(0) = \int_{\rm Min} \hspace{-.3cm}
{\rm d}^4k \frac{1}{(k-q)^2 - m^2 + i \epsilon}
(2 k^\mu - q^\mu) \frac{1}{k^2 - m^2 + i \epsilon} \;
\frac{1}{(P-k)^2 - m^2 + i \epsilon},
\end{equation}
where the kinematics are given in Fig.~\ref{figEMform}a.
All time-orderings corresponding to this diagram
are given in Fig.~\ref{figEMform}b.

An essential difference between the instant-form and
the light-front occurs if we write the Feynman propagator in terms of
the poles in the energy plane. In terms of IF coordinates we find:
\begin{eqnarray}
k^2-m^2+i \epsilon = \left(k^0 - \sqrt{\vec{k}^2 + m^2 - i \epsilon}\,\right)
                       \left(k^0 + \sqrt{\vec{k}^2 + m^2 - i \epsilon}\,\right) ,
\end{eqnarray}
and on the light-front we have:
\begin{eqnarray}
k^2-m^2+i \epsilon
= 2 k^+ \left(k^- - \frac{{k^\perp}^2 + m^2 - i \epsilon}{2 k^+}\right).
\end{eqnarray}

We see that the Feynman propagator is quadratic in the IF
energy $k^0$ but only linear in the LF energy $k^-$. 
In the former case it leads to the presence of both positive and negative energy eigenstates,
whereas on the light-front only positive energy states occur.
In the instant-form, half of the poles occur above the real axis, and 
the other half below. Therefore
contour integration will always give a nonvanishing result.
In contrast to this, on the light-front the poles can cross 
the real axis. If all poles are on the same side of the real axis,
the contour can be closed in the other half of the complex plane, and
contour integration gives a vanishing result. Because of this effect, four
of the six time-ordering in Fig.~\ref{figEMform} disappear. Only
the first two remain. This is another manifestation of the spectrum
condition. If we then turn to the Breit-frame ($q^-,q^+,q^\perp) =
(0,0,q^\perp$), also the second diagram of Fig.~\ref{figEMform}
vanishes, as will follow from the analysis we present below.

\begin{figure}
\begin{center}
(a) $\;\;k^+ < 0$ \hspace{1.5cm}
(b) $\;\;0 < k^+ < P^+$ \hspace{1.5cm}
(c) $\;\;P^+ < k^+$ 
\end{center}
\vspace{-.7cm}
\[
\epsfxsize=12cm \raisebox{0cm}{\epsffile{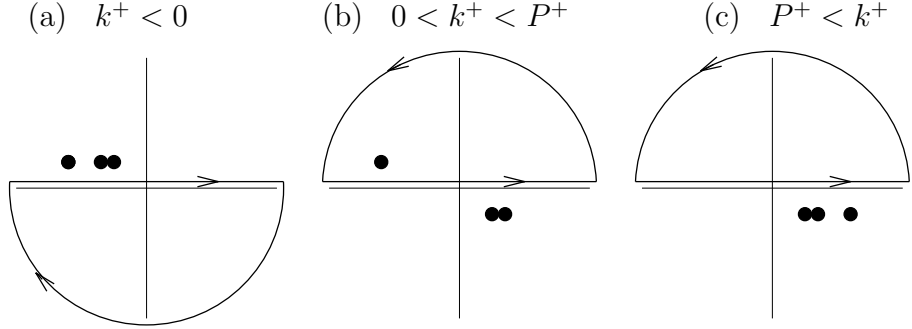}}
\]
\caption{\label{figpoles}The position of the double dot (first and second
propagator in Eq.~\r{EMformcov}) and the dot (third propagator in 
Eq.~\r{EMformcov})
indicate in which quadrant of the complex $k^-$-plane the corresponding
poles are located. }
\end{figure}

Most important in our analysis is the sign of the imaginary part of
the poles. Because of our choice of the Breit-frame, these are identical
for the first and the second Feynman propagator in Eq.~\r{EMformcov},
namely $-\epsilon/2k^+$. The imaginary part of the third Feynman
propagator is $\epsilon/2(P^+ - k^+)$. In Fig.~\ref{figpoles} we show
the location of these poles for different $k^+$ intervals.

We see in (a) and (c) of Fig.~\ref{figpoles} that the contour can be
closed in such a way that no poles are inside the contour, and
therefore contour integration leads to a vanishing result. In case that
we calculate the component $J^-$ of the current, application of
Cauchy's theorem is not valid because there is a contribution to the
integral from a pole at infinity, i.e., for large absolute values of $k^-$ the
integrand goes as $1/k^-$. Therefore we restrict ourselves in
this example to the components $J^+$ and $J^\perp$. Only one LF
time-ordered diagram contributes to the current:

\begin{eqnarray}
\label{EMformto}
J^\mu(0)&=&
\epsfxsize=2cm \raisebox{-.5cm}{\epsffile{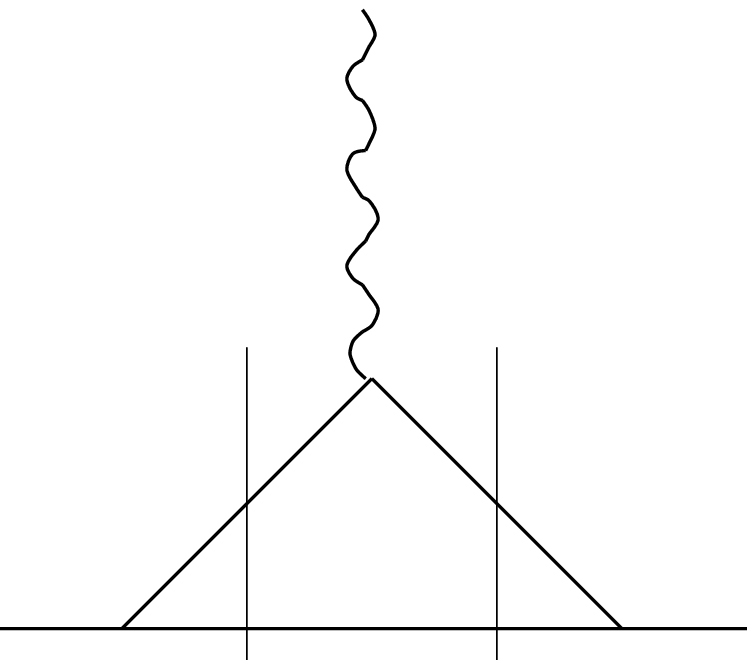}}
= 2 \pi i
\int {\rm d}^2k^\perp
\int_0^{P^+} \frac{{\rm d}k^+ }{8 {k^+}^2 (P^+-k^+) } \;
(2 k^\mu - q^\mu) \\ 
&\times&\frac{1}{P^- - \frac{{k^\perp}^2 + m^2}{2k^+} - 
               \frac{{(P^\perp-k^\perp)}^2 + m^2}{2(P^+-k^+)}} \;
\frac{1}{P^- - \frac{{(k^\perp-q^\perp)}^2 + m^2}{2k^+} - 
               \frac{{(P^\perp-k^\perp)}^2 + m^2}{2(P^+-k^+)}}, \nonumber
\end{eqnarray}
where we have drawn vertical lines in the LF time-ordered diagram
to indicate the energy denominators and to avoid confusion with the
covariant diagram. The kinematics are given in Fig.~\ref{figEMform}a.
The photon line is vertical to indicate that we are in the
Breit-frame.  The imaginary parts have been omitted.

For the result above to be correct three assumptions are
essential:
\begin{enumerate}
\item Interchange of the limit $q^+ \rightarrow 0$ and $k^-$-integration
is valid,
\item There is no contribution of poles at infinity upon doing the
$k^-$-integration,
\item The amplitude is well-defined and finite.
\end{enumerate}
All three assumptions can be justified in this case. De Melo {\em et
al.} \cite{MSFS98} have shown that the interchange mentioned under
assumption~1 may cause pair creation or annihilation contributions to
become nonvanishing.  In \sec~\ref{secpairc} of Chapter \ref{chap2}
we show that this effect may also occur in the Yukawa model. However,
it is not a violation of the spectrum condition.

The second assumption can be justified by looking at
Eq.~\r{EMformcov}.  As $k^2$ is linear in $k^-$, we see that the
integration over the minus component is well-defined for each component
of the current. In a theory with fermions, this integration can be
ill-defined, leading to longitudinal divergences and the occurrence of
so-called forced instantaneous loops (FILs). Divergences for the Yukawa
model are classified in \sec \ref{secdiv} of Chapter \ref{chap2} and
the longitudinal ones are dealt with in Chapter~\ref{chap3}, where it
is shown that the FILs vanish upon using an appropriate regularization
method: ``minus regularization'' \cite{LB95a}.

The third assumption is also satisfied, since the superficial degree of
divergence for integration over the perpendicular components is smaller
than zero for all components of the current. If any transverse
divergences occur, they can be attacked with the method of extended
minus regularization presented in Chapter~\ref{chap4}.  The phase space
factor contains endpoint singularities in $k^+$. However, these are
canceled by identical factors in the energy denominators.

\chapter{\label{chap2}The Yukawa model}
In particle physics several models are used to
describe existing elementary particles, interactions and bound states.
Many of these models are just used to highlight certain properties, or
to make exact or numerical calculation possible. Although the latter
are referred to as toy models, they are helpful because they are
stripped from those properties that are of no concern to the
investigation that is done. In this thesis we are going to ``play
around'' with two models. One of them we already met in the
introductory chapter:  $\phi^3$~theory. In this model one can very nicely demonstrate
that higher Fock states are much more suppressed on the light-front
than in instant-form Hamiltonian dynamics, as will be done in
Chapter~\ref{chap5}. This model only contains scalar particles. The
simplest model including fermions is the Yukawa model, which has the
following Lagrangian:  
\begin{equation}
 {\cal L} = \bar{\psi}(i \partial_\mu \gamma^\mu - m)\psi +
	    \phi ( \partial_\mu \partial^\mu + \mu^2 ) \phi + g
	    \bar{\psi} \psi \phi.
 \label{lag} 
\end{equation} 
The field $\psi$ describes the fermions and the field $\phi$ describes
the scalar particles, from now on referred to as bosons.  The last term
is the interaction between the fermion--anti-fermion field and the
boson field. 
Yukawa introduced this model to describe the interaction of nucleons (fermions)
via pions (bosons). The strength of the interaction is given by $g$.
In our calculations we limit ourselves to a scalar coupling.

\section{Feynman rules}

Using perturbation theory one can deduce from the Lagrangian the
well-known rules for Feynman diagrams. Summing over these diagrams one then
finds the $S$-matrix. 

The first term of the Lagrangian~\r{lag} leads to 
the following propagator:
\begin{equation}
\label{fermpropc}
\setlength{\unitlength}{0.008500in}
\begin{picture}(160,14)(120,560)
\thicklines
\put(120,560){\line( 1, 0){100}}
\put(170,565){\makebox(0,0)[lb]{\raisebox{0pt}[0pt][0pt]{$k$}}}
\end{picture}
= \frac{i(\slash{k} + m)}{k^2 - m^2 + i \epsilon},
\end{equation}
for a fermion with momentum $k^\mu$ and mass $m$. For a (scalar) boson
with momentum $k^\mu$ and mass $\mu$ we have the following
Feynman rule:
\begin{equation}
\label{bosonprop}
\setlength{\unitlength}{0.008500in}
\begin{picture}(160,14)(120,560)
\thicklines
\multiput(120,560)(11.76471,0.00000){9}{\line( 1, 0){  5.882}}
\put(170,565){\makebox(0,0)[lb]{\raisebox{0pt}[0pt][0pt]{$k$}}}
\end{picture}
= \frac{i}{k^2 - \mu^2 + i \epsilon}.
\end{equation}
The full set of Feynman rules to compute the scattering amplitude in the
Yukawa model can be found in many text books such as Itzykson and Zuber
\cite{IZ85}. Our goal is to translate these rules to rules for diagrams
that one uses in LFD.


In Chapter~\ref{chap1} we introduced the $k^-$-integration to obtain
the rules for the LF time-ordered diagrams. 
A complication in this procedure was already
mentioned there: the integration over $k^-$ may be
ill-defined, and the resulting integral may be divergent.
Before solving these problems, we first classify the divergences.

\section{\label{secdiv}Divergences in the Yukawa model}
In the previous subsection we described how to construct each covariant
Feynman diagram. Covariant diagrams may contain infrared and
ultraviolet divergences. Therefore we are not surprised that both in
the process of constructing the LF time-ordered diagrams as in the
diagrams themselves divergences can be encountered. The first
type can be classified as longitudinal divergences, and the second
as transverse divergences.

\subsection{Longitudinal divergences}

We can deduce what (superficial) divergences we
are going to encounter upon integration over $k^-$.  We denote the 
longitudinal degree of divergence by $D^-$.
Suppose we have a truncated one-loop diagram containing $b$ bosons and
$f$ fermions. In the fermion propagator Eq.~\r{fermpropc} the
factor $k^-$ occurs both in the numerator and in the denominator, and
therefore it does not contribute to $D^-$. Each boson will,
according to Eq.~\r{bosonprop} contribute $-1$ to the degree of divergence, and
the measure ${\rm d}^4k$ of the loop contributes $1$, resulting in
\begin{equation}
D^- = 1-b.
\end{equation}
Longitudinally divergent diagrams, i.e., $D^- \geq 0$, contain one boson
in the loop, or none.  Since every loop contains at least two lines, a
longitudinally divergent diagram contains at least one fermion. For the
model we discuss, the Yukawa model with a scalar coupling, the
degree of divergence is reduced. For scalar coupling $g$ it turns out that
$\gamma^+ g \gamma^+ = 0$ and therefore two instantaneous parts
cannot be neighbors. The longitudinal  degree of divergence for the
Yukawa model with scalar coupling is

\begin{equation}
D_{\rm Yuk}^- = 1 - b - \left[ \frac{1\! +\! f\! -\! b}{2} \right]_{\rm entier}
= 1 - \left[ \frac{1\! +\! f\! +\! b}{2} \right]_{\rm entier}  ,
\end{equation}
where the subscript ``entier'' denotes that we take the largest integer
not greater than the value between square brackets.

\subsection{Transverse divergences}
The transverse degree of divergence $D^\perp$ of a LF time-ordered
diagram is the divergence one encounters upon integrating over the
perpendicular components.  In most cases this degree of divergence is
the same as what is known in covariant perturbation theory as the
superficial degree of divergence $D$ of a diagram.  In that case it is
the divergence one finds if in the covariant amplitude odd terms are
removed and Wick rotation is applied.  For a one-loop Feynman diagram
in four space-time dimensions with $f$ internal fermion lines and $b$
internal boson lines the transverse degree of divergence is
\begin{equation}
D^\perp = 4 - f - 2b.
\end{equation}
In case of $d$ space-time dimensions we have to replace the term $4$ by $d$.

\begin{table}
\begin{center}
\begin{tabular}{|l|r|r|r|}
\hline
   & $D_{\rm Yuk}^-=0$ &     $D_{\rm Yuk}^-=-1$ &      $D_{\rm Yuk}^-=-1$ \\
&&&\\
$b=1$ &
\epsfxsize=3cm \epsffile[  5 00 165 60]{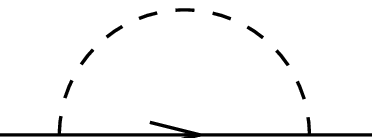} &
\epsfxsize=3.5cm \epsffile[ 05 20 165 80]{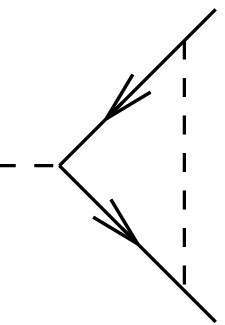} &
\epsfxsize=3.8cm \epsffile[-10 0 150 60]{b4.eps} \\
&&&\\
                  & $D^\perp=1$ & $D^\perp=0$ & $D^\perp\!=\!-1$ \\
\hline
   & $D_{\rm Yuk}^-=0$ &    $D_{\rm Yuk}^-\!=\!-1$ &     $D_{\rm Yuk}^-\!=\!-1$
\\
&&&\\
$b=0$ &
\epsfxsize=3cm \epsffile[  5 15 165 75]{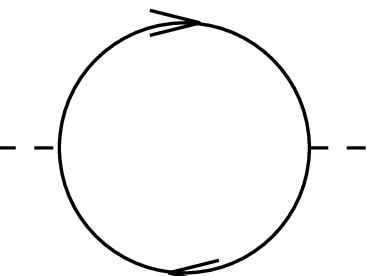} &
\epsfxsize=3.5cm \epsffile[ 05 20 165 80]{f3.eps}  &
\epsfxsize=3.8cm \epsffile[-10 00 150 60]{f4.eps} \\
&&&\\
                  & $D^\perp=2$ &    $D^\perp= 1$ &     $D^\perp= 0$ \\
\hline
\end{tabular}
\caption{Longitudinal ($D^-$) and transverse ($D^\perp$) degrees of 
         divergence in the Yukawa model.
\label{tabYukdiv}}
\end{center}
\end{table}

In Table~\ref{tabYukdiv} all one-loop diagrams up to order $g^4$ that
are candidates to be divergent have been listed with their longitudinal
and transverse degree of divergence.

\section{Instantaneous terms and blinks}
As was already illustrated in the introduction, in the case of
fermions we have to differentiate between propagating and instantaneous
parts. Therefore this distinction plays an important role in the Yukawa model.
The covariant propagator in momentum representation for an off-shell
spin-1/2 particle can be written analogously to Eq.~\r{fermprop3}:

\begin{equation}
\label{abcd}
\frac{i(\slash{k} + m)}{k^2 - m^2 + i \epsilon}
= \frac{i ( \slash{k}_{\rm on} + m)}{k^2 - m^2 + i \epsilon}
+ \frac{i \gamma^+}{2 k^+} .
\end{equation}

The first term on the right-hand side is the propagating part.
The second one is the instantaneous part.  The splitting of the
covariant propagator corresponds to a similar splitting of LF
time-ordered diagrams. For any fermion line in a covariant diagram two
LF time-ordered diagrams occur, one containing the propagating part of
the covariant propagator, the other containing the instantaneous part.
For obvious reasons we call the corresponding lines in the LF
time-ordered diagrams propagating and instantaneous respectively.  For a
general covariant diagram the $1/k^+$-singularity in the propagating
part cancels a similar singularity in the instantaneous part. Therefore
the LF time-ordered diagrams with instantaneous lines are necessary;
they are usually well-defined.

If the $1/k^+$-singularities are inside the area of integration we
may find it necessary to combine the propagating and the instantaneous
contribution again into the so-called blink, introduced by Ligterink
and Bakker~\cite{LB95b}, such that there is a cancellation of the
singularities:

\begin{equation}
\label{abcde}
\raisebox{-.86cm}{\epsfxsize=8.7cm \epsffile[ 0 0 504 108]{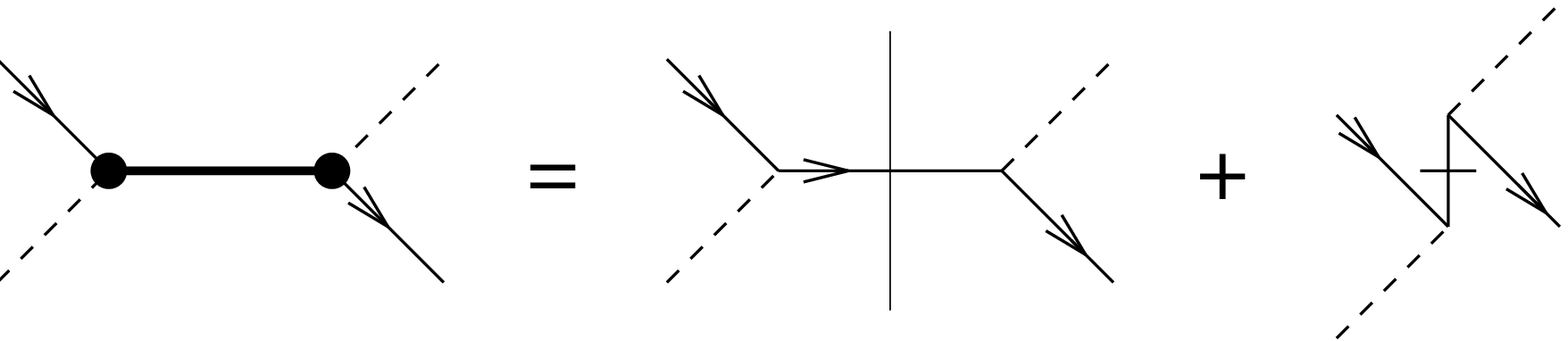}}.
\end{equation}

The thick straight line between fat dots is a blink.  The bar in
the internal line of the third diagram is the common way to denote an
instantaneous fermion.  When a LF time-ordered diagram resembles a
covariant diagram,  we draw a vertical line as in the second diagram of
Eq.~\r{abcde}. If no confusion is possible, we omit it in the remainder
of this thesis.  The difference between Eqs.~\r{abcd} and \r{abcde} lies in
the fact that the former uses covariant propagators, and the latter has
energy denominators. In this case the difference is only formal.
However, in more complicated diagrams there is a big difference, as we
will see later. Examples of blinks are discussed in the next section,
and in \sec \ref{secobeiii} of Chapter~\ref{chap3} where we discuss the
one-boson exchange correction to the vertex.

\section{\label{secpairc}Pair contributions in the Breit-frame}
In Chapter~\ref{chap1} we found that for massive particles the
spectrum condition applies: there can be no creation from or
annihilation into the vacuum.  This gives a significant reduction of
the number of diagrams that one has to incorporate in a light-front
calculation. In any frame where the particles have positive
plus-momentum this is valid. In Leutwyler and Stern \cite{LS78} it was
already noted that on the light-front the regions $p^+ < 0$, $p^+ = 0$
and $p^+ > 0$ are kinematically separated, another manifestation of the
spectrum condition. This fact should already make us aware that the
Breit-frame, where one takes the limit of the plus-momentum of the
incoming virtual photon going to zero, is dangerous.

Indeed one finds that pair creation or annihilation contributions play
a role in this limit. This was first found by De Melo {\em et al.}
\cite{MSFS98} and later by Choi and Ji \cite{CJ98}. They discuss as an
example the electro-magnetic current in $\phi^3$~theory, and find a pair
creation contribution for the component $J^-$ of the current.
We have shown in Chapter~\ref{chap1} that for the other components
$J^+$ and $J^\perp$ pair creation contributions vanish. Because De Melo
{\em et al.} discuss a scalar theory, we infer that this effect is not
related to the presence of fermions in the theory.

In a theory with fermions, such as the Yukawa model, we now show that
the pair creation/annihilation term is also nonvanishing, and that its 
omission leads to a breaking of covariance and rotational invariance.

In the presence of fermions, the individual time-ordered diagrams may
contain $1/k^+$ singularities that cancel in the full sum. 
Because this cancellation has nothing to do with the time-ordering of the
diagrams, we combine the LF time-ordered diagrams into blink diagrams. 
After that, we have a clear view on the point we want to discuss.

Again, we use kinematics as in Fig.~\ref{figEMform}a.
Two blinks contribute to the current, provided we have chosen the
plus-component of the momentum of the outgoing boson $q^+\geq 0$.
\begin{equation}
\label{blink0}
\epsfxsize=2.7cm \raisebox{-.45cm}{\epsffile{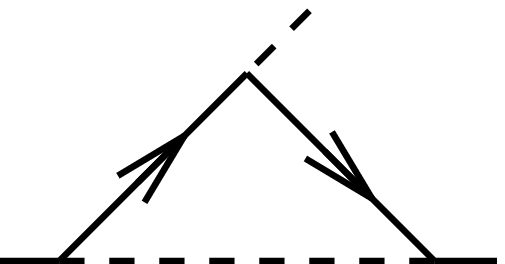}}=
\epsfxsize=2.73cm \raisebox{-.5cm}{\epsffile{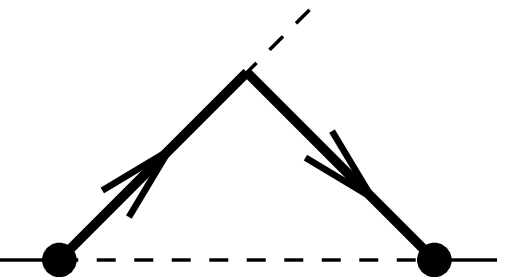}}+
\epsfxsize=2.53cm \raisebox{-.5cm}{\epsffile{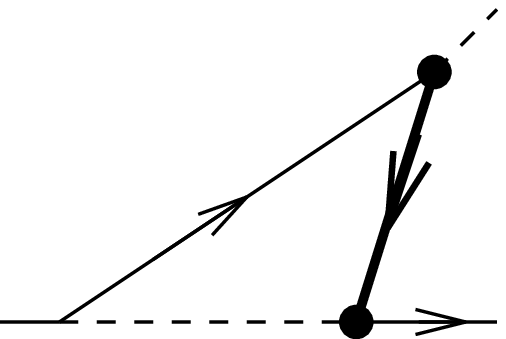}},
\end{equation}
where the diagrams containing blinks are given by:
\begin{eqnarray}
\label{blink1}
\epsfxsize=2.73cm \raisebox{-.5cm}{\epsffile{blink1.eps}}=
- 2 \pi i \;\int {\rm d}^2k^\perp
\int_{q^+}^{P^+} \frac{{\rm d}k^+ }{8 {k^+}(k^+-q^+)(P^+-k^+) }
\nonumber\\
\times \frac{- \left( \slash{P} - \slash{k}\right)_{\rm on} + \slash{P} +m}
{P^- - \frac{{k^\perp}^2 + m^2}{2k^+} -
       \frac{{(P^\perp-k^\perp)}^2 + m^2}{2(P^+-k^+)}} \;
\frac{- \left( \slash{P} - \slash{k}\right)_{\rm on} + \slash{P} -\slash{q}+m}
{P^- - \frac{{(k^\perp-q^\perp)}^2 + m^2}{2(k^+-q^+)} -
       \frac{{(P^\perp-k^\perp)}^2 + m^2}{2(P^+-k^+)} - q^-},
\end{eqnarray}
\begin{eqnarray}
\label{blink2}
\epsfxsize=2.53cm \raisebox{-.5cm}{\epsffile{blink2.eps}}=
- 2 \pi i \;\int {\rm d}^2k^\perp
\int_{0}^{q^+} \frac{{\rm d}k^+ }{8 {k^+}(q^+-k^+)(P^+-k^+) }
\nonumber\\
\times \frac{\slash{k}_{\rm on} +m}
{P^- - \frac{{k^\perp}^2 + m^2}{2k^+} -
       \frac{{(P^\perp-k^\perp)}^2 + m^2}{2(P^+-k^+)}} \;
\frac{\slash{k}_{\rm on} - \slash{q} +m}
{q^- - \frac{{k^\perp}^2 + m^2}{2k^+} -
       \frac{{(k^\perp-q^\perp)}^2 + m^2}{2(q^+-k^+)}}.
\end{eqnarray} The diagram \r{blink1} is an example of a
``double'' blink. The total blink is the thick line between the two 
fat dots. 
For both blinks we see that the energy denominators are the same as for
usual LF time-ordered diagrams. However, the numerators are different. 
In the next subsection we show how the numerator of the blink is constructed.

\subsection{Construction of the blink}
As an example, we show how we can construct the blink~\r{blink2}.
For the fat line we have to substitute the propagating and instantaneous
part.
\begin{equation}
\label{blinkexp}
\epsfxsize=2.53cm \raisebox{-.5cm}{\epsffile{blink2.eps}}
=
\epsfxsize=2.53cm \raisebox{-.45cm}{\epsffile{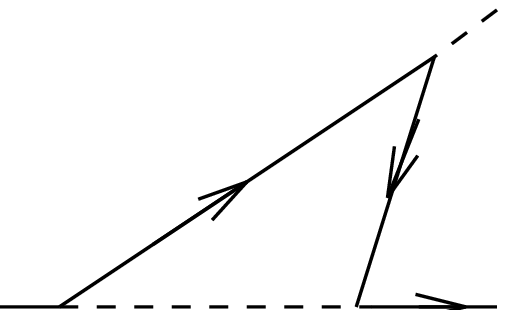}}
+
\epsfxsize=2.53cm \raisebox{-.45cm}{\epsffile{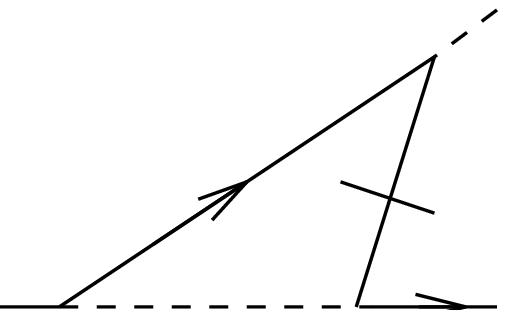}}.
\end{equation}
The propagating contribution is
\begin{eqnarray}
\label{blink2p}
\epsfxsize=2.53cm \raisebox{-.45cm}{\epsffile{curp.eps}}=
- 2 \pi i \;\int {\rm d}^2k^\perp
\int_{0}^{q^+} \frac{{\rm d}k^+ }{8 {k^+}(q^+-k^+)(P^+-k^+)}
\nonumber\\
\times \frac{\slash{k}_{\rm on} +m}
{P^- - \frac{{k^\perp}^2 + m^2}{2k^+} -
       \frac{{(P^\perp-k^\perp)}^2 + m^2}{2(P^+-k^+)}} \;
\frac{(\slash{k} - \slash{q})_{\rm on} +m}
{q^- - \frac{{k^\perp}^2 + m^2}{2k^+} -
       \frac{{(k^\perp-q^\perp)}^2 + m^2}{2(q^+-k^+)}},
\end{eqnarray}
and the instantaneous contribution, denoted by the perpendicular tag, is
\begin{eqnarray}
\label{blink2i}
\epsfxsize=2.53cm \raisebox{-.45cm}{\epsffile{curi.eps}}=
- 2 \pi i \;\int {\rm d}^2k^\perp
\int_{0}^{q^+} \frac{{\rm d}k^+ }{8 {k^+}(q^+-k^+)(P^+-k^+)}       
\nonumber\\
\times \frac{\slash{k}_{\rm on} +m}
{P^- - \frac{{k^\perp}^2 + m^2}{2k^+} -
       \frac{{(P^\perp-k^\perp)}^2 + m^2}{2(P^+-k^+)}} \;
       {\gamma^+}.
\end{eqnarray}
We see that both have a singularity at the upper boundary $q^+$ of the
integration interval over $k^+$.  These cancel in the sum: the blink
Eq.~\r{blink2}. It is obtained by making the denominators common for
the two diagrams. We can verify, using the relation $\gamma^+ \gamma^+ =
0$, that the lower boundary at $k^+=0$ does not cause any problems, neither
for the LF time-ordered diagrams, nor for the blink.

In an analogous way the double blink is constructed. It
 consists out of the following LF time-ordered diagrams:
\begin{equation}
\label{jul33}
\epsfxsize=2.53cm \raisebox{-.45cm}{\epsffile{blink1.eps}}=
\epsfxsize=2.73cm \raisebox{-.55cm}{\epsffile{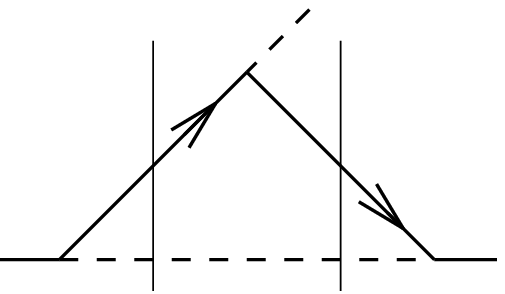}} +
\epsfxsize=1.903cm \raisebox{-.39cm}{\epsffile{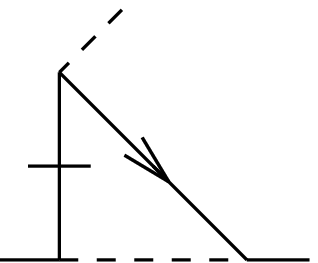}} +
\epsfxsize=1.903cm \raisebox{-.39cm}{\epsffile{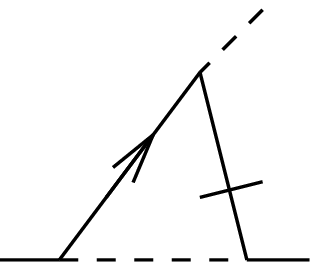}}.
\end{equation}
We see that one diagram is missing: the diagram with two instantaneous
fermions. Because it contains two neighboring $\gamma^+$ matrices, it
vanishes. It is an example of a forced instantaneous loop (FIL), which
is related to longitudinal divergences and will be discussed in the
next chapter. The LF time-ordered diagrams have the same integration
interval as the double blink~\r{blink1}. The last diagram on the
right-hand side is the same as the diagram in Eq.~\r{blink2i}, the only
difference being the integration range. The instantaneous fermions have been
'tilted' a little in these two diagrams, to indicate that the
integration interval is such that the instantaneous fermions carry
positive plus-momentum. The second diagram on the right-hand side of
Eq.~\r{jul33} has no tilted instantaneous fermion, because the 
integration range has not been split as in the previous case.
We will not give the formulas for the LF
time-ordered diagrams in Eq.~\r{jul33}, but we have verified that their
end-point singularities are removed when the diagrams are combined into
the double blink.

\subsection{The Breit-frame} 
What happens to the current in Eq.~\r{blink0} in the limit of $q^+
\downarrow 0$? Relying on the spectrum condition, one may expect that
diagrams like \r{blink2} disappear, and that only the double blink
\r{blink1} contributes.  This is confirmed by the fact that the
integration area of the single blink \r{blink2} goes to zero.  However,
it could be that the integrand obtains singularities in the limit $q^+
\downarrow 0$ that cause a nonzero result. We denote this limit in the
diagrams by drawing the line of the outgoing boson vertically.

For the numerator of the blink \r{blink2} we use the relation
\begin{equation}
\left(\slash{k}_{\rm on} +m \right)\left(\slash{k}_{\rm on} - \slash{q} +m
\right) = 2 m^2 + \slash{k}_{\rm on} ( 2 m - \slash{q}).
\end{equation}
The integral is dominated by factors $k^+$ and $(k^+ - q^+)$. 
Identifying these factors in \r{blink2} we find:
\begin{equation}
\label{Blink2}
\epsfxsize=2.53cm \raisebox{-.5cm}{\epsffile{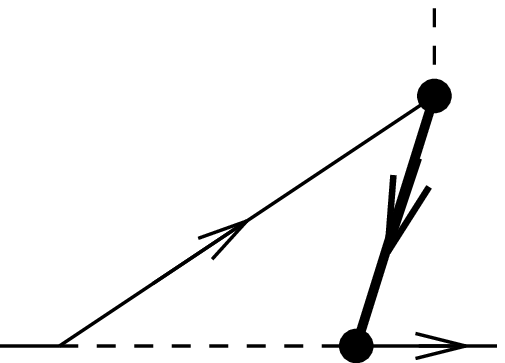}}=
-2 \pi i \;\int {\rm d}^2k^\perp
\int_{0}^{q^+} \frac{{\rm d}k^+ }{8 {k^+}(k^+-q^+)P^+}
\frac{\gamma^+ ( 2 m - \slash{q})}
{q^- - \frac{{k^\perp}^2 + m^2}{2k^+} -
       \frac{{(k^\perp-q^\perp)}^2 + m^2}{2(q^+-k^+)}}.
\end{equation}
We write Eq.~\r{Blink2} in internal coordinates $x=k^+/q^+$, and find that
the $q^+$ dependence on the integration range drops. Moreover, the
integration contains no singularities in the internal variable~$x$.
\begin{equation}
\epsfxsize=2.53cm \raisebox{-.5cm}{\epsffile{Blink2.eps}}=
- \pi i \;\int {\rm d}^2k^\perp
\int_{0}^{1} {\rm d}x \;
\frac{\gamma^+}{2 P^+} \;
\frac{ 2 m - \slash{q}}
{(k^\perp - x q^\perp)^2 + m^2 - x (1-x) q^2}.
\end{equation}
If we disregard for a moment the transverse integration, we see that
in the Breit-frame there is a finite contribution of 
pair-creation/annihilation to the current. This agrees with the result of
De Melo {\em et al.} \cite{MSFS98}. Furthermore, we see that it is not
covariant, and therefore its omission will not only lead to the wrong 
amplitude, but also to breaking of Lorentz covariance and rotational invariance.

\chapter{\label{chap3}Longitudinal divergences in the Yukawa model}
\def \fse{\epsfxsize=3cm \epsffile[-10 10 150 70]{fse.eps} }
\def \fsex{\epsfxsize=3cm \epsffile[-10 10 150 70]{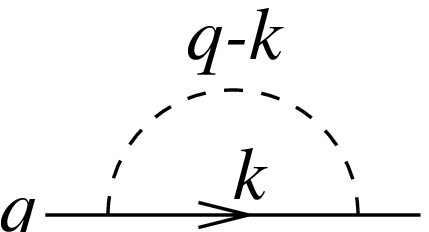} }
\def \bse{\epsfxsize=3cm \epsffile[-30 35 130 95]{bse.eps} }
\def \bsex{\epsfxsize=3cm \epsffile[-30 35 130 95]{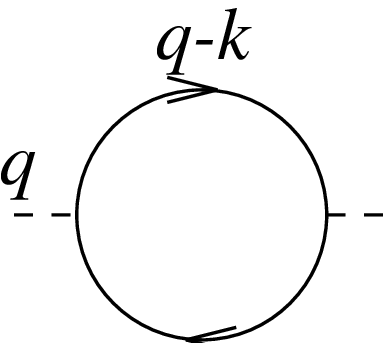} }
\def \bsep{\epsfxsize=3cm \epsffile[-30 45 130 105]{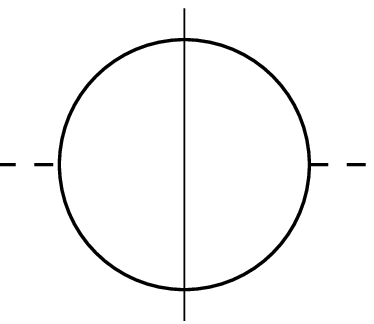} }
\def \bsei{\epsfxsize=3cm \epsffile[-30 25 130 085]{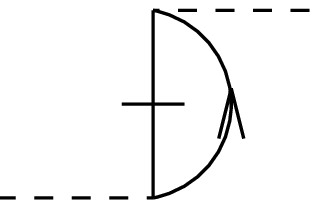} }
\def \obe{\epsfxsize=3cm \epsffile[ -10 40 150 110]{obe.eps}}
\def \obex{\epsfxsize=1.6cm \raisebox{-.97cm}{\epsffile{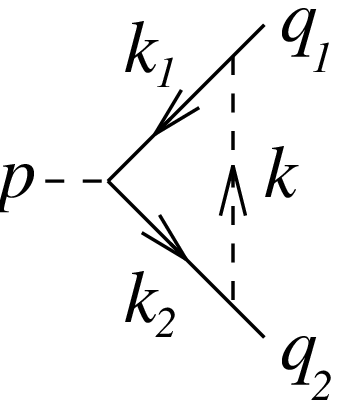}}}
\def \obepa{\epsfxsize=3cm \epsffile[ -10 50 150 120]{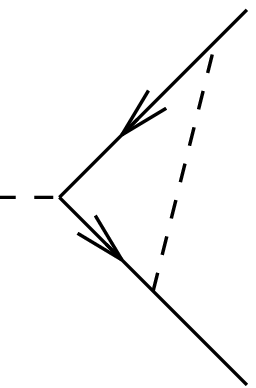} \hspace{-.8cm}}
\def \obepb{\epsfxsize=3cm \epsffile[ -10 50 150 120]{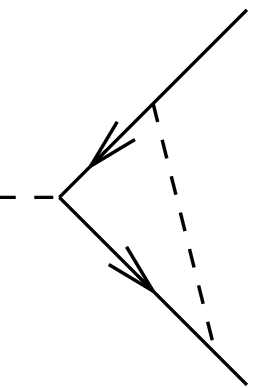} \hspace{-.8cm}}
\def \obeia{\epsfxsize=3cm \epsffile[ -10 50 150 120]{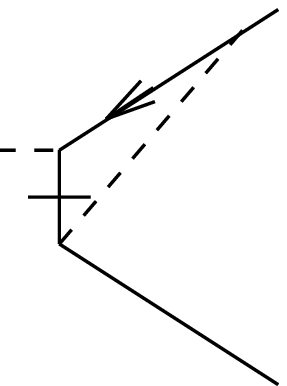} \hspace{-.8cm}}
\def \obeib{\epsfxsize=3cm \epsffile[ -10 50 150 120]{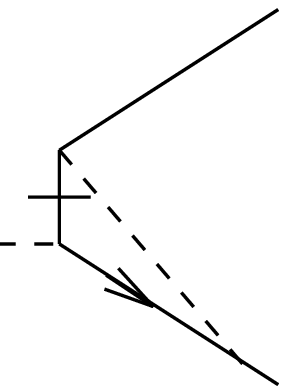} \hspace{-.8cm}}
\def \obeii{\epsfxsize=3cm \epsffile[ -10 50 150 120]{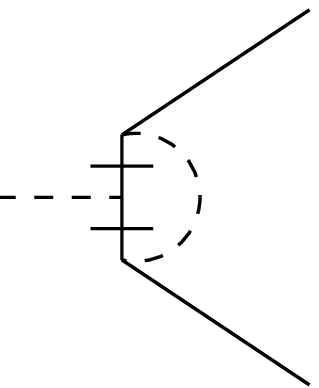} \hspace{-.8cm}}
\def \obebb{\epsfxsize=3cm \epsffile[ -10 50 150 120]{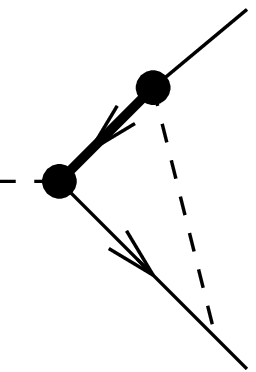} \hspace{-.8cm}}
\def \hfse{\epsfxsize=3cm \epsffile[-10 10 150 70]{hfse.eps} }
\def \htri{\epsfxsize=3cm \epsffile[-10 10 150 70]{htri.eps} }
\def \htria{\epsfxsize=3cm \epsffile[-10 10 150 70]{htria.eps} }
\def \htrib{\epsfxsize=3cm \epsffile[-10 10 150 70]{htrib.eps} }
\def \htric{\epsfxsize=3cm \epsffile[-10 10 150 70]{htric.eps} }
\def \fseprop{\epsfxsize=3cm \epsffile[-10 20 150 80]{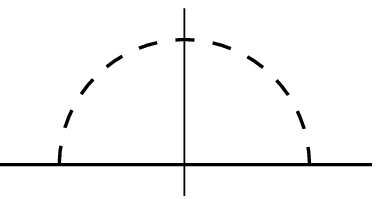} }
\def \fseinst{\raisebox{.12cm}{\epsfxsize=3cm \epsffile[-10 20 150 80]{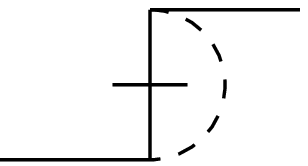} }}

\begin{quote}
{\em If the doors of perception were cleansed everything would appear
as it is, infinite.}

\raggedleft William Blake, The marriage of heaven and hell \cite{Ryd85}
\end{quote}

For a number of reasons mentioned in the previous chapters,
quantization on the light-front is nontrivial.
Subtleties arise that have no counterpart in ordinary time-ordered
theories. We will encounter some of them in this chapter and show
how to deal with them in such a way that covariance of the perturbation
series is maintained.

In LFD, or any other Hamiltonian theory, covariance is not manifest.
Burkardt and Langnau \cite{BL91} claimed that, even for scattering
amplitudes, rotational invariance is broken in naive light-cone
quantization (NLCQ).  In the case they studied, two types of infinities
occur: longitudinal and transverse divergences.  They regulate the
longitudinal divergences by introducing noncovariant counterterms. In
doing so, they restore at the same time rotational invariance.  The
transverse divergences are dealt with by dimensional regularization.

We would like to maintain the covariant structure of the Lagrangian and
take the path of Ligterink and Bakker \cite{LB95b}. 
Following Kogut and Soper~\cite{KS70} they derive rules
for LFD by integrating covariant Feynman diagrams over the LF energy
$k^-$.  For covariant diagrams where the $k^-$-integration is
well-defined this procedure is straightforward and the rules
constructed are, in essence, equal to the ones of NLCQ. However, when
the $k^-$-integration diverges the integral over $k^-$ must be
regulated first.  We stress  that it is important to do this in
such a way that covariance is maintained.

In this chapter, we will show that the occurrence of longitudinal
divergences is related to the so-called forced instantaneous loops
(FILs).  If these diagrams are included and renormalized in a proper
way we can give an analytic proof of covariance.  FILs were discussed
before by Mustaki {\em et al.} \cite{MPSW91}, in the
context of QED. They refer to them as {\em seagulls}. There are,
however, some subtle differences between their treatment of
longitudinal divergences and ours, which are explained in
\sec \ref{longfse}.

Transverse divergences have a different origin. However, they can
be treated with the same renormalization method as longitudinal
divergences.  We shall present an analytic proof of the equivalence of the
renormalized covariant amplitude and the sum of renormalized LF
time-ordered amplitudes in two cases, the fermion and the boson 
self-energy.  In the other cases we have to use numerical techniques. They
will be dealt with in Chapter~\ref{chap4}.

\section{Introduction}
In the previous chapters we already introduced instantaneous fermions.
For a discussion on longitudinal divergences they play an important
role. Without fermions there are no longitudinal divergences! The
longitudinal divergences can be both seen from a ``pictorial'' and
a mathematical point of view.  

\begin{figure}
\[
\epsfxsize=8cm \epsffile[-20 20 373 186]{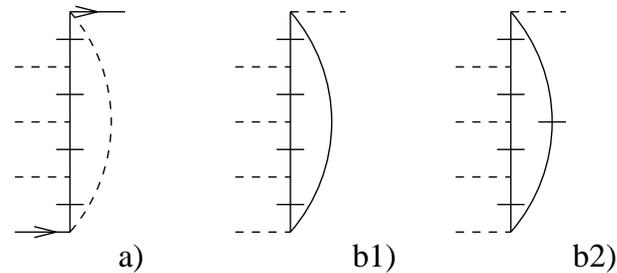}
\]
\caption{\label{fils}Examples of FILs. In (a) a boson in the
loop is forced to be instantaneous. In (b1) a fermion is obstructed in its
propagation. In (b2) all fermions are instantaneous.}
\end{figure}

The pictorial view is the following.
When a diagram contains a loop where all particles but one are
instantaneous, a conceptual problem occurs.  Should the remaining boson
or fermion be  interpreted as propagating or as instantaneous?  Loops
with this property are referred to as forced instantaneous loops (FILs).
Loops where all fermions are instantaneous are also considered as FILs.
However, they do not occur in the Yukawa model with (pseudo-)scalar
coupling. Examples of these three types of FILs are given in Fig.~\ref{fils}.

Mathematically this problem also shows up. The FILs
correspond to the part of the covariant amplitude where the
$k^-$-integration is ill-defined. The problem is solved in the
following way. First we do not count FILs as
LF time-ordered diagrams.  Second we find that this special type of
diagram disappears upon regularization if we use the method of 
Ligterink and Bakker~\cite{LB95a}: minus regularization.

\subsection{Minus regularization}
The minus-regularization scheme was developed for the 
purpose of maintaining the symmetries of the theory such that the
amplitude is covariant order by order.  It can be applied to Feynman
diagrams as well as to ordinary time-ordered or to LF time-ordered
diagrams. Owing to the fact that minus regularization is a linear
operation, it commutes with the splitting of Feynman
diagrams into LF time-ordered diagrams.

We explain very briefly how the method works. Consider a diagram defined by a
divergent integral. Then the integrand is differentiated with respect
to the external energy, say $q^-$, until the integral is well defined.  Next
the integration over the internal momenta is performed.  Finally the
result is integrated over $q^-$ as many times as it was differentiated before. 
This operation is the same as removing the lowest orders in the Taylor
expansion in $q^-$.  For example, if the two lowest orders of the Taylor
expansion with respect to the external momentum $q$ of a LF time-ordered diagram 
$\int {\rm d}^3 k {\cal F}(q, k)$ are
divergent, minus regularization is the following operation:
\begin{equation} \int_{\frac{q_\perp^2}{2 q^+}}^{q^-} {\rm d}q'
\int_{\frac{q_\perp^2}{2 q^+}}^{q'^-} {\rm d}q'' \int {\rm
d}^{2}k^\perp {\rm d}k^+ \left( \frac{\partial}{\partial
q''^-} \right)^2 {\cal F}(k, q'').
\end{equation}
The point $q^2=0$ is chosen in this example as the renormalization point.
This regularization method of subtracting the lowest order terms in the
Taylor expansion is similar to what is known in covariant perturbation
theory as BPHZ (Bogoliubov-Parasiuk-Hepp-Zimmermann) \cite{Col84}.
Some advantages of the minus-regularization scheme are preservation of
covariance and local counterterms.  Another advantage is that
longitudinal as well as transverse divergences are treated in the same
way.  A more thorough discussion on minus regularization can be found
in the next chapter.

\subsection{Proof of equivalence for the Yukawa model}

The proof of equivalence will not only hold order by order in the
perturbation series, but also for every covariant diagram separately.
In order to allow for a meaningful comparison with the method of
Burkardt and Langnau we apply our method to the same model as they
discuss, the Yukawa model, as introduced in Chapter~\ref{chap2}.  

\begin{figure}
\[
\epsfxsize=11.5cm \epsffile[-30 0 570 450]{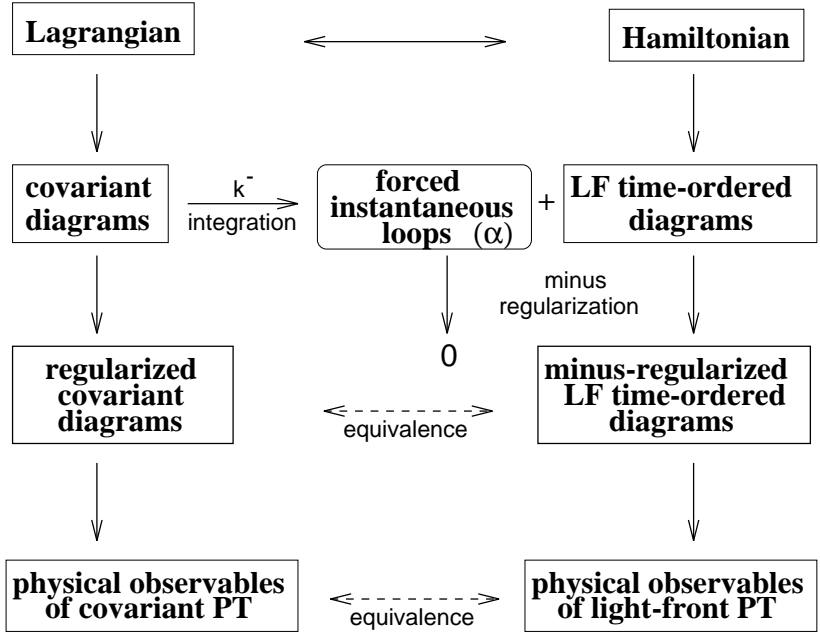}
\]
\vspace{-1cm}
\caption{\label{schema}Outline of the proof of equivalence for diagrams
with longitudinal divergences.}
\end{figure}

In this model we have to distinguish four  types of diagrams,
according to their longitudinal ($D^-$) and transverse degrees 
($D^\perp$) of divergence.
These divergences were classified also in
Table~\ref{tabYukdiv} on page~\pageref{tabYukdiv}. The
proof of equivalence is illustrated in Fig.~\ref{schema}.

We integrate an arbitrary covariant diagram over LF
energy.  For longitudinally divergent diagrams this integration is
ill-defined and results in FILs.  A regulator
$\alpha$ is introduced which formally restores equivalence.  Upon minus
regularization the $\alpha$-dependence is lost and the transverse
divergences are removed.
We can distinguish

\begin{itemize}
\item 
{\em Longitudinally and transversely convergent diagrams} ($D^-<0, D^\perp<0$).
No  FILs will be generated. No regularization is needed. The LF time-ordered
diagrams may contain $1/k^+$-poles, but these can be removed using blinks.
A rigorous proof of equivalence for this class
of diagrams is given by Ligterink and Bakker~\cite{LB95b}. 

\item
{\em Longitudinally convergent  diagrams} ($D^-<0$) {\em with a transverse
divergence} ($D^\perp \geq 0$).
In the Yukawa model there are three such diagrams: the four fermion box,
the fermion triangle and the one-boson exchange correction. Again, no
FILs occur. Their transverse divergences and therefore the proof of
equivalence will be postponed until Chapter~\ref{chap4}. However,
because the one-boson exchange correction illustrates the concept of
$k^-$-integration, the occurrence of instantaneous fermions and the
construction of blinks, it will be discussed as an example in
\sec \ref{secobeiii}. In Chapter~\ref{chap1} we gave an example for a 
longitudinally convergent diagram in $\phi^3$ theory: the
electro-magnetic current.

\item
{\em Longitudinally divergent diagrams} ($D^-=0$) {\em with a
logarithmic transverse divergence} ($D^\perp=0$).
In the Yukawa model with a scalar coupling there is one such diagram:
the fermion self-energy.  Upon splitting the fermion propagator two
diagrams are found. The troublesome one is the diagram containing the
instantaneous part of the fermion propagator.  According to our 
definition it is a FIL and needs a regulator. In \sec \ref{longfse} we show
how to determine the regulator $\alpha$ that restores covariance
formally.  Since $\alpha$ can be chosen such that it does not depend on
the LF energy, the FIL will vanish upon minus regularization.

\item 
{\em Longitudinally divergent diagrams with a quadratic transverse
divergence} ($D^\perp=2$).
In the Yukawa model only the boson self-energy is in this class.  We
are not able to give an explicit expression for $\alpha$.  However, in
\sec \ref{secbse} it is shown that the renormalized boson self-energy
is equal to the corresponding series of renormalized LF time-ordered
diagrams. This implies that the contribution of FILs has again
disappeared after minus regularization.

\end{itemize}

\section{Example: the one-boson exchange correction}
\label{secobeiii}
We will give an example of the construction of the LF time-ordered
diagrams, the occurrence of instantaneous fermions and the construction
of blinks.
It concerns the correction to the boson--fermion--anti-fermion
vertex due to the exchange of a boson by the two outgoing fermions.
Here, and in the sequel, we drop the dependence on the coupling constant
and numerical factors related to the symmetry of the Feynman diagrams.

A boson of mass $\mu$ with momentum $p$ decays into a fermion anti-fermion 
pair with momenta $q_1$ and $q_2$ respectively. 
The covariant amplitude for the boson exchange correction can be written as

\begin{equation}
\label{obe1i}
\obex  = \int_{\rm Min}  
\frac{{\rm d}^4k \;\; (\slash{k}_1 + m)(\slash{k}_2 + m)}
{(k_1^2-m^2- i \epsilon) (k_2^2  - \! m^2  -  \!i \epsilon)
(k^2  -  \mu^2  -  i \epsilon)} .
\end{equation}
The subscript Min denotes that the integration is over Minkowski space.
The momenta $k_1$ and $k_2$ indicated in the diagram are given by
\begin{equation}
k_1 = k - q_1 , \;\;\; k_2 = k + q_2 .
\end{equation}
We can rewrite Eq.~\r{obe1i} in terms of LF coordinates
\vspace{-.5cm}
\begin{equation} \label{obe2i} \obe  \hspace{-1.4cm} = \int
 \frac{{\rm d}^2k^\perp {\rm d}k^+ {\rm d}k^- (\slash{k}_1 +
m)(\slash{k}_2 + m)}
{8 k_1^+ \! k_2^+ {k}^+
(k^-\!- H_1^-)(k^-\!- H_2^-)(k^-\!- {H}^-)},
\end{equation}

\noindent
where the poles in the complex $k^-$-plane are given by 
\begin{eqnarray}
\label{pole1bla}
H^-    &=& \frac{{k^\perp}^2 + \mu^2 - i \epsilon}{2k^+} , \\
\label{pole2bla}
H^-_1  &=& q^-_1-\frac{{k_1^\perp}^2 + m^2 - i \epsilon}{2k^+_1} , \\
\label{pole3bla}
H^-_2  &=& -q^-_2+\frac{{k_2^\perp}^2 + m^2 - i \epsilon}{2k^+_2} . 
\end{eqnarray}
We now show how the LF time-ordered diagrams, including
those containing instantaneous terms, can be constructed. 
The LF time-ordered diagrams contain on-shell spin projections
in the numerator. They are
\begin{equation}
\slash{k}_{\!\!\!\!i\;\rm on} = k_{i\;\rm on}^- \gamma^+ + k_i^+ \gamma^- 
- k_i^\perp \gamma^\perp .
\end{equation}
We also use the following relation:
\begin{equation}
k^- - H_i^- = k^-_i - k_{i\;\rm on}^- .
\end{equation}
We rewrite the numerator
\begin{eqnarray}
(\slash{k}_{1} + m)(\slash{k}_{2} + m) &=&
((k^-\!-\!H^-_1) \gamma^+ + (\slash{k}_{1\rm on} + m))\nonumber\\
&\times&
((k^-\!-\!H^-_2) \gamma^+ + (\slash{k}_{2\rm on} + m)) .
\end{eqnarray}
This separation allows us to write Eq.~\r{obe2i} as
\begin{eqnarray}
\hspace{-.9cm} \obe \hspace{-1.5cm}  = \int
\frac{{\rm d}^2k^\perp {\rm d}k^+ {\rm d}k^-}
{8 k_1^+ \! k_2^+ {k}^+}\hspace{-.5cm}
&&\left\{ 
\frac{\gamma^+ \gamma^+}{(k^-\! \!-\! H^-)}
+ 
\frac{(\slash{k}_{1\rm on} + m)
(\slash{k}_{2\rm on} + m) }
{(k^-\! \!-\! H_1^-)(k^-\! \!-\! H_2^-)(k^-\! \!-\! {H}^-)} 
\right.\nonumber\\ 
&&\left. 
+
\frac{\gamma^+ (\slash{k}_{2\rm on} + m) }
{(k^-\! \!-\! H_2^-)(k^-\! \!-\! H^-)}
+
\frac{ (\slash{k}_{1\rm on} + m) \gamma^+}
{(k^-\! \!-\! H_1^-)(k^-\! \!-\! H^-)} 
\label{obe3}
\right\}.
\end{eqnarray}
The splitting corresponds to the splitting of the covariant amplitude
into LF time-ordered diagrams.  The numerators are written in such a form
that  Cauchy's formula can be applied easily to the $k^-$-integration.
Only for the first term of Eq.~\r{obe3} can $k^-$ contour integration 
not be applied because the semi-circle at infinity gives a nonvanishing
contribution. Such a singularity corresponds to a pole at infinity.
However, we are saved by the fact that $\gamma^+\gamma^+=0$.  Therefore
we obtain for  the first term of Eq.~\r{obe3}

\vspace{-.3cm} 
\begin{equation}
\obeii = \; 0 .
\end{equation}
\vspace{.3cm} 

\noindent
Here the bars in the two internal fermion lines again denote
instantaneous terms. This forces the boson line to be instantaneous
too.  We see that this diagram is a FIL according to the definition we
gave in the previous section.  The longitudinal divergences which occur
due to such diagrams are discussed in the next sections. Since FILs are
not LF time-ordered diagrams, the rules given by NLCQ do not apply.

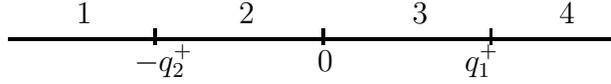
\begin{figure}
\[
\setlength{\unitlength}{0.005500in}
\begin{picture}(580,34)(100,520)
\thicklines
\put(100,540){\line( 1, 0){580}}
\put(240,550){\line( 0,-1){ 15}}
\put(400,550){\line( 0,-1){ 15}}
\put(560,550){\line( 0,-1){ 15}}
\put(220,510){$- q^+_2$}
\put(395,510){$0$}
\put(535,510){$q^+_1$}
\put(165,555){1}
\put(320,555){2}
\put(485,555){3}
\put(625,555){4}
\end{picture}
\]
\caption{\label{intervals}Regions for the $k^+$-integration. At the boundaries 
a pole crosses the real $k^-$-axis. }
\end{figure}

The second term of Eq.~\r{obe3} contains only propagating parts. It has
three poles \r{pole1bla}-\r{pole3bla}. We are free to close
the contour either in the lower or in the upper half plane. The poles
do not always lie on the same side of the real $k^-$-axis. For example,
the pole given in Eq.~\r{pole1bla} is in the upper half plane for
$k^+<0$.  At $k^+=0$ it changes side. In Fig.~\ref{intervals} we show
the four intervals that can be distinguished.

In region 1 all poles lie above the real $k^-$-axis. By closing the contour
in the lower half plane we see that the integral vanishes. 
At $k^+=-q^+$ the pole \r{pole3bla} crosses the real axis. 
In interval 2 the integral is proportional to its residue: 
\begin{eqnarray}
\label{obepb}
\obepb \hspace{-.4cm} = 
 -2 \pi i \int {\rm d}^2k^\perp \int_{-q_2^+}^0
\frac{ {\rm d}k^+ }
{8 k_1^+ \! k_2^+ {k}^+}
\; \frac{(\slash{k}_{1\rm on} + m)
(\slash{k}_{2\rm on} + m) }
{(H_1^-\! \!-\! H_2^-)(H^-\! \!-\! H_2^-)} .\\ \nonumber
\end{eqnarray}
No vertical lines are drawn since this is clearly a LF time-ordered
diagram. The factor $(H_1^- \!- H_2^-)^{-1}$ is the energy
denominator corresponding to the fermion--anti-fermion state between the
moment in LF time that the boson decays and the moment that the
exchanged boson is emitted.  $(H^-\!- H_2^-)^{-1}$ is the energy
denominator corresponding to the state in the period that the exchanged
boson exists.

At $k^+=0$ a second pole crosses the real axis. For positive $k^+$ we
close the contour in the upper half plane. Here only one pole \r{pole2bla}
is present.  The result is
\begin{eqnarray}
\label{obepa} 
\obepa \hspace{-.4cm} = 
 2 \pi i \int {\rm d}^2k^\perp \int_0^{q_1^+}
\frac{ {\rm d}k^+ }
{8 k_1^+ \! k_2^+ {k}^+}
\; \frac{(\slash{k}_{1\rm on} + m)
(\slash{k}_{2\rm on} + m) }
{(H_1^-\! \!-\! H_2^-)(H_1^-\! \!-\! H^-)} .\\ \nonumber
\end{eqnarray}
Only the second energy denominator differs from the one in Eq.~\r{obepb}. 

The terms of Eq.~\r{obe3} with one instantaneous term are easier to
determine.  There are two poles and a contribution only occurs if
the poles are on different sides of the real $k^-$-axis.  The third
term of Eq.~\r{obe3} is

\begin{equation}
\label{obeib}
\obeib = 
  - 2 \pi i \int {\rm d}^2k^\perp \int_{-q_2^+}^0
\frac{ {\rm d}k^+ }
{8 k_1^+ \! k_2^+ {k}^+}
\frac{\gamma^+ 
(\slash{k}_{2\rm on} + m) }
{H^-\! \!-\! H_2^-} .
\end{equation}
\vspace{.5cm}

For the fourth and last term of Eq.~\r{obe3} we have
\begin{equation}
\label{obeia}
\obeia = 
  2 \pi i \int {\rm d}^2k^\perp \int_0^{q^+_1}
\frac{ {\rm d}k^+ }
{8 k_1^+ \! k_2^+ {k}^+}
\frac{
(\slash{k}_{1\rm on} + m) \gamma^+ }
{H_1^-\! \!-\! H^-} .
\end{equation}
\vspace{.5cm}

The possible $1/k^+$ poles inside the integration area can be removed using
the blinks. 

\begin{equation}
\obebb = \obepb + \obeib.
\end{equation}
\vspace{.5cm}

Using Eqs.~\r{obepb} and \r{obeib} we get 
\begin{eqnarray}
\label{obebbi}
\obebb \hspace{-.5cm} =
- 2 \pi i \int {\rm d}^2k^\perp \int_{-q_2^+}^0
\frac{ {\rm d}k^+ }
{8 k_1^+ \! k_2^+ {k}^+}
\; \frac{(\slash{k}_{2\rm on} - \slash{p} + m)
(\slash{k}_{2\rm on} + m) }
{(H_1^-\! \!-\! H_2^-)(H^-\! \!-\! H_2^-)} . \\ \nonumber
\end{eqnarray}
The other blink is constructed in the same way.

We have now succeeded in doing the $k^-$-integration and have rewritten
the covariant expression for the one-boson exchange correction \r{obe1i}
in terms of LF time-ordered diagrams. The result is
\vspace{-.3cm} 
\begin{equation}
\obe \hspace{-.9cm} = \obepb + \obepa + \obeib + \obeia.
 \label{eq20}
\end{equation}
\vspace{.2cm} 
  
Diagrams with instantaneous parts are typical for LFD. 
There is another difference with equal-time
PT. Of the six possible time-orderings of the triangle diagram two have
survived, which give rise to two diagrams each, upon splitting the
fermion propagators into instantaneous and propagating parts.  This
reduction of the number of LF time-ordered diagrams compared to ordinary
time-ordered ones is well known in LFD, and explained in detail 
in Ref.~\cite{LB95b}.

All the calculations in this section were purely algebraic. The
formulas for the LF time-ordered diagram we derived are the same as those
given by NLCQ. The integrals that remain are logarithmically divergent
in the transverse direction and must be regularized. 
This calculation will be done in Chapter~\ref{chap4} in which
we discuss transverse divergences. 

\section{Equivalence of the fermion self-energy}
\label{longfse}
There are two longitudinally divergent diagrams in the Yukawa model. We
first discuss the fermion self-energy. For our discussion the location
of the poles is not relevant and therefore we ignore the $i \epsilon$
term.  For a fermion momentum $q$ we have the following self-energy
amplitude:
\begin{equation}
\label{fse1}
\raisebox{-.2cm}{\fsex} \hspace{-.5cm}
= \int_{\rm Min}  \frac{{\rm d}^4k \;\;\; (\slash{k} + m)}
{(k^2 - m^2 ) ((q\!-\!k)^2 - \mu^2 )} .
\end{equation}
\subsection{Covariant calculation}
We introduce a Feynman parameter $x$ and change the integration
variable to $k'$ given by $k=k'+xq$ in order to complete the square in
the denominator. This gives
\begin{equation}
\label{fse2}
\fse \hspace{-.5cm}=
\int_0^1 \!{\rm d}x\! \int_{\rm Min} 
\frac{ {\rm d}^4k' \;\;\; ( \slash{k}' + x\!\! \slash{q} + m)}
{\left( k'^2 - (1\!-\!x) m^2 - x \mu^2 + x(1\!-\!x) q^2 \right)^2} .
\end{equation}
The integral \r{fse2} is ill-defined. 
The appearance of $\not\!\!{k'}$ in the numerator
causes the integral to be divergent in the minus-direction 
and obstructs the Wick rotation. However, this term is odd
and is removed in accordance with common practice \cite{Col84}.
Wick rotation gives then
\begin{equation}
\label{fse3}
\fse \hspace{-.5cm}= i
\int_0^1 \!{\rm d}x\! \int
\frac{ {\rm d}^4k' \;\;\; ( x\!\! \slash{q} + m)}
{\left( k'^2 + (1\!-\!x) m^2 + x \mu^2 - x(1\!-\!x) q^2 \right)^2} .
\end{equation}
The subscript Min is dropped denoting that the integration is over Euclidean
space.  From Eq.~(\ref{fse3}) we can immediately infer that the fermion
self-energy has the covariant structure
\begin{equation}
\fse \hspace{-.5cm}= \slash{q} \; F_1(q^2) + m \;  F_2(q^2) .
\end{equation}
\subsection{Residue calculation}
To obtain the LF time-ordered diagram and the FIL corresponding to
the fermion self-energy we perform the $k^-$-integration by 
doing the contour integration:
\begin{equation}
\label{fse4}
\fse \hspace{-.5cm} =
\int \frac{{\rm d}^2k^\perp {\rm d}k^+ {\rm d}k^-}{4 k^+ (q^+\!-\!k^+)}
\frac{ k^- \gamma^+ + k^+ \gamma^- - k^\perp \gamma^\perp + m}
{(k^- - H_1^-)(k^- - H_2^-)} , 
\end{equation}
with the following poles:
\begin{eqnarray}
\label{h1}
H_1^- &=& \frac{{k^\perp}^2 + m^2 }{2 k^+} , \\
\label{h2}
H_2^- &=& q^- - \frac{(q^\perp - k^\perp)^2 + \mu^2 }{2 (q^+ - k^+)} .
\end{eqnarray}
We rewrite Eq.~\r{fse4} as
\begin{eqnarray}
\fse \hspace{-.2cm} \hspace{-.4cm}
&=& \int \frac{{\rm d}^2k^\perp {\rm d}k^+ {\rm d}k^-}{4 k^+ (q^+\!-\!k^+)}\;
\frac{ H_1^- \gamma^+ \! + k^+ \gamma^- \! - k^\perp \gamma^\perp \! + m}
{(k^- - H_1^-)(k^- - H_2^-)}\nonumber\\
&+& \int \frac{{\rm d}^2k^\perp {\rm d}k^+ {\rm d}k^-}{4 k^+ (q^+\!-\!k^+)}\;
\frac{ \gamma^+ (k^- - H_1^-)}{(k^- - H_1^-)(k^- - H_2^-)} \label{fseres} .
\end{eqnarray}
The first term of Eq.~\r{fseres} is the part that gives a convergent
$k^-$-integration. 
The second term contains the divergent part. This separation can
also be written in terms of diagrams: 
\begin{equation}
\label{fselftoexp}
\fse \hspace{-.3cm} = \fseprop \hspace{-.3cm} + \fseinst \hspace{-1cm}.
\end{equation}
\vspace{.4cm}

\noindent
The propagating diagram is
\begin{eqnarray}
\label{fseprop}
\hspace{-.6cm }\fseprop \hspace{-.8cm} =
2 \pi i \int {\rm d}^2k^\perp 
\int_0^{q^+} \frac{{\rm d}k^+}{4 k^+ (q^+ - k^+)}
\; \frac{  \frac{m^2 + {k^\perp}^2}{2 k^+} \gamma^+ 
+ k^+ \gamma^- - k^\perp \gamma^\perp + m}
{H_2^- - H_1^-} . \\ \nonumber
\end{eqnarray}
It has the usual form for a LF time-ordered diagram. It is divergent because of 
the $1/k^+$ singularity in the numerator. To shed more light on the structure
of this formula we introduce internal variables $x$ and $k'^\perp$:
\begin{equation}
\label{intvar}
x = \frac{k^+}{q^+} , \;\;
k'^{\perp} = k^\perp - x q^\perp .
\end{equation}
The denominator is now a complete square and we drop as usual the terms odd in 
$k'^\perp$ in the numerator. Then we find
\begin{eqnarray}
\label{fseprop2}
\fseprop \hspace{-.8cm} =
\pi i \int {\rm d}^2k'^\perp 
\int_0^1 {\rm d}x
\; \frac{\frac{m^2 + {k'^\perp}^2 - x^2 q^2}{2 x q^+}\gamma^+ 
+ x \!\!\slash{q} + m}
{{k'^\perp}^2 + (1\!-\!x) m^2 + x \mu^2 - x(1\!-\!x) q^2 } .
\\ \nonumber
\end{eqnarray}
The FIL  is
\begin{equation}
\label{fseinst}
\fseinst \hspace{-.3cm} = \int \frac{{\rm d}^2k^\perp {\rm d}k^+ {\rm d}k^-}
{4 k^+ (q^+\!-\!k^+)} \;
\frac{ \gamma^+}{k^- - H_2^-} .
\end{equation}
It contains the divergent part of the $k^-$-integration and a $1/k^+$
singularity too. The single bar in
Eq.~\r{fseinst} stands for an instantaneous part. The diagram is
instantaneous because it does not depend on the external energy $q^-$.
In order to demonstrate this we shift $k^-$ by $q^-$. Then we see that
the dependence on $q^-$ disappears.  However, this way of reasoning
is dangerous since the integral is divergent.
We make the integral well-defined by
inserting a function ${\cal R}$ containing a regulator $\alpha$:

\begin{equation}
\label{regulator}
{\cal R} = 
\left( \frac{\alpha(k^+)}{1 - i \delta q^+ k^-} +
       \frac{1 - \alpha(k^+)}{1 + i \delta q^+ k^-} \right) .
\end{equation}
If we choose $\alpha = 1$ for $k^+<0$ and $\alpha = 0$ for
$k^+ > q^+$, the extra pole only contributes
for $0<k^+<q^+$. In other words, then the spectrum condition is also satisfied
for all lines in the FIL.  This is convenient,
but not necessary. Mustaki {\em et al.}~\cite{MPSW91} do not require the spectrum condition to be
fulfilled for instantaneous particles.  
They have as integration boundaries for the FIL $0<k^+<\infty$.

We perform the $k^-$-integration and take the limit $\delta 
\rightarrow 0$.  This gives
\begin{equation}
\label{fseinst2}
\fseinst \hspace{-.3cm} = 2 \pi i \int {\rm d}^2k^\perp \int_0^{q^+}{\rm d}k^+ 
\frac{\gamma^+ \alpha(k^+)}{4 k^+ (q^+\!-\!k^+)} .
\end{equation}
Using internal variables (\ref{intvar}) we obtain
\begin{equation}
\label{fseinst3}
\fseinst \hspace{-.3cm} = \pi i \frac{\gamma^+}{2 q^+}
 \int {\rm d}^2k'^\perp \int_0^1{\rm d}x
\frac{\alpha(x)}{x (1-x)} .
\end{equation}

\subsection{Equivalence}
\label{III.C}
The FIL is not a LF time-ordered diagram. We think it is
a remnant of the problems encountered in quantization on the
light-front.  We require it to satisfy two conditions:

\begin{enumerate}
\item \label{cond1}
      the FIL has to restore covariance and equivalence
      of the full series of LF time-ordered diagrams;
\item \label{cond2}
      the FIL has to be a polynomial in $q^-$. 
\end{enumerate}
The first condition will also ensure that the FIL
contains a $1/k^+$ singularity that cancels a similar singularity in
the propagating diagram.  The second condition is that the
FIL is truly instantaneous; i.e., it does not contain
$q^-$ in the denominator like a propagating diagram.  To find the
form of the FIL that satisfies these conditions we
calculate
\begin{equation} \label{e1200} 
\fse \hspace{-.3cm} - \hspace{.2cm} \fseprop .
\end{equation} 
where we take for the covariant diagram Eq.~\r{fse3}.  This is a
strictly formal operation.  The covariant diagram is a $4$-dimensional
integral, whereas the propagating diagram has only 2 dimensions (not
counting the $x$-integration).  We can calculate Eq.~\r{e1200} without
evaluation of the integrals.  In Appendix~\ref{appeuclint} useful relations
are derived between $d$- and $(d\!-\!2)$-dimensional integrals.  Upon
using them we obtain
\begin{eqnarray}
\fse \hspace{-.5cm} - \fseprop \hspace{-.4cm}
= - \pi i \frac{\gamma^+}{2 q^+}
\int {\rm d}^2k'^\perp \int_0^1{\rm d}x \nonumber\\
\times \frac{m^2 + {k'^\perp}^2 - x^2 q^2}
{x \left({k'^\perp}^2 + (1\!\!-\!\!x) m^2 + x \mu^2 - x(1\!\!-\!\!x) q^2
\right)} .
\end{eqnarray}
This can be rewritten as
\begin{eqnarray}
\fse \hspace{-.5cm} - \fseprop \hspace{-.4cm}
= - \pi i \frac{\gamma^+}{2 q^+}
\int {\rm d}^2k'^\perp \int_0^1{\rm d}x \nonumber\\
\times \left( \frac{1}{x} + \frac{m^2 - \mu^2 + (1-2x) q^2}
{{k'^\perp}^2 + (1\!\!-\!\!x) m^2 + x \mu^2 - x(1\!\!-\!\!x) q^2 } 
\label{e1347}
\right).
\end{eqnarray}
The dependence on $q^2$ is limited to the second term. The integral
over $x$ of the latter can be done explicitly, whence one finds that
the integral is independent of $q^2$.  Therefore we can take $q^2=0$ in
Eq.~\r{e1347}.
\begin{eqnarray}
\fse \hspace{-.5cm} - \fseprop \hspace{-.4cm}
= - \pi i \frac{\gamma^+}{2 q^+}
\int {\rm d}^2k'^\perp \int_0^1{\rm d}x \nonumber\\
\times \left( \frac{1}{x} + \frac{m^2 - \mu^2}
{{k'^\perp}^2 + (1\!\!-\!\!x) m^2 + x \mu^2} 
\right) .
 \label{e1348}
\end{eqnarray}
This is a good moment to see if we can satisfy the two conditions
we put forward in the beginning of this subsection. 

The first condition is satisfied if the right-hand sides of Eqs.~\r{e1348} and
\r{fseinst3} are equal.
We can verify that there is an infinite number of solutions for
$\alpha$ to make this happen.  We are free to choose $\alpha$ to be
$q^-$-independent. This will make formula \r{fseinst3} also independent
of $q^-$. Then the second condition is trivially satisfied.

\subsection{Conclusions}

\begin{figure}
\[
\epsfxsize=10.5cm \epsffile[130 600 380 700]{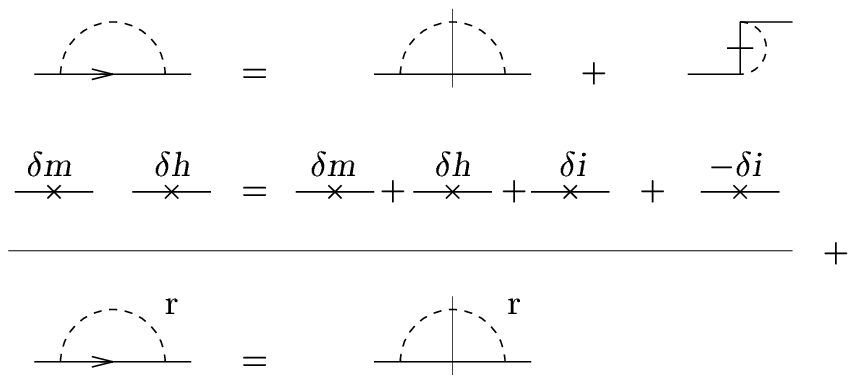}
\]
\caption{\label{fig1}Addition of the counterterms. The result is the minus-regularized
fermion self-energy.}
\end{figure}

Our renormalization method is visualized in Fig.~\ref{fig1}. 

There are two
noncovariant counterterms ($\delta i$). One of them occurs in the 
LF time-ordered part; the other one is associated with a self-induced 
inertia. Minus regularization
guarantees that they cancel provided the regulator $\alpha$ is chosen
appropriately. The other counterterms $\delta m$ and $\delta h$ are
covariant.  After the (infinite) counter\-terms have been added the
renormalized amplitude (denoted by the superscript $r$) remains. An
illustration of the full procedure of minus regularization is given in
the next section. 

We take another look at Fig. \ref{fig1}.  The first line contains three
ill-defined objects. The covariant amplitude \r{fse1} has a Minkowskian
measure and contains odd terms.  Divergent odd terms are dropped as
part of the regularization procedure.  To calculate the LF time-ordered
diagram \r{fseprop} we also dropped surface terms.  Can these
assumptions be justified? Would another set of assumptions give
different physical amplitudes? We conjecture that any set of
assumptions corresponds to a certain class of choices for $\alpha$. The
$\alpha$-dependence is only present in the FILs. In
the process of minus regularization the $\alpha$-dependence is lost, as
we see for the fermion self-energy in Fig. \ref{fig1}. Therefore the
physical observables do not depend on the assumptions we started out
with.

Finally we give the result for the fermion self-energy.
\begin{eqnarray}
\fseprop \put(-32,11){r} \hspace{-.4cm} = - \pi^2 i \int_0^1 {\rm d}x
\; (x \!\!\slash{q} + m)
\log \left(1 - \frac{x (1\!-\!x) q^2}{(1\!-\!x)m^2 + x \mu^2} \right).
\label{fsepropr} \\ \nonumber
\end{eqnarray}

\begin{figure}
\[
\epsfxsize=16cm \epsffile[165 300 535 450]{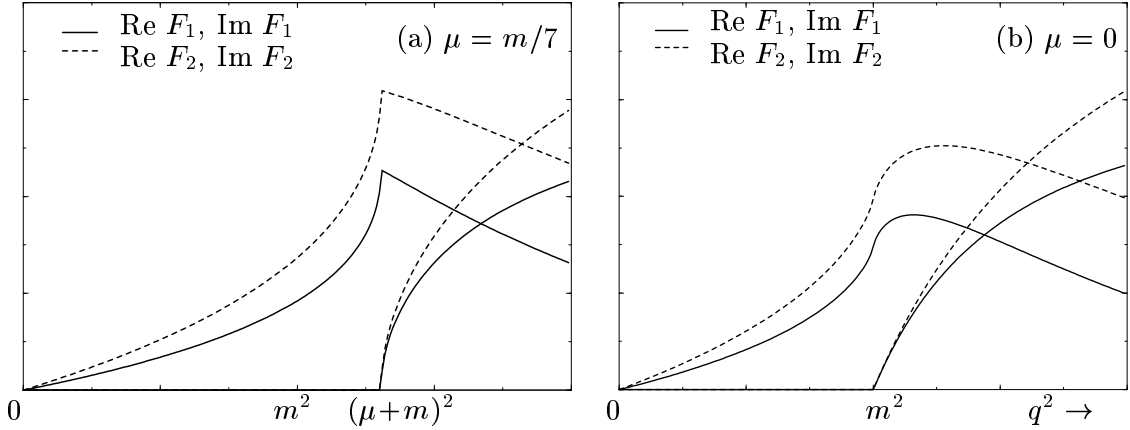}
\]
\caption{The renormalized fermion self-energy.
         The left-hand panel (a) shows the case $\mu = m/7$; 
         the right-hand panel (b) is for $\mu = 0$.}
 \label{plotfse}
\end{figure}

This integral can be done analytically, but the result is a rather long
formula, which we give in Appendix~\ref{exactfse}. Here we display the
result in pictorial form. Fig.~\ref{plotfse} shows $F_1$ and $F_2$ for
values of the fermion momentum squared in the range $q^2 \in [0, 2m^2]$
for the case of a massless boson and the case where $\mu = m/7$, corresponding 
to the self-energy correction for a nucleon due to a scalar pion. The case 
$\mu = 0$ is included because it was calculated before by 
Ligterink and Bakker~\cite{LB95a}.

The threshold behavior in the two cases is clearly seen in this figure.
Above threshold, $q^2 > (m + \mu)^2$, the self-energy 
becomes complex.

We have verified that
our result is in agreement with the result given by dimensional regularization 
and the result given by Bjorken and Drell \cite{BD64}, using
Pauli-Villars regularization. 

For the following reasons our analysis differs essentially from the
analysis of Mustaki {\em et al.}~\cite{MPSW91}. First of all, we make an
explicit distinction between LF time-ordered diagrams and FILs.
Second, we make the integration over the longitudinal coordinates
well-defined by introducing a regulator $\alpha(k^+)$.  Mustaki {\em et al.}
make the $k^+$-integration well-defined by using cutoffs. The form of
the cutoffs depends on the regularization scheme of the divergences in
the transverse directions. In our calculation the form of $\alpha(k^+)$
is determined by requiring equivalence to the covariant calculation.
In our opinion, this is the most important constraint on the FIL.  We
do not think that the cutoffs can always be determined from an analysis
of the transverse divergences. For example, in two dimensions ($D=1+1$)
there are no transverse divergences, but longitudinal divergences are
still present and $\alpha(k^+)$ has to ensure that covariance is
maintained. Moreover, in $D=1+1$ the covariant calculation of the
fermion self-energy gives a finite result.  Our choice of
$\alpha(k^+)$, independent of $k^\perp$, ensures also in this case that
the LF time-ordered calculation reproduces the covariant result.  The
same is true for the calculation by Mustaki {\em et al.} if they make a
particular choice for the cutoffs.

\section{Equivalence of the boson self-energy}
\label{secbse}
Our analysis of the boson self-energy serves two purposes. First of all
it illustrates in detail the concept of minus regularization.
Second it concludes our proof of equivalence for one-loop diagrams
with longitudinal divergences
in the scalar Yukawa model.  The covariant expression for the boson
self-energy at one-loop level is
\begin{equation}
\label{bse1}
\bsex \hspace{-.2cm}
= \int_{\rm Min}  \frac{{\rm d}^4k 
\; {\rm Tr} \left[ (\slash{k} + m) (\slash{k}\!\! -\!\slash{q} + m) \right]}
{(k^2 - m^2 ) ((q\!-\!k)^2 - m^2 )} .
\end{equation}
\vspace{.4cm}

\noindent
The momenta are chosen in the same way as for the fermion self-energy.
The location of the poles is given by Eqs.~\r{h1} and \r{h2} with
$\mu$ replaced by $m$.  In order to do the $k^-$-integration we
separate the numerator into three parts.  We find

\vspace{-.2cm}
\begin{equation}
\bse = \bsep + \; 2 \bsei \hspace{-.5cm}.
\end{equation}
\vspace{+.2cm}

\noindent
The second term on the right-hand side are the two FILs, which
are identical.  The first term is the LF time-ordered boson self-energy.
It can be rewritten as
\begin{eqnarray}
\hspace{-.8cm} \bsep \hspace{-.2cm}
= 
2 \pi i \int {\rm d}^2k^\perp 
\int_0^{q^+} \frac{{\rm d}k^+}{4 k^+ (q^+ - k^+)}
\; \frac{ {\rm Tr} \left[ \left( \slash{k}_{\rm on} + m \right)
\left( (\slash{k} -\!\slash{q})_{\rm on} + m \right) \right]}
{H_2^- - H_1^-} . \\ \nonumber
\label{bsep}
\end{eqnarray}
The FIL is given by
\vspace{-.3cm}
\begin{equation}
\label{bseinst}
\hspace{-.8cm}
\bsei \hspace{-.5cm} = \int \frac{{\rm d}^2k^\perp {\rm d}k^+ {\rm d}k^-}
{4 k^+ (q^+\!-\!k^+)}
\frac{ {\rm Tr} \left[
\gamma^+ \left( (\slash{k} -\!\slash{q})_{\rm on} + m \right) \right]}
{k^- - H_2^-} .
\end{equation}
We have seen in our discussion of the fermion self-energy that it is
possible to determine the exact form of the FIL that
maintains covariance.  However, we have also seen that taking this step
is not necessary, since upon minus regularization the FILs
disappear.  An analysis along lines similar to those in
\sec \ref{III.C} will show that the FIL
is also in this case independent of $q^-$.  Therefore we limit
ourselves to the calculation and renormalization of the propagating
diagram.

\subsection{Minus regularization}
We will now apply the minus regularization scheme to the LF time-ordered
boson self-energy. For a self-energy diagram the following ten steps
can be used to find the regularized diagram.  Some steps are explained
in more detail for the boson self-energy.
\begin{enumerate}
\setcounter{enumi}{0}
\item Write the denominator in LF coordinates.
\label{stap1}
\item Complete the squares in the denominator by introducing internal
      variables ($k'^\perp$ and $x$).
\item Write the numerator in terms of internal and external LF coordinates.
\item Remove terms odd in $k'^\perp$ in the numerator.
\end{enumerate}
These steps were also taken in  our discussion of the fermion self
energy. Next we diverge.
\begin{enumerate}
\setcounter{enumi}{4}
\item Subtraction of the lowest order in the Taylor expansion is equivalent
      to inserting a multiplier $X$. Construct the multiplier.
\item Compensate for the subtraction by adding
      counterterms. Verify that they are infinite. If they are not,
      the corresponding divergence was only apparent and we should not
      subtract it. We do not allow for finite renormalizations.
\end{enumerate}
For the boson self-energy all terms have the same
denominator. For them we can write the expansion
\begin{equation}
\frac{1}{ {k'^\perp}^2 + m^2 -x(1-x) q^2} = \frac{1}{{k'^\perp}^2 + m^2}
\sum_{j=0}^{\infty} X^j ,
\end{equation}
where the multiplier $X$ has the form
\begin{equation}
X = \frac{x (1-x) q^2}{{k'^\perp}^2 + m^2 } .
\end{equation}
\begin{enumerate}
\setcounter{enumi}{6}
\item Identify, term by term, the degree of divergence and
      insert the corresponding
      multiplier. To compensate for this, add a polynomial of 
      the appropriate degree with infinite coefficients. 
\label{stap5}
\end{enumerate}
Steps \ref{stap1}-\ref{stap5} lead to the following result for the boson self-energy:
\begin{eqnarray}
\bsep = \pi i \int {\rm d}^2k'^\perp \left. 
\int_0^1 {\rm d}x X \;{\rm Tr} \right[ \phantom{abcdefghijk}
\hspace{2cm}
\nonumber\\
\left(
\frac{X {k'^\perp}^2   +   x^2         {q^\perp}^2   +   m^2}
{2xq^+}         \gamma^+ 
+ x         (q^+ \gamma^- \!\!-\! q^\perp \gamma^\perp)\!+\! m
\right)\hspace{.6cm}
\nonumber\\
\left(
\frac{X {k'^\perp}^2\!\!+\!(x\!-\!1)^2 {q^\perp}^2\!\!+\! m^2}
{2(x\!-\!1)q^+} \gamma^+
+ (x\!-\!1) (q^+ \gamma^- \!\!-\! q^\perp \gamma^\perp)\!+\! m
\right)
\nonumber\\
+ X (k'^\perp \gamma^\perp)^2 \left] \;
\left( {k'^\perp}^2 + m^2 -x(1-x) q^2\right)^{-1} \right.
+ A + B q^2. \hspace{.5cm} 
\label{eqnstap5}
\end{eqnarray}
Longitudinal divergences appear as $1/x$ singularities. Transverse
divergences appear as ultraviolet $k'^\perp$ divergences. Since every
term in the boson self-energy is at least logarithmically divergent,
there is an overall factor~$X$. Some of the terms are quadratically
divergent in $k'^\perp$ and have an extra factor $X$. We use the
fact that terms containing the factor $\gamma^+ \gamma^+ $ vanish.  We
are not interested in the exact form of the counterterms $A$ and~$B$.
We can verify that they are infinite.  They are included to allow for
comparison with other regularization schemes.
\begin{enumerate}
\setcounter{enumi}{7}
\item Rewrite the numerator in terms of objects having either covariant
      or $\gamma^+/q^+$ structure.
\end{enumerate}
For our integral we use the following relation
\begin{equation}
\frac{x^2 {q^\perp}^2\!\!+\!m^2}{2 x q^+} \gamma^+ 
+ x( q^+ \gamma^- \!\! -\!  q^\perp
\gamma^\perp) = \frac{x^2 q^2\! +\! m^2}{2 x} 
\frac{\gamma^+}{q^+} + x \! \slash{q} .
\end{equation}
\begin{enumerate}
\setcounter{enumi}{8}
\item Perform the trace, if present.
\item Do the $x$ and $k'^\perp$ integrations.
\end{enumerate}
Application of the last two steps gives
\begin{eqnarray}
\hspace{-.6cm} \bsep \hspace{-.5cm} =  \! A +\! B q^2 - 2  \pi^2 i
\left( 3 q^2 - 8 m^2 \nonumber
+ 2 (4m^2\!-\!q^2)
\sqrt{\frac{4 m^2\! -\! q^2}{q^2}} \arctan
\sqrt{\frac{q^2}{4 m^2 \!-\! q^2}} \; \right). \hspace{-1cm} \phantom{a} \\ 
\label{bseminusreg}
\end{eqnarray}

\subsection{Equivalence}
We will now compare the result of the minus regularization scheme applied
to the LF time-ordered boson self-energy with dimensional regularization
applied to the covariant diagram. Using the standard rules of dimensional
regularization given by Collins~\cite{Col84} we obtain
\begin{eqnarray}
\label{bsedimreg}
\hspace{-.6cm} \bse \hspace{-.3cm} = A' + B' q^2   - 4 \pi^2 i (4 m^2 - q^2)
\sqrt{\frac{4 m^2 - q^2}{q^2}} \arctan \sqrt{\frac{q^2}{4 m^2 - q^2}} .
\\ \nonumber
\end{eqnarray}
The constants $A'$ and $B'$ contain $1/\varepsilon$, where $\varepsilon$
is the dimensional regulator. In the
limit of $\varepsilon \rightarrow 0$ they diverge. Of course, $A'$ and
~$B'$ can not be related to the infinite constants generated by minus
regularization. However, this is not necessary. Both schemes are
equivalent if the same physical amplitudes are generated.  To calculate
them we have to construct the counterterms or, equivalently, fix the
amplitude and its first derivative at the renormalization point.  For
the unrenormalized amplitudes \r{bseminusreg} and \r{bsedimreg} the
coefficients~$A$ or $A'$ of the constant term are used to determine the
physical mass $\mu_{\rm ph}$ of the boson. The coefficients $B$ or $B'$
determine the fermion wave function renormalization. Only the $q^4$ and
higher order terms can be used to make predictions. These coefficients
must be the same for the two methods. We see that Eqs.~\r{bseminusreg}
and \r{bsedimreg} only differ in the first two coefficients of the
polynomial in $q^2$.  Therefore the two methods generate the same
physical amplitudes.

\section{Conclusions}
\label{secconc3}
We discussed in this chapter the problem of covariance, which includes
the problem of nonmanifest rotational invariance, in LFD.

For diagrams which are both longitudinally and transversely convergent
one can give a rigorous demonstration of equivalence, without
discussing renormalization explicitly.  It is given by Ligterink and
Bakker~\cite{LB95b}.

For longitudinally divergent diagrams such a proof is not possible
because the integration over LF energy is ill-defined.  Still, LF
time-ordered diagrams can be  constructed applying the rules of NLCQ.
However, FILs have to be included to make the full series add up to the
covariant diagram.  These FILs contain the ambiguity related to the
ill-defined integration, as can be shown by our analysis involving the
regulator~$\alpha$.

We conjecture that the FILs are remnants of the difficulty of
quantizing on the light-front. Just like NLCQ, we are not able to
provide general rules to construct them.  However, we can identify the
conditions for their occurrence.  We show that it is not necessary to
find an explicit expression for the FILs. Upon minus regularization
they vanish.  Therefore the $\alpha$-dependence drops too. The
remaining series of regularized LF time-ordered diagrams is again
covariant.

The main difficulty we encountered was to show that the FILs are
instantaneous indeed.  This can be shown by proving that the regulator
$\alpha$ does not depend on the LF energy, as we did for the fermion
self-energy. Another way is to show that the regularized covariant
amplitude equals the corresponding series of minus-regularized LF
time-ordered diagrams. We used this technique for the boson
self-energy.

This concludes our proof of equivalence of renormalized covariant and
LF perturbation theory for longitudinally divergent diagrams in the
Yukawa model. Three diagrams with transverse divergences remain.  They
require a more elaborate analysis of minus regularization and numerical
implementation of the method.  This subject is treated in the next
chapter.

\chapter{\label{chap4}Transverse divergences in the Yukawa model}
\def \ftri{\raisebox{-.95cm}{\epsfxsize=1.4cm \epsffile[0 0 63 90 ]{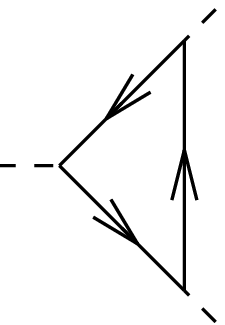}}}
\def \ftrix{\raisebox{-1.14cm}{\epsfxsize=1.9cm \epsffile{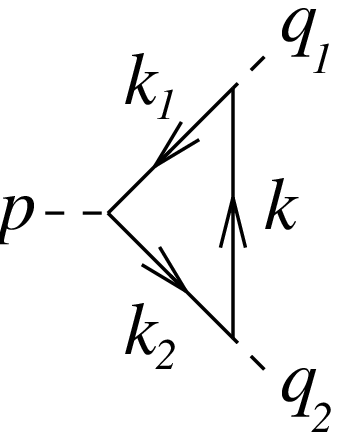}}}
\def \ftriapp{\raisebox{-.55cm}{\epsfxsize=1.8cm \epsffile{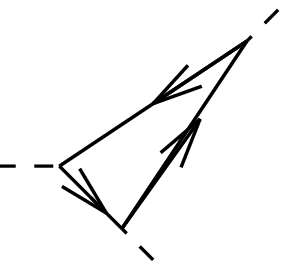}}}
\def \ftriaip{\raisebox{-.72cm}{\epsfxsize=1.8cm \epsffile{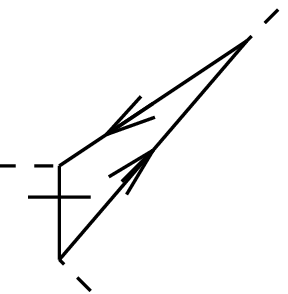}}}
\def \ftriapi{\raisebox{-.57cm}{\epsfxsize=1.6cm \epsffile{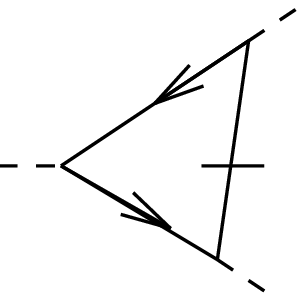}}}
\def \ftribpp{\raisebox{-.92cm}{\epsfxsize=1.8cm \epsffile{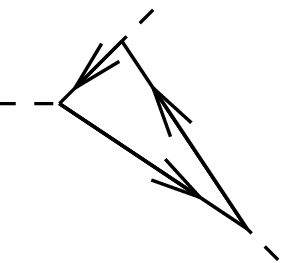}}}
\def \ftribip{\raisebox{-.92cm}{\epsfxsize=1.8cm \epsffile{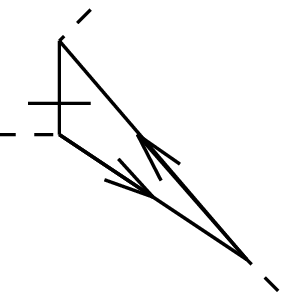}}}
\def \ftribpi{\raisebox{-.80cm}{\epsfxsize=1.6cm \epsffile{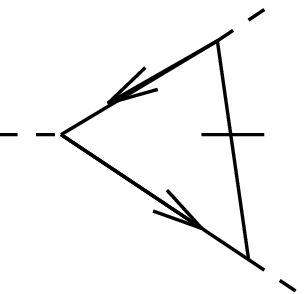}}}
\def \obe{\epsfxsize=1.4cm \raisebox{-.98cm}{\epsffile{obe.eps}}}
\def \obex{\epsfxsize=1.95cm \raisebox{-1.2cm}{\epsffile{obex.eps}}}
\def \obeba{\epsfxsize=1.39cm\raisebox{-1.00cm}{\epsffile{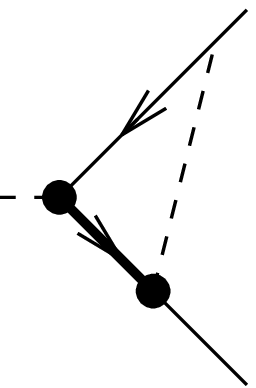}}}
\def \obebb{\epsfxsize=1.39cm\raisebox{-1.00cm}{\epsffile{obeb2.eps}}}
Light-front quantization has
found many applications since it was conceived.  Still, some problems
of a fundamental nature remained.  One that we are particularly
interested in is the question of whether full covariance can be maintained
in the Hamiltonian formulation, which is of course not manifestly
covariant.  A partial answer can be obtained in perturbation theory.
Then the  problem can be reformulated as follows:  can one prove that
LF perturbation theory produces the same values of the
$S$-matrix elements as covariant perturbation theory?  If the answer to
this question is affirmative, then we use the word equivalent to
describe the situation.

This chapter is concerned with one aspect of this problem, viz the
treatment of transverse divergences in a simple model: the Yukawa model
with spin-1/2 fermions, spin-0 bosons and a scalar coupling.

\section{Formulation of the problem}
\label{secintro}

In the previous chapters, we used the method of Kogut and Soper
\cite{KS70} to define LF perturbation theory. This method
defines LF time-ordered ($x^+$-ordered) amplitudes by
integration of the integrand of a covariant diagram, say

\begin{equation} \label{1500} F(q) = \int {\rm d}^4k\;\;
I(q;k), \end{equation}

\noindent
over the LF energy variable $k^-$.  In
this chapter, $q$ always denotes the external momenta and $k$ the loop
momentum. We can also write Eq.~\r{1500} using LF coordinates:
\begin{equation} 
F(q)=\int \! {\rm d}k^+ {\rm d}^2k^\perp \int
{\rm d}k^- I(q^-\!,q^+\!,q^\perp;k^-\!,k^+\!,k^\perp).
\end{equation}
Next, one expresses the integral over $k^-$, using Cauchy's formula, as
a sum of residues.  One arrives in this way at an expression that can
be interpreted, possibly after recombination of the terms in this sum,
as the splitting of the covariant amplitude $F(q)$ into a sum
of noncovariant but LF time-ordered amplitudes.

\begin{figure}
\[
\epsfxsize=9.5cm \epsffile[ 0 0 531 189   ]{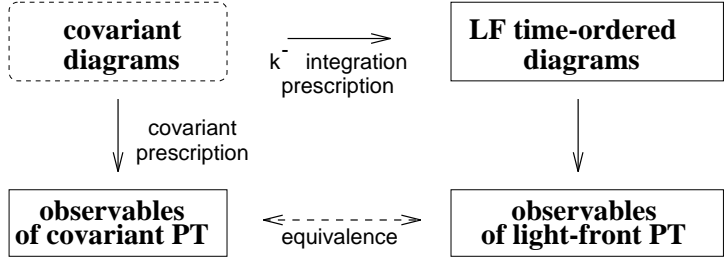}
\]
\vspace{-1cm}
\caption{The ``ideal'' case:
\label{figeen}
Outline of our proof of equivalence of LF and covariant
perturbation theory (PT) for convergent diagrams. The dashed
box indicates an ill-defined object.}
\end{figure}

This procedure, sometimes called naive light-cone quantization,
has been in principle known since the early work of Kogut and Soper~\cite{KS70}.
For convergent diagrams, it is nicely pictured in Fig.~\ref{figeen}.
 
The covariant diagram in Fig.~\ref{figeen} is an ill-defined object
and needs some prescription to give it a definite meaning.
For example, the measure of the Minkowski integration
is not positive definite.  The covariant prescription involves
introduction of Feynman parameters to complete the squares in the
denominator, the removal of terms odd in the loop momentum $k$ and the
Wick rotation to obtain a Euclidian integral.

It has been the work of Ligterink and Bakker \cite{LB95b} to prove in detail
that the rules for constructing
LF time-ordered diagrams, explained in many articles
\cite{KS70,LB80}, are correct upon using the $k^-$-integration
prescription.  They were the first authors to give a systematic
derivation of all the different time-ordered diagrams corresponding to
a given covariant amplitude, for any number of particles involved.  If
the $k^-$-integral is convergent and the corresponding covariant
diagram is also superficially convergent, then what remains can be
written in terms of well-defined, convergent Euclidian integrals.

When the $k^-$-integration is divergent, the prescription must be altered.
Naive light-front quantization fails in this case and one must first find a way
to regulate the $k^-$-integrals. We proposed in Chapter~\ref{chap3}
a regularization that maintains covariance. There we showed that the
longitudinal divergences give rise to so-called forced instantaneous loops
(FILs) and we showed how to deal with them such that covariance is maintained.
This method was also applied
to the Yukawa model containing spin-$1/2$ and spin-$0$ particles.
We were able to regularize the $k^-$-integrals for the diagrams with one
loop. However, in order to show full equivalence to the
covariant calculation one needs to compute the full integral including
the integrations over $k^+$ and $k^\perp$.
 
\subsection{Ultraviolet and transverse divergences}
 
\begin{figure}
\[
\epsfxsize=9.5cm \epsffile[ 0 0 526 193]{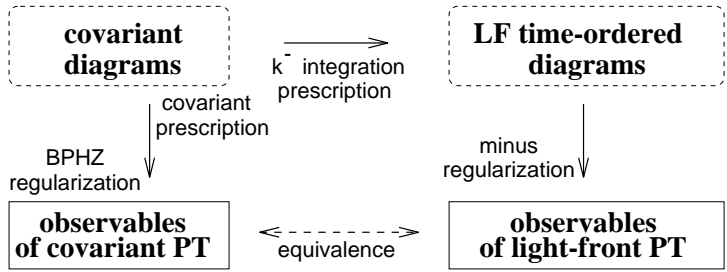}
\]
\vspace{-1cm}
\caption{Outline of our proof of equivalence for diagrams with ultraviolet
divergences. \label{figtwee} Dashed boxes indicate ambiguously 
defined objects.}
\end{figure}

Even after the usual procedure has been followed, the covariant integral
can still be
ultraviolet divergent. Ligterink and Bakker did not only discuss diagrams that
are superficially convergent, but also what to do in cases
where the covariant diagram is divergent. Their method of regularizing
divergent diagrams, minus regularization \cite{LB95a}, is also used in this
chapter.  A scheme for the equivalence of ultraviolet
divergent diagrams is given in Fig.~\ref{figtwee}.  
 
Several techniques are available to remove the ultraviolet divergences,
not involving the $k^-$- integration. They remain
in the LF time-ordered diagrams as divergences of the
integrals over the transverse momenta.  Therefore these diagrams are
also ill-defined, as indicated by the dashed box in Fig.~\ref{figtwee}.
A problem is that many of the techniques which are used to regularize
covariant diagrams have limited use for LF time-ordered diagrams. For
example, one cannot use dimensional regularization for the
longitudinal divergences.  Still, it is common to apply it to the
transverse divergences.  The strength of the regularization scheme we
use, minus regularization, is that it does not discriminate between
transverse and longitudinal divergences.  Minus regularization
is based on the Bogoliubov-Parasiuk-Hepp-Zimmerman method of
regularization \cite{Hep66,HZ68,Zim68,Zim69,CK82}.  In their paper,
Ligterink and Bakker applied minus regularization to three self-energy diagrams.
Our contribution is to extend their method to more complicated diagrams
and prove that there is a one-to-one relation between minus
and BPHZ regularization, such that the physical observables found using
LF perturbation theory exactly match those found in covariant
perturbation theory.

In the Yukawa model there are five covariant diagrams with ultraviolet
divergences.  The boson and the fermion self-energy were treated in
Chapter~\ref{chap3} in which longitudinal divergences were discussed.
Minus regularization was applied and simultaneously removed the
longitudinal and the transverse divergences. Equivalence was
established.

In two cases we were not able to either find an answer in the
literature or produce ourselves full analytic results for the
integrals involved; so we had to resort to numerical integration. 
In this chapter we discuss
these two diagrams: the one-boson exchange correction to the
boson-fermion-fermion vertex and the fermion loop with three
external boson lines. 
The first one was considered by Burkardt and Langnau \cite{BL91}, who 
stated that naive light-cone quantization leads to a violation of 
rotational invariance of the corresponding $S$-matrix elements and
found that invariant results can be obtained using noncovariant counterterms.
Here we show that no violation of rotational invariance occurs
if our method of regularization is applied. Furthermore, our results
for the LF time-ordered diagrams sum up to the covariant amplitude,
calculated using conventional methods.

\subsection{Light-front structure functions}

The two triangle diagrams, presented before in Table~\ref{tabYukdiv},
 can be written in the form of a sum of tensors in the
external momenta, multiplied by scalar functions, which we call (covariant)
structure functions. After splitting a covariant diagram in LF 
time-ordered ones, these can be written again in terms of tensors multiplied by
functions of the external momenta. The latter are called LF structure
functions. They are not invariant  as they are not defined by four-dimensional
invariant integrals, but rather by three-dimensional integrals. The different
structure functions have different divergences and they must be treated
according to their types of divergence, which we enumerate.

\paragraph{Type 1: LF structure functions without transverse
divergences.}  Neither the covariant nor the LF formulation
contains any divergences.  Integration over $k^-$ suffices to prove
equivalence. Minus regularization is not allowed.

\paragraph{Type 2: LF structure functions with cancelling transverse
divergences.} The individual LF time-ordered diagrams contain
divergences not present in the covariant amplitude.  Application of
minus regularization to the time-ordered diagrams is not allowed. We
show that the divergences cancel if all the time-ordered diagrams are
added, and that their sum equals the corresponding covariant
amplitude.

\paragraph{Type 3: LF structure functions with overall transverse
divergences.}  Divergences appear in the covariant amplitude as well as
in the LF time-ordered diagrams. We apply BPHZ regularization
to the covariant amplitude and minus regularization to the 
time-ordered diagrams.  

\paragraph{}

For the first two cases one can prove equivalence using analytic methods alone.
This proof is found in Ref.~\cite{LB95b} and in Chapter~\ref{chap3}.
For the structure functions with overall transverse divergences we have to use
numerical techniques. We show that for the decay of a boson at rest, for both
triangle diagrams, one obtains a rotational invariant amplitude, identical to
the covariant calculation using BPHZ regularization.  The fifth diagram
with transverse divergences, the fermion box, will not be discussed.


The setup for this chapter is as follows.  In \sec \ref{secmr} we
introduce minus regularization.  In \sex \ref{secftri} and~\ref{secobe} 
we discuss the equivalence of covariant and LF
perturbation theory for the fermion triangle and the one-boson exchange
correction. In both cases we start with the covariant calculation and
do the BPHZ regularization if necessary. Then we calculate the
LF time-ordered diagrams and apply the method mentioned above.
In both cases we conclude by giving a numerical example of rotational
invariance.

\section{Minus regularization}
\label{secmr}

Minus regularization is inspired by the BPHZ method of regularization,
which gives finite and covariant results. By construction, we ensure
that minus regularization does the same. First we sketch the method
in the case of one-loop diagrams with one independent
external momentum (self-energies),
and next when two independent external momenta (triangle diagrams) are present.
We conclude by generalizing this to a one-loop diagram with $n$
external momenta. For convenience, we shall assume in the latter case
that only logarithmic and
linear divergences are present, such that only the first term of the
Taylor expansion around the renormalization point needs to be subtracted.

Wherever we use the word ``amplitude'' in this chapter, we refer to an invariant
function of the external momenta. It is understood that the integrals defining
the invariant functions are formally written down in terms of four-dimensional
integrals, which are split into time-ordered pieces by integration over $k^-$.

\subsection{One external momentum}
\label{ssec1}

First we discuss the simple case of one external momentum, which
can be applied for self-energy diagrams.

\subsubsection{BPHZ regularization}
We start with the BPHZ regularization method, which can be applied to
covariant diagrams.
The amplitude has the following form
\begin{eqnarray}
F(q^2) = \int {\rm d}^4k\; I_{\rm cov}(q^2;k)
\label{Fcone}
       = F(0) + q^2 F'(0) + \dots
\end{eqnarray}
where $I_{\rm cov}(q^2;k)$ is the covariant integrand generated by
applying standard Feynman rules.
BPHZ regularization renders the amplitude finite by subtracting the infinite parts.
We choose the point $q^2=0$ as the renormalization point, around
which we expand the amplitude in a Taylor series. The higher
orders in the expansion \r{Fcone} are denoted by the ellipsis. The regularized
amplitude is then
\begin{equation}
F^{\rm R}(q^2) = F(q^2) - F(0).
\end{equation}
However, this is a purely formal operation, since we are subtracting
two infinite quantities. It is better to write:
\begin{eqnarray}
F^{\rm R}(q^2) = \int {\rm d}^4k\;
\left( I_{\rm cov}(q^2;k) - I_{\rm cov}(0;k)\right)
\label{Fcrone}
= \int_0^{q^2}{\rm d} {q'^2} \; \int {\rm d}^4k\;
\frac{\partial}{\partial q'^2} I_{\rm cov}(q'^2;k).
\end{eqnarray}
This guarantees that the amplitude becomes finite.
 
\subsubsection{Minus regularization}
 
Typical for minus regularization is that one writes the amplitude, as well
as the renormalization point, in LF coordinates.
The covariant choice $q^2=0$ corresponds to $q^- = {{q^\perp}^2}/({2q^+})$.
A time-ordered amplitude corresponding to the covariant form
\r{Fcone} can be written in LF coordinates as follows
\begin{eqnarray}
F(q^-\!,q^+\!,q^\perp) &=& \int {\rm d}^3k\; 
I_{\rm lfto}(q^-\!,q^+\!,q^\perp;k)
\label{Ftone}
\\
 &=& F(\frac{{q^\perp}^2}{2q^+},q^+\!,q^\perp) +
2q^+(q^-\!\! - \frac{{q^\perp}^2}{2q^+})
F'(\frac{{q^\perp}^2}{2q^+},q^+\!,q^\perp) + \dots
\nonumber
\end{eqnarray}
where $I_{\rm lfto}$ is the integrand of the LF time-ordered
diagram, which was generated by integrating the covariant integrand
$I_{\rm cov}$ over $k^-$ as is explained in Ref.~\cite{LB95b}.
The prime denotes differentiation with respect to $q^-$.
Similar to Eq.~\r{Fcrone} we can write the regularized amplitude as
\begin{equation}
F^{\rm MR}(q^-\!\!,q^+\!\!,q^\perp) =\hspace{-.12cm}
\int_\frac{{q^\perp}^2}{2q^+}^{q^-} {\rm d}q'^- \hspace{-.12cm}
\int {\rm d}^3k\;
\frac{\partial}{\partial q'^-} I_{\rm lfto}(q'^-\!,q^+\!,q^\perp;k).
\end{equation}
So far we have described the minus regularization method introduced
by Ligterink and Bakker \cite{LB95a}.

\subsection{Two external momenta}
\label{ssec2}
In Ref.~\cite{LB95a} three self-energy diagrams were discussed. For
the triangle diagram the minus regularization technique needs 
to be extended. We suggest the name MR$^+$. We will tune
the technique by comparing it to BHPZ regularization.
\subsubsection{BPHZ regularization}
The amplitude has the following covariant form
\begin{eqnarray}
F(q_1^2,q_2^2,q_1\!\cdot\!q_2) &=& \int {\rm d}^4k\;
I_{\rm cov}(q_1^2,q_2^2,q_1\!\cdot\!q_2;k)\nonumber\\
\label{Fctwo}
       &=& F(\tilde{0}) + q_1^2 F'_{1}(\tilde{0}) + q_2^2 F'_{2}(\tilde{0})
+ q_1\!\cdot\!q_2 F'_{3}(\tilde{0}) + \dots
\end{eqnarray}
where $\tilde{0}$ is the
renormalization point $q_1^2=q_2^2=q_1\!\cdot\!q_2=0$ and $F'_i$ is the 
derivative of $F$ with respect to the $i$th argument.
\begin{equation}
F^{\rm R}(q_1^2,q_2^2,q_1\!\cdot\!q_2) = F(q_1^2,q_2^2,q_1\!\cdot\!q_2)
- F({\tilde{0}}).
\end{equation}
Again, this is a purely formal operation, since we are subtracting
two infinite quantities. It is better to write
\begin{eqnarray}
F^{\rm R}(q_1^2,q_2^2,q_1\!\cdot\!q_2) = \int {\rm d}^4k
\left( I_{\rm cov}(q_1^2,q_2^2,q_1\!\cdot\!q_2;k) - 
I_{\rm cov}(\tilde{0};k)\right).
\end{eqnarray}
We cannot, as in the previous section, differentiate with respect to
all external momenta. We would then subtract finite parts from
the Taylor series, containing physical information. This can be
circumvented by introducing a dummy variable $\lambda$, which parametrizes
a straight line in the space of the invariants
between the actual values $q_1^2,q_2^2,q_1\!\cdot q_2$ and
the renormalization point:
\begin{eqnarray}
\label{bphz}
F^{\rm R}(q_1^2,q_2^2,q_1\!\cdot\!q_2) = \int_0^1 {\rm d} \lambda
\int {\rm d}^4k 
\frac{\partial}{\partial \lambda}
I_{\rm cov}(\lambda q_1^2,\lambda q_2^2, \lambda q_1\!\cdot\!q_2;k).
\end{eqnarray}
We have verified that the $\lambda$-method gives the correct result for the
case where one independent external momentum occurs.
 
\subsubsection{Minus regularization}
 
Again, we write the amplitude in the LF
time-ordered case as a three-dimensional integral:
\begin{equation}
F(q_i^-\!,q_i^+\!,q_i^\perp) = \int {\rm d}^3k\;
I_{\rm lfto}(q_i^-\!,q_i^+\!,q_i^\perp;k).
\end{equation}
The regularized amplitude is
\begin{equation}
F^{\rm R}(q_i^-\!,q_i^+\!,q_i^\perp)
= F(q_i^-\!,q_i^+\!,q_i^\perp) - F(r_i^-\!,r_i^+\!,r_i^\perp),
\end{equation}
where the four-vector $r$ defines the renormalization surface. It is a hypersurface
determined by the following conditions:
\begin{eqnarray}
r_1^2 &=& 2 r_1^- r_1^+ - {r_1^\perp}^2 = 0, \nonumber\\
r_2^2 &=& 2 r_2^- r_2^+ - {r_2^\perp}^2 = 0, \\
r_1\!\cdot\!r_2 &=& r_1^- r_2^+ + r_1^+ r_2^- 
           - r_1^\perp\!\cdot r_2^\perp = 0.  \nonumber
\end{eqnarray}
This set of equations is equivalent to
\begin{equation}
r_1^2 = 0, \;\;\;
r_2  = \chi r_1. \\
\end{equation}
The $r_i^+$ enter in the integration boundaries; therefore we would like them to
remain unaffected by regularization ($r_i^+ = q_i^+$). This implies that
$\chi$ can be found from 
\begin{equation}
\chi= \frac{q_2^+}{q_1^+}.
\end{equation}
The only freedom that remains is the choice for $r_1^\perp$.
Two choices come
easily to mind:  $r_1^\perp=0$ (method MR0) and $r_1^\perp=q_1^\perp$
(method MR1).
\begin{eqnarray}
({\rm MR0}) \hspace{2cm} r_1^\perp &=& 0^\perp
\Rightarrow r_2^\perp = 0^\perp,\\
({\rm MR1}) \hspace{2cm} r_1^\perp &=& q_1^\perp \Rightarrow r_2^\perp =
\chi {q_1^\perp}.
\end{eqnarray}

\begin{table}
\[
\begin{tabular}{|c|c|c|}
\hline
&MR0&MR1\\
\hline
$\;(r_1^-,r_1^+,r_1^\perp)\;$ &
$\;\phantom{\chi}(0,q_1^+,0^\perp)\;$ &
$\;\phantom{\chi}(
{{q_1^\perp}^2}/({2q_1^+}), q_1^+,
q_1^\perp)\;$ \\
\hline
$\;(r_2^-,r_2^+,r_2^\perp)\;$ &
$\;\chi \;(0,q_1^+,0^\perp)\;$ &
$\;\chi \;( {{q_1^\perp}^2}/({2q_1^+}), q_1^+, q_1^\perp)\;$ \\
\hline
\end{tabular}
\]
\vspace{-.3cm}
\caption{The LF parametrization of the renormalization point $r^\mu$ 
for two equivalent choices of minus regularization, MR0 and MR1.}
\label{tab1}
\end{table}

The final results are given in Table~\ref{tab1}.
The LF coordinates of the renormalization point are used in
the following way to find the regularized LF amplitude:
\begin{eqnarray}
\hspace{-.6cm}
F^{\rm MR}(q_i^-,q_i^+,q_i^\perp) = \int_0^1 {\rm d} \lambda
\int {\rm d}^3k \;
\frac{\partial}{\partial \lambda}
\label{mr} 
I_{\rm lfto}\left(\lambda (q_i^-\!\!-r_i^-)\! +\! r_i^-,
q_i^+\!,
\lambda (q_i^\perp\!\!-r_i^\perp)\! +\! r_i^\perp;k\right)\!.
\end{eqnarray}
In this formula we recognize our choice $r_i^+ = q_i^+$.

\subsection{Several external momenta}
\label{ssecn}
 
The method just described can be generalized to the case of a loop with
an arbitrary number of external lines.
The procedure is almost the same as for two external momenta. 
The renormalization surface is given by
\begin{eqnarray}
\label{ncond1}
 r^2_i & = & 2 r^-_i r^+_i - r^{\perp 2}_i = 0, \\
\label{ncond2}
r_i\cdot r_j&  = & r_i^- r_j^+ + r_i^+ r_j^- - r_i^\perp\!\cdot r_j^\perp = 0
\;\;\;(i \not= j).
\end{eqnarray}
These equations are equivalent to
\begin{equation}
r^2_1 = 0, \;\;\;r_i = \chi_{\scriptscriptstyle i} r_1.
\end{equation}
Again, we make the choice to leave the plus-components of
the momenta unaffected by regularization: $r^+_i = q^+_i$.
This implies that the $\chi_{\scriptscriptstyle i}$ are
fractional longitudinal LF momenta.
\begin{equation}
\chi_{\scriptscriptstyle i} = \frac{q_i^+}{q_1^+}.
\end{equation}
Two choices for $r_1^\perp$ are listed below. This
then determines all other $r_i^\perp$.
\begin{eqnarray}
({\rm MR0}) \hspace{2cm} r_1^\perp &=& 0^\perp
\Rightarrow r_i^\perp=0^\perp,\\
({\rm MR1}) \hspace{2cm} r_1^\perp &=& q_1^\perp \Rightarrow r_i^\perp =
\chi_{\scriptscriptstyle i} {q_1^\perp}.
\end{eqnarray}
 
\subsection{Summary}

The way we setup minus regularization does not rely on the structure
of the covariant or the time-ordered diagrams, but works on the level
of the external momenta only.  If an amplitude has a covariant
structure before regularization, minus regularization guarantees that
it remains covariant.  In our implementation of BPHZ regularization, the
renormalization point corresponds to all invariants connected to the
external momenta being equal to zero. These conditions allow minus
regularization to take on a number of forms.  Of these, we shall apply
MR0 and MR1. The main difference between them is that MR0 does not
choose one of the momenta as a preferred direction, and therefore it
explicitly maintains all symmetries in the external momenta.
Furthermore, MR0 gives rise to shorter formulas for the regularized
integrands.

In the next two sections both  methods are being applied to
the parts of two LF time-ordered triangle diagrams
in the Yukawa model containing transverse divergences, viz the fermion
triangle and the one-boson exchange correction.
 
\section{Equivalence for the fermion triangle}
 
\label{secftri}
In the Yukawa model there is an effective three-boson interaction,
because to a fermion loop with a scalar coupling Furry's theorem does not
apply. The leading order contribution to this process is the fermion 
triangle. A scalar boson of mass $\mu$ and momentum $p$ comes in and
decays into two bosons of momentum $q_1$ and $q_2$ respectively.
The fermions in the triangle have mass $m$. The covariant expression for the
amplitude is
\begin{equation}
\label{ftri1}
\ftrix = \int_{\rm Min}{\rm d}^4k\;
\frac{ {\rm Tr}
\left [(\slash{k}_1 + m)(\slash{k}_2 + m) (\slash{k} + m)\right]} 
{(k_1^2-m^2) (k_2^2  -  m^2) (k^2  -  m^2)} .
\end{equation}
The usual imaginary parts of the Feynman propagators are not written 
explicitly.
We have omitted numerical factors and have set the coupling constant to unity.
The momenta $k_1$ and $k_2$ indicated in the diagram are given by
\begin{equation}
k_1 = k - q_1 , \;\;\; k_2 = k + q_2 .
\end{equation}
Of course, by momentum conservation we have
\begin{equation}
p = q_1 + q_2.
\end{equation}
We evaluate the integral \r{ftri1} first in the usual covariant way,
and subsequently carry out $k^-$-integration to produce the LF time-ordered
diagrams.
Note that integral \r{ftri1} is an ill-defined object. In
both methods mentioned we have to define what we mean by this integral.

\subsection{Covariant calculation}
The following method is usually applied to calculate the fermion triangle
in a covariant way.
First, one introduces Feynman parameters $x_1$ and $x_2$, and then one
shifts the loop variable $k$ to complete the squares in the denominator.
The result is
\begin{equation}
\label{ftri2} \hspace{-.5cm}
\ftri=8 \int_0^1 {\rm d}x_1
\int_0^{1-x_1} {\rm d}x_2 \int_{\rm Min} {\rm d}^4k \;
\frac{ m^3 + m \left( 3{k}^2 + {\cal P}^2 \right) + {\rm terms\; odd\;in\;}k}
{\left( k^2 - m^2 + {\cal Q}^2 \right)^3},
\end{equation}
with
\begin{eqnarray}
{\cal Q}^2 &=&x_1(1\!-\!x_1)\; q_1^2 + x_2(1\!-\!x_2)\; 
q_2^2 + 2 x_1 x_2\;q_1\!\cdot\!q_2,\\
{\cal P}^2 &=& x_1 (3 x_1\!-\!2) q_1^2 + x_2 (3 x_2\! -\! 2 ) q_2^2
+ \left( 2 (x_1\! +\! x_2) - 6 x_1 x_2 - 1 \right) q_1\!\cdot\!q_2.
\end{eqnarray}
As a last step, we remove the terms odd in $k$.

\subsection{BPHZ regularization}
The regularized fermion triangle can be found by
applying the BPHZ regularization scheme \r{bphz} to the covariant 
formula \r{ftri2}.
The integral is now finite; so we can do the Wick rotation and
perform the $k$ integrations.
The result is
\begin{eqnarray}
\ftri^{{\rm \; R}} =
 {-4 \pi^2 i} \!\int_0^1 \!{\rm d}x_1\!
\int_0^{1-x_1}\!\!{\rm d}x_2\!
\int_0^1 {\rm d}\lambda
\frac{ m \left(m^2 ( 5 {\cal Q}^2 - {\cal P}^2) - 6 \lambda {\cal Q}^4\right)}
{\left(m^2 - \lambda {\cal Q}^2 \right)^2}.
\end{eqnarray}
The superscript R indicates an integral regularized according to the
BPHZ method.

\subsection{Light-front calculation}
Using the method given in Ref.~\cite{LB95b} we proceed as follows.
The $k^-$ dependence of a spin projection in the numerator is removed by 
separating it into an on-shell spin projection and an instantaneous part:
\begin{equation}
\slash{k}_i + m =(\slash{k}_{i\; \rm on}+m)+(k^- - k^-_{i\; \rm on})\gamma^+,
\end{equation}
where the vector $k^\mu_{i\; \rm on}$ is given by
\begin{equation}
\left(k^-_i,k^+_i,k^\perp_i\right)_{\rm on}=
\left(\frac{{k^\perp_i}^2 + m^2}{2 k^+_i},k^+_i,{k^\perp_i} \right).
\end{equation}
Factors like $(k^-\! - k^-_{i\; \rm on})$ can be divided out against
propagators and this cancellation gives rise to instantaneous fermions.
The integration over $k^-$ is performed by contour 
integration. The poles of the propagators are given by
\begin{eqnarray}
\label{pole1a}
H^-    &=& \frac{{k^\perp}^2 + m^2 }{2k^+} , \\
\label{pole2}
H^-_1  &=& q^-_1-\frac{{k_1^\perp}^2 + m^2 }{2k^+_1} , \\
\label{pole3}
H^-_2  &=& -q^-_2+\frac{{k_2^\perp}^2 + m^2 }{2k^+_2} .
\end{eqnarray}
This integration gives rise to the different time-ordered diagrams, as 
explained in more detail in Chapter~\ref{chap1}.
The result is
\begin{eqnarray}
\ftri &=& \ftriapp + \ftriaip + \ftriapi \nonumber\\
\vspace{.2cm}
      &+& \ftribpp + \ftribip + \ftribpi.
\end{eqnarray}
The diagrams on the right-hand side are LF time-ordered diagrams.
Time goes from left to right. The pictures can be recognized as
time-ordered diagrams because of the time-ordering of the
vertices and the occurrence of instantaneous fermions, indicated by
a horizontal tag. Explicitly;
\begin{eqnarray}
\ftriapp &=& 2 \pi i \int {\rm d}^2k^\perp \int_0^{q_1^+}
\frac{ {\rm d}k^+ }
{8 k_1^+ \! k_2^+ {k}^+}
\label{ftriapp}
\frac{{\rm Tr}\left[(\slash{k}_{1\rm on} + m) (\slash{k}_{2\rm on} + m) 
(\slash{k}_{\rm on} + m)\right] }
{(H_1^-\! - H_2^-)(H_1^-\! - H^-)},
\\ \nonumber \\ \nonumber \\
\vspace{.5cm}
\ftriaip &=&  2 \pi i \int {\rm d}^2k^\perp \int_0^{q^+_1}
\frac{ {\rm d}k^+ }
{8 k_1^+ \! k_2^+ {k}^+}
\label{ftriaip}
{}
\frac{{\rm Tr}\left[(\slash{k}_{1\rm on} + m) \gamma^+
(\slash{k}_{\rm on} + m)\right]} {H_1^- \!- H^-} ,
\\ \nonumber \\ \nonumber \\  
\ftriapi &=&  2 \pi i \int {\rm d}^2k^\perp \int_0^{q^+_1}
\frac{ {\rm d}k^+ }
{8 k_1^+ \! k_2^+ {k}^+}
\label{ftriapi}
{}
\frac{{\rm Tr}\left[(\slash{k}_{1\rm on} + m) (\slash{k}_{2\rm on} + m)
\gamma^+\right]} {H_1^- \!- H_2^-},
\\ \nonumber \\ \nonumber \\  
\ftribpp &=&\!\!\!-2 \pi i \int {\rm d}^2k^\perp \int_{-q_2^+}^0
\frac{ {\rm d}k^+ }
{8 k_1^+ \! k_2^+ {k}^+}
\label{ftribpp}
\frac{{\rm Tr}\left[(\slash{k}_{1\rm on} + m) (\slash{k}_{2\rm on} + m)
(\slash{k}_{\rm on} + m)\right] }
{(H_1^-\! - H_2^-)(H^- \!- H_2^-)}, \phantom{blabla}
\\ \nonumber \\  
\ftribip &=&\!\!\!  -2 \pi i \int {\rm d}^2k^\perp \int_{-q_2^+}^0
\frac{ {\rm d}k^+ }
{8 k_1^+ \! k_2^+ {k}^+}
\label{ftribip}
\frac{{\rm Tr}\left[\gamma^+ (\slash{k}_{2\rm on} + m) 
(\slash{k}_{\rm on} + m)\right]} {H^- \!- H_2^-} ,
\\ \nonumber \\  
\ftribpi &=&\!\!\!  -2 \pi i \int {\rm d}^2k^\perp \int_{-q_2^+}^0
\frac{ {\rm d}k^+ }
{8 k_1^+ \! k_2^+ {k}^+}
\label{ftribpi}
\frac{{\rm Tr}\left[(\slash{k}_{1\rm on} + m) (\slash{k}_{2\rm on} + m)
\gamma^+\right]} {H_1^-\!- H_2^-}.
\end{eqnarray}
Note that the diagrams \r{ftriapi} and \r{ftribpi}
with the instantaneous exchanged fermions have
the same integrand. However, the longitudinal momentum 
$k^+$ has a different sign. We encountered such a similarity before
in \sec \ref{secpairc} of Chapter~\ref{chap2} for the current.

Although we could have expected diagrams with two instantaneous
fermions, we see that they are not present. This is so because we use a
scalar coupling and therefore two $\gamma^+$ matrices becoming
neighbors give~$0$.  No so-called forced instantaneous loops 
are present.  These FILs obscure the equivalence of LF and
covariant perturbation theory and have been analyzed in Chapter~\ref{chap3}.

The traces can be calculated. We obtain
\begin{eqnarray}
&{\rm Tr}&\left[(\slash{k}_{1\rm on} + m) (\slash{k}_{2\rm on} + m)
(\slash{k}_{\rm on} + m)\right]\nonumber\\ &&
= 4 m (m^2 
+ {k}_{1\rm on}\!\cdot\!{k}_{\rm on} + {k}_{2\rm on}\!\cdot\!{k}_{\rm on}
+ {k}_{1\rm on}\!\cdot\!{k}_{2\rm on}),
\\
&{\rm Tr}&\left[(\slash{k}_{1\rm on} + m) (\slash{k}_{2\rm on} + m)
\gamma^+\right] = 4 m \left( 2 k^+\!\!- q_1^+\!\! 
                   + q_2^+ \right)\!,\hspace{-.3cm}
\\
&{\rm Tr}&\left[(\slash{k}_{1\rm on} + m) \gamma^+
(\slash{k}_{\rm on} + m)\right]\; = 4 m \left( 2 k^+\! - q_1^+ \right),
\\
&{\rm Tr}&\left[\gamma^+ (\slash{k}_{2\rm on} + m)
(\slash{k}_{\rm on} + m)\right]\; =  4 m \left( 2 k^+\! + q_2^+ \right).
\end{eqnarray}
We see that the high orders in $k^\perp$ have disappeared in the traces. 
However, logarithmic divergences remain in all LF time-ordered
diagrams \r{ftriapp}-\r{ftribpi}. We tackle them with minus 
regularization, as introduced in the previous subsection.

\subsection{Equivalence}
 
As the fermion triangle is a scalar amplitude,
there is only one structure function present. It belongs to the
third category we mentioned in \sec \ref{secintro}: it is logarithmically 
divergent, but has no longitudinal divergences.

\subsubsection{Type 3: LF structure functions with overall transverse divergences} 

We applied minus regularization to the integrands of the six
LF time-ordered diagrams,  using both the MR0 and MR1 methods.
We used {\sc mathematica} to do the substitution and the differentiation with
respect to $\lambda$, given by Eq.~\r{mr}. 
However, {\sc mathematica} was not able to do the
integration, neither analytically nor numerically. Therefore the
integrand was implemented in {\sc fortran} which was well capable of doing
the four-dimensional integration using {\sc imsl} routines based on Gaussian
integration.
 
Because  the integrations cannot be done exactly, we saw no
possibility of giving a rigorous proof of the equivalence of
LF and covariant perturbation theory. Instead we make a
choice for the parameters, such as the masses and the external momenta,
and show that our method gives the same result as the
covariant calculation with BPHZ regularization.  

\begin{figure}
\[
\epsfxsize=4cm \epsffile[ 0 0 205 227]{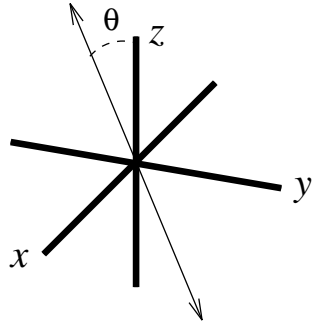} 
\]
\vspace{-.9cm}
\caption{A boson is at rest and decays into two particles flying off
\label{figtheta}
in opposite directions. The angle $\theta$ is the angle between the momentum of
one of the fermions and the $z$-axis. }
\end{figure}

We calculated the
decay amplitude of a scalar boson at rest, as is pictured in
Fig.~\ref{figtheta}. 
From a physical point of view, there is no preferred direction, and
therefore we demand that our choice of the coordinates of the
light-front have no influence on the outcome of the calculation. The
decay amplitude, which is a scalar quantity, should give the same
result for each possible direction in which the bosons can fly off.

Of the  six minus-regularized LF time-ordered fermion triangle diagrams
contributing to the boson decay, each one has a manifest rotational
invariance in the $x$-$y$-plane, and therefore we expect the same for
the sum.  However, since LF perturbation theory discriminates between
the $z$-direction and the other space directions, the LF time-ordered
diagrams can (and should) differ as a function of the angle, $\theta$,
between the momentum of one of the particles flying off and the
$z$-axis. The absolute value of the momentum is kept fixed.  It is not
immediately clear that the sum should be invariant.  This investigation
becomes more interesting since it is believed \cite{BL91} that
rotational invariance is broken in naive light-cone quantization of the
Yukawa model.

\begin{figure} 
\[
\epsfxsize=11.7cm \epsffile[75 480 505 795 ]{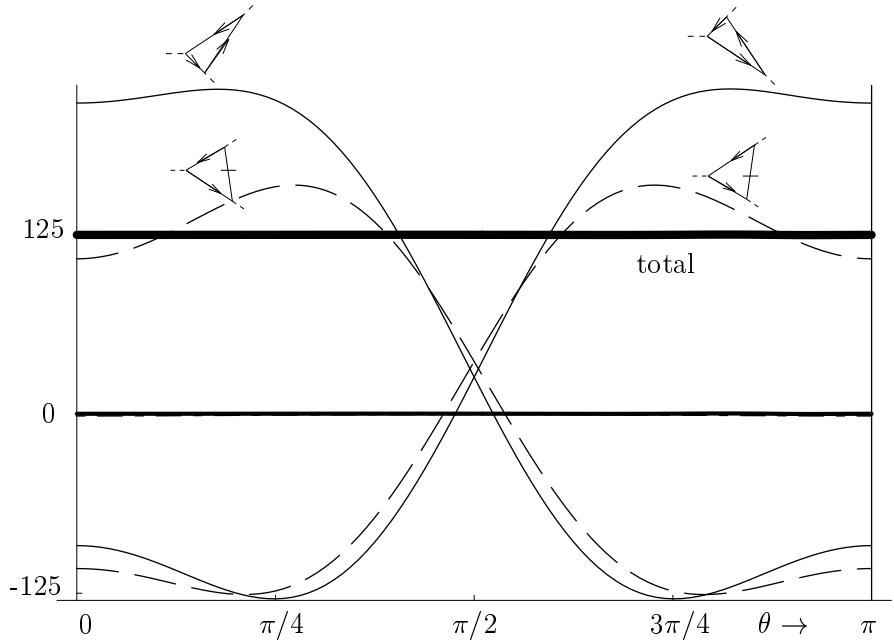} 
\]
\vspace{-1.2cm}
\caption{The thick line at a value of 125 represents the sum
of the six LF time-ordered amplitudes. It is independent of the
angle $\theta$, defined in Fig.~\ref{figtheta}.
The four largest contributions come from the diagrams without instantaneous 
parts (solid lines) and the diagrams with an instantaneous exchanged
\label{figdrie} 
fermion (dashed lines), as indicated by the diagrams.} 
\end{figure}

\begin{figure}
\vspace{-.5cm}
\[
\epsfxsize=11.8cm \epsffile[75 480 500 750 ]{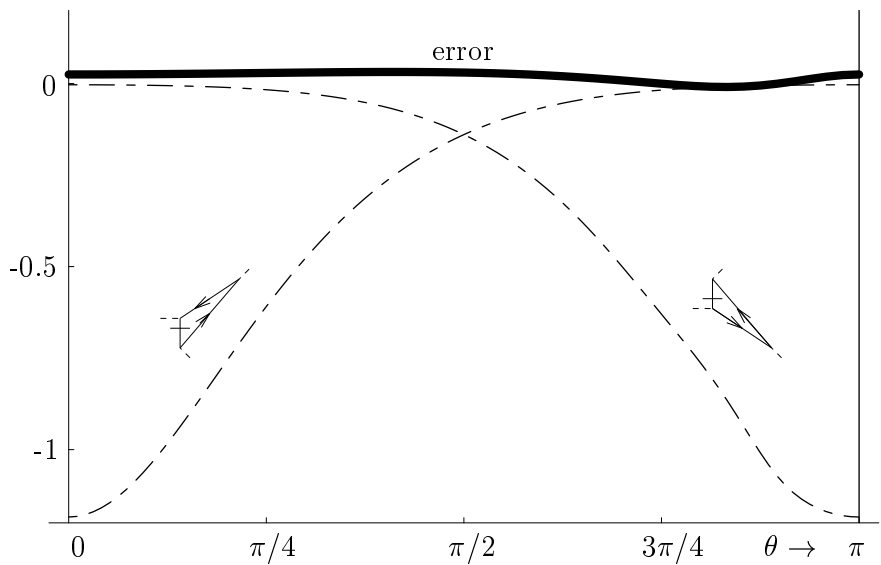}  
\]
\vspace{-1.2cm}
\caption{The amplitudes of the two small contributions (double-dashed lines) and
the difference between the sum of the six LF time-ordered diagrams
\label{figvier}
and the covariant amplitude (thick solid line).}
\end{figure}

The results are shown in
\mbox{Figs.~\ref{figdrie}-\ref{figvijf}}. They demonstrate that rotational
invariance is not broken.  Note that we have dropped the factor~$-i$
common to all diagrams.
Two LF time-ordered diagrams \r{ftriaip} and \r{ftribip}
contributing to the boson decay and indicated by double-dashed lines
are so small they can hardly be identified in Fig.~\ref{figdrie}. 
In Fig.~\ref{figvier} we depict these two on a scale that is a factor
100 larger. In the same figure we show 
 the difference of the sum of the six LF time-ordered diagrams 
(using MR1 and 128 points in every integration variable)
and the covariant result. It has a maximum of 0.03\%.
In Figs.~\ref{figdrie} and~\ref{figvier} we see that interchanging the outgoing
bosons is the same as replacing  $\theta$ by $\pi - \theta$. This property
is expected because of Bose-Einstein symmetry.
We verified that the individual diagrams are rotational invariant
around the $z$-axis. We illustrate this in Fig~\ref{figvijf}.
\begin{figure}
\[
{\epsfxsize=11.7cm \epsffile[190 420 430 780]{figure6.eps}}
\]
\vspace{-1.2cm}
\caption{Commutative diagram of the boson decay amplitude.
The boson is at rest in the origin and decays.
The outgoing bosons fly off in opposite directions.
Points on the surfaces have polar coordinates ($A, \theta, \phi)$,
where $A$ the is magnitude of the amplitude and $\theta$ and $\phi$ are
the polar angles of the momentum of one of the outgoing particles, as
defined in Fig.~\ref{figtheta}.  Because the diagrams on
the second line are very small, the scale  has been enlarged by a factor of 100.
For the LF time-ordered diagrams on the first three lines 
minus regularization (both MR0 and MR1) is used, 
\label{figvijf}
for the covariant diagram on the last line we used BPHZ regularization.}
\end{figure}

Summing up, we find that the sum of the minus-regularized 
LF time-ordered diagrams is rotational invariant.
The deviation from the covariant result is very small.
We checked, by varying the number
of integration points, that the deviations are due to numerical
inaccuracies only. We conclude that, for the fermion 
triangle, the covariant calculation in combination with the BPHZ regularization
scheme gives the same result as the LF calculation in combination
with minus regularization.

\section{Equivalence for the one-boson exchange diagram}
\label{secobe}
The second process under investigation was studied before by
Burkardt and Langnau~\cite{BL91}, and in \sec \ref{secobeiii} of 
Chapter~\ref{chap3}
as an example of how different LF time-orderings are constructed. 
A scalar boson of mass $\mu$ and momentum $p$ decays into two
fermions of mass $m$ and momentum $q_1$ and $q_2$ respectively.
The lowest order correction to this process is the
one-boson exchange correction. The amplitude is given by the integral
\begin{equation}
\label{obe1}
\obex   = \int_{\rm Min} {\rm d}^4k \;
\frac{(\slash{k}_1 + m)(\slash{k}_2 + m)}
{(k_1^2-m^2) (k_2^2  -  m^2)
(k^2  -  \mu^2)} .
\end{equation}
Again, this equation is undefined as it stands. First we have to make
it a well-defined object. In \sec \ref{1207} we apply the 
covariant method and in \sec \ref{1208} we use LF coordinates.
\subsection{Covariant calculation}
\label{1207}
Using Feynman parametrization the one-boson exchange correction
can be rewritten as
\begin{eqnarray}
\label{obe2} 
&&\hspace{-.5cm} \obe=2 \int_0^1 {\rm d}x_1
\int_0^{1-x_1} {\rm d}x_2 \int_{\rm Min} {\rm d}^4k 
\nonumber\\
&\times&\!\! \frac{ \slash{k}^2 +
\left( (1\!-\!x_1) \slash{q}_1\! +\!x_2) \slash{q}_2\! +\! m \right)
\left( -\!x_1 \slash{q}_1\! -\! (1\!-\!x_2) \slash{q}_2 \!+\! m \right)
+ { \rm \; terms \; odd \; in \;}k}
{\left( k^2 - {\cal M}^2 + {\cal Q}^2 \right)^3}, \phantom{blabla}
\end{eqnarray}
with
\begin{eqnarray}
{\cal M}^2 &=&(x_1 + x_2) m^2 + (1-x_1-x_2) \mu^2,\\
{\cal Q}^2 &=& x_1(1-x_1) q_1^2 + x_2(1-x_2) q_2^2 + 2 x_1 x_2q_1\!\cdot\!q_2,
\end{eqnarray}
and where terms odd in $k$ in the numerator are not specified, since they
will be removed according to the covariant prescription. We also define
\begin{equation}
{\cal P}^2 = {\cal Q}^2 + (1 - x_1 - x_2) q_1\! \cdot\! q_2.
\end{equation}

From Eq.~\r{obe2} we can infer that the Dirac structure of the diagram is
\begin{equation}
\label{xyz5}
\vspace{-.2cm}
\obe= F^1 + F^{2\mu} \gamma_\mu
+ F^{3\mu \nu} \frac{1}{2} \left[ \gamma_\mu, \gamma_\nu \right].
\vspace{.2cm}
\end{equation}
where the vector part contains a symmetric and an anti-symmetric part,
\begin{equation}
F^{2\mu} = F^{2\rm s}(q_1^{\mu} + q_2^{\mu})
        + F^{2\rm a}(q_{1}^{\mu} - q_{2}^{\mu}),
\end{equation}
and the tensor part has the form
\begin{equation}
F^{3\mu \nu} = (q_{1}^{\mu} q_{2}^{\nu} - q_{1}^{\nu} q_{2}^{\mu}) F^3.
\end{equation}
The functions $F^i$ depend on the masses and the
external momenta $q_1^2$, $q_2^2$ and $q_1\!\cdot\!q_2$. 
If we define the integral operator
\begin{equation}
\label{xyz1}
I[f] =2 \int_0^1 {\rm d}x_1 \int_0^{1-x_1} {\rm d}x_2 \int_{\rm Min}{\rm d}^4k
\left( k^2 -{\cal M}^2\!+ {\cal Q}^2 \right)^{-3}\!\! f,
\end{equation}
then we have, using ${\slash{q}_1}^2 = q_1^2$, et cetera.
\begin{eqnarray}
\label{F1eq}
F^1 &=& I \left[ k^2+ m^2- {\cal P}^2\right],\\
F^{2\rm a} &=& 2 m I \left[ 1 - x_1 - x_2 \right], \\
F^{2\rm s} &=& 2 m I \left[ - x_1 + x_2 \right], \\
F^3&=& I \left[ 1 - x_1 - x_2\right] .
\end{eqnarray}
We see that the only function which needs to be regularized is $F^1$.
The functions $F^2$ and $F^3$ are convergent and do not require
regularization in a covariant calculation.
\subsection{BPHZ regularization}
The regularized structure function $F^{1 \rm R}$ can be found by 
applying the BPHZ regularization scheme \r{bphz} to the structure
function~\r{F1eq}.
The integral is now finite; so we can do the Wick rotation and
perform the $k$ integrations.
\begin{eqnarray}
\hspace{-.7cm} F^{1{\rm R}}(q_1^2,q_2^2,q_1\! \cdot \!q_2) =\!
-2 \pi^2 i \int_0^1\!\!{\rm d}x_1
\int_0^{1-x_1}\!\!\!\!\!\!{\rm d}x_2
\int_0^1\!\! {\rm d}\lambda
\left( \frac{{\cal Q}^2 (\lambda {\cal P}^2 - m^2)}
            {2 ({\cal M}^2 \!-\! \lambda {\cal Q}^2)^2} +
       \frac{{\cal Q}^2 + \frac{1}{2}{\cal P}^2}
            {{\cal M}^2 \!-\! \lambda {\cal Q}^2} \right).
\end{eqnarray}
We have not been able to do all three integrations exactly.  The
$\lambda$ integration and one of the $x$ integrations can be done
analytically, and the remaining integration numerically.  As $F^{2\mu}$
and $F^3$ do not need to be regularized, this concludes the covariant
calculation of the one-boson exchange correction.

\subsection{Light-front calculation}
\label{1208}
In the previous chapters it was shown how to derive the
LF time-ordered diagrams corresponding to the covariant 
diagram \r{obe1} using $k^-$-integration.  One can write the
time-ordered diagrams individually, or one can combine propagating and
instantaneous parts into so-called blinks. Blinks
have the advantage that the
$1/k^+$-singularities cancel and the number of diagrams is reduced.

In the two triangle diagrams studied here it makes no difference
whether blinks are used or not. In the case of the fermion
triangle we calculated LF time-ordered diagrams.
Here we use blinks, to demonstrate that our technique also works in
this case.  The one-boson exchange correction has two blink diagrams:

\vspace{-.2cm}
\begin{equation}
\obe \;=\; \obebb \;+ \;\obeba.
{}
\end{equation}

\vspace{.3cm} 
\noindent
The poles of the two fermion propagators in the triangle are 
given by Eqs.~\r{pole2} and
\r{pole3}. The pole of the boson propagator is given by
\begin{equation}
\label{pole1b}
H^-    = \frac{{k^\perp}^2 + \mu^2 }{2k^+} . \\
\end{equation}
The amplitudes including blinks are 
\begin{eqnarray}
\label{obebb}
{}
\obebb =
- 2 \pi i \int {\rm d}^2k^\perp \int_{-q_2^+}^0
\frac{ {\rm d}k^+ }
{8 k_1^+  k_2^+ {k}^+}
\; \frac{(\slash{k}_{2\rm on} - \slash{p} + m)
(\slash{k}_{2\rm on} + m) }
{(H_1^- \!-\! H_2^-)(H^- \!-\! H_2^-)} ,
\\
\label{obeba}
{}
\obeba =
2 \pi i \int {\rm d}^2k^\perp \int_0^{q_1^+}
\frac{ {\rm d}k^+ }
{8 k_1^+ k_2^+ {k}^+}
\; \frac{(\slash{k}_{1\rm on} + m)
(\slash{k}_{1\rm on} + \slash{p} + m) }
{(H_1^- \!-\! H_2^-)(H_1^- \!-\! H^-)} .
\end{eqnarray}
We focus on the blink in Eq.~\r{obeba}.
It  simplifies  because we can use
\begin{equation}
\slash{k}_{1\rm on} \slash{k}_{1\rm on} = {k}_{1\rm on} \cdot {k}_{1\rm on}
= m^2.
\end{equation}
Therefore we obtain
\begin{equation}
\label{obeba2}
\obeba =
2 \pi i \int {\rm d}^2k^\perp \int_0^{q_1^+}
\frac{ {\rm d}k^+ }
{8 k_1^+  k_2^+ {k}^+}
\; \frac{2 m^2 + \slash{k}_{1\rm on} (\slash{p} + 2 m)}
{(H_1^- \!-\! H_2^-)(H_1^- \!-\! H^-)} .
\end{equation}
In the same way as we did for the covariant amplitude we can identify
the different Dirac structures

\begin{eqnarray}
\label{obeba3}
\vspace{-.4cm}
\obeba &=&
F^1_1 + F^{2\mu}_{1} \gamma_\mu
+ F^{3\mu\nu}_{1} \frac{1}{2} \left[ \gamma_\mu, \gamma_\nu \right].
\\ \nonumber
\end{eqnarray}

Although at first sight it looks as if the diagram in Eq.~\r{obeba2}
has a covariant structure, covariance is spoiled by the integration
boundaries for $k^+$.  Therefore these functions are not covariant
objects.  We have to investigate equivalence for the structure
functions separately.

The LF structure function $F^1_1$ can be  
found by taking the trace of Eq.~\r{obeba2}, since
all the other structures are traceless. Carrying out the traces one finds
\begin{eqnarray}
\label{F11}
F^1_1 =
  2 \pi i \int {\rm d}^2k^\perp \int_0^{q_1^+}
\frac{ {\rm d}k^+ }
{8 k_1^+  k_2^+ {k}^+}
\; \frac{2 m^2 + {k}_{1\rm on}\!\cdot\!p}
{(H_1^- \!-\! H_2^-)(H_1^- \!-\! H^-)} .\\ \nonumber
\end{eqnarray}
The other structures of the blink diagram \r{obeba2} are:
\begin{eqnarray}
\label{F21}
F_{1}^{2\mu} &=&
  2 \pi i \int {\rm d}^2k^\perp \int_0^{q_1^+}
\frac{ {\rm d}k^+ }
{8 k_1^+  k_2^+ {k}^+}
\; \frac{2 m \; ({k}_{1\rm on})^\mu}
{(H_1^-\! -\! H_2^-)(H_1^- \!-\! H^-)} , \\ 
\label{F31}
F_{1}^{3\mu\nu} &=&
  2 \pi i \int {\rm d}^2k^\perp \int_0^{q_1^+}
\frac{ {\rm d}k^+ }
{8 k_1^+  k_2^+ {k}^+}
\; \frac{({k}_{1\rm on})^\mu \; p^\nu}
{(H_1^-\! -\! H_2^-)(H_1^- \!-\! H^-)} .
\end{eqnarray}

In a similar way we can derive the structure functions corresponding
to the other blink diagram. 

\subsection{Equivalence}

We can identify the different types of divergences, as explained
in \sec \ref{secintro}:

\subsubsection{Type 1: LF structure functions without transverse
divergences}
The parts of the blink diagrams without any ultraviolet divergences
are $F^{2\mu}_{i}$ and $F^{3\mu\nu}_{i}$, except for $\mu$ being $-$.
No cancellations need to be found and no regularization is necessary.

\subsubsection{Type 2: LF structure functions with cancelling
transverse divergences}
In the last two structure functions we
see something odd happening. Both $F_{i}^{2\mu}$ and $F_{i}^{3\mu\nu}$
are divergent for $\mu$ being $-$. However, these divergences
are not present in the covariant structure functions $F^{2\mu}$
and $F^{3\mu\nu}$. It would we illegal to apply minus regularization,
since the covariant amplitude does not need to be regularized. 
We found that the divergences corresponding to the first blink
cancel exactly against those of the second blink. 
To simplify the calculation
we use internal variables $x'$ and $k^\perp$ 
and external variables $\chi$, $q_i^-$ and $q_i^\perp$. These
are introduced in Appendix~\ref{app1}. 

We have to verify the following relation of equivalence
\begin{equation}
F^{2-} = F^{2-}_1 + F^{2-}_2.
\end{equation}
For the reasons mentioned above we have to
demand that the divergent parts in the right-hand side cancel.
We find that only the highest order contribution in $k^\perp$
contributes to a divergent integral, because we can write
\begin{eqnarray}
F^{2-}_i = \int {\rm d}^2k^\perp \left( \frac{f^{2-}_i}{{k^\perp}^2}
  + g^{2-}_i(k^\perp) \right),
\end{eqnarray}
where $g^{2-}_i(k^\perp)$ is the part of the integrand without 
ultraviolet divergences, and the term with $f^{2-}_i$ gives rise
to a logarithmically divergent integral. We have to check if 
\begin{equation}
\label{cancel}
f^{2-}_1 + f^{2-}_2 = 0.
\end{equation}
In Appendix~\ref{app1} the full formulas for the functions $f^{2-}_i$ are given,
from which it follows that condition~\r{cancel} holds. 
For $\mu$ being $-$ in the
structure function $F^{3\mu\nu}_{1}$ one can apply the same method.

\subsubsection{Type 3: LF structure functions with overall transverse
divergences}

The structure function $F^1$ in the covariant
calculation contains an ultraviolet divergence. In the LF
structure functions $F^1_i$ these appear as divergences in the
transverse direction. 
The equation under investigation is the following:
\begin{equation}
\label{eq52}
F^{1 \rm MR}_1 + F^{1 \rm MR}_2 = F^{1 \rm R}.
\end{equation}
For the same reason as for the fermion triangle, an analytic proof of
this equation is not possible.  We investigated rotational invariance
of the left-hand side of this equation, and furthermore we checked if
it gives the same result as the covariant calculation on the right-hand
side.  

A boson is at rest and decays into two fermions as indicated in
Fig.~\ref{figtheta}. The fermion mass is taken to be the same as the
boson mass. Therefore there can be no on-shell singularities of
intermediate states.  
The contributions of the two blink diagrams are given in the commutative
diagram of Fig.~\ref{figzes}.  We made the rather arbitrary choice of
applying minus regularization MR1, and used 128 points in every
integration variable.

The error, i.e., the difference between the covariant calculation with
BPHZ regularization and the sum of minus-regularized blink diagrams, has a
maximum of 0.02\%.  This deviation results from numerical inaccuracies,
as was checked by varying the number of integration points.

We conclude that no significant deviation from a rotational invariant
amplitude is found. Moreover, we found that the sum of the LF
time-ordered diagrams is the same as the covariant amplitude for the
one-boson exchange correction.  Again, the procedure of
$k^-$-integration and minus regularization proved to be a valid
method.

\begin{figure}
\[
{\epsfxsize=11.7cm \epsffile[200 500 420 790]{figure7.eps}}
\]
\vspace{-.2cm}
\caption{Commutative diagram of the one-boson exchange correction.
A boson decays at rest.
The outgoing fermions fly off in opposite directions. As in Fig.~\ref{figvijf},
 the radial coordinate gives the amplitude of the regularized diagram
for the fermion flying off in this direction. 
For the LF structure functions on the first two lines,
minus regularization (MR1) is used; for the covariant structure 
\label{figzes}
function on the last line, we used BPHZ regularization.}
\end{figure}

\section{Conclusions} \label{secconc4}

In the Yukawa model with a scalar coupling there are five single-loop
diagrams with transverse divergences, of which two also contain
longitudinal divergences. For all other one-loop diagrams and all
multiple-loop diagrams that do not contain subdivergences, the proof of
the equivalence of covariant and LF perturbation theory was given
by Ligterink~\cite{Lig95} upon using the $k^-$-integration
prescription.  For the two single-loop diagrams with longitudinal
divergences this integration is ill-defined. This problem was dealt
with in the previous chapter.

Of the three remaining diagrams two are thoroughly analyzed in this
chapter. For the parts of these diagrams without transverse divergences
the $k^-$-integration recipe applies.  For the parts with
transverse divergences a proof of equivalence is complicated by the
fact that the amplitudes depend on three independent scalar products of
the external momenta.  We applied an extended version of the method of
minus regularization.  It is on a friendly
footing with the light-front, because it can be applied to both
longitudinal and transverse divergences.  Moreover, it has strong
similarities to BPHZ regularization, which is suitable for covariant
perturbation theory. We were able to tune the regularization in such a
way that minus regularization is analogous to BPHZ regularization. Therefore, we
expect an exact equality between the covariant and the LF
amplitudes.  We showed that rotational invariance is maintained and we
expect that other nonmanifest symmetries on the light-front, such as
boosts in the $x$-$y$-plane, are also conserved.

The final formulas obtained did not yield to analytic integration.
Therefore we had to resort to multidimensional numerical integration.
As rotational invariance was shown previously to be violated in naive
light-cone quantization \cite{BL91}, we investigated rotational invariance,
which is one of the nonmanifest symmetries on the light-front. 
Our results demonstrate, within the errors due to the numerical methods
used, that covariant and LF time-ordered perturbation
theory give the same physical matrix elements.

One diagram with transverse divergences is not discussed in 
this thesis, namely the fermion box with four external
boson lines. It is a scalar object, similar to the fermion triangle. 
The results obtained for the latter convinced us that upon 
minus regularization we shall find a covariant result. As there
are more time-orderings, and because one cannot test
for rotational invariance as easily as for the triangle diagrams,
we did not investigate this much more complicated situation. 

We trust that with our elaborate discussion of 
divergent diagrams in the Yukawa model we have illustrated the 
power of minus regularization and taken away doubts about
the covariance of LF perturbation theory.

%
%
%
%
%
%
%
%
%
%
%
%
%
%
%

\def \inp#1#2{#1\!\cdot\!#2}
\def \inperp#1#2{#1^\perp\!\!\cdot\!#2^\perp}
\def \cd{\makebox[0.08cm]{$\cdot$}}
\def \Mone  {\raisebox{-0.35cm}{\epsfxsize=2.0cm \epsffile{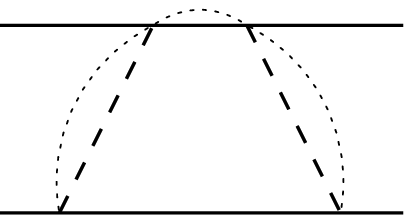}}}
\def \Mtwo  {\raisebox{-0.35cm}{\epsfxsize=2.0cm \epsffile{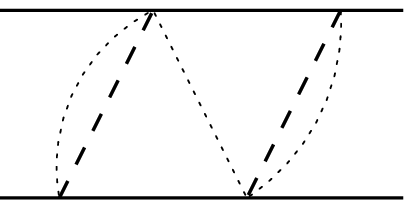}}}
\def \Mthree{\raisebox{-0.35cm}{\epsfxsize=2.0cm \epsffile{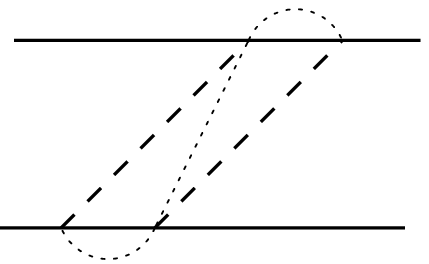}}}

\chapter{\label{chap5}Entanglement of Fock-space expansion and covariance}

In Chapter~\ref{chap1} we made the comparison between LF en IF
quantization and we saw that LFD has a number of advantages, one
being that it has the lowest number of dynamical
operators, namely three. However, two of them involve rotations around
the $x$ or $y$-axis.  Therefore LF time-ordered amplitudes are not
invariant under such rotations.

The question of rotations in LFD was discussed before
with the aim of constructing the angular momentum operators; see, e.g.,
Fuda~\cite{Fud91} and the review by Carbonell {\em et al.} \cite{CDKM98}.
While these authors emphasize the algebraic properties of the
generators of the Poincar\'{e} group, we stress 
the connection between expansions in Fock-space and covariance. It has
been remarked before by Brodsky {\em et al.} \cite{BJS85} that the
higher components in Fock space contribute to the difference between
the Bethe-Salpeter equation and the evolution equation in LFD. These
authors do not give numerical estimates of the corrections. The latter
has been done by Mangin-Brinet and Carbonell \cite{MC97}, and by 
Frederico~\cite{SFC98}, who studied the same model and found the
effect of higher Fock states on the binding energy to be small.
In a calculation of positronium, Trittmann and Pauli \cite{TP97}  used
an effective theory, where the effects of all Fock states are included
in the interaction. They found rotational symmetry to be restored in the
solution.

\section{Formulation of the problem}
In this chapter we consider first standard LF quantization and discuss
the problem of noncovariance, which includes violation of rotational
invariance, in the framework of LF time-ordered perturbation theory. We
give numerical results for the simplified case of two heavy scalars
exchanging light scalar particles. This choice is motivated by the
popular meson-exchange models in nuclear physics. We do not include the
internal spin degrees of freedom, as this is a complication that may
obscure the main point of our investigation: the connection between the
breaking of covariance and a truncation of the expansion in Fock
space.  In two interesting papers, Fuda \cite{Fud95f,Fud95i} reported
on detailed calculations of realistic one-meson exchange models in both
LF and IF dynamics. There the emphasis is on the comparison between the
two, when in both cases the ladder approximation is made. It is the
purpose of this chapter to show to what extent the ladder approximation
itself may violate covariance.

\subsection{Suppression of higher Fock states}
A reason why LFD is often preferred is that higher Fock states are
said to be more strongly suppressed in this form of dynamics. The
reason for this suppression is believed to be the spectrum condition
discussed before in \sec \ref{secspec} of Chapter~\ref{chap1}. As a
disadvantage, the lack of manifest rotational invariance, and therefore
covariance, is mentioned.  We call a symmetry manifest when it is
connected to a kinematical operator. Then all time-ordered diagrams
exhibit this symmetry.  Equal-time ordered diagrams lack boost
invariance, whereas on the light-front the longitudinal boost $P^+$ is a
kinematical operator.  Therefore, if one refers to a lack of
manifest covariance, one should include not only rotational invariance
but also other nonmanifest symmetries.  One reason why scientists have
rather stressed rotational invariance comes easily to mind: in many
cases it is easy to convince oneself by inspection whether a matrix
element is rotational invariant, viz when the amplitude can be
expressed in terms of scalar products of three-vectors. On the other
hand, it is not more difficult to test numerically for invariance under
boost transformations than for invariance under rotations. Indeed, the
method used in this chapter can easily be extended to check for boost
invariance.

A way to test for covariance is to compare the LF time-ordered diagrams
to the covariant amplitude, since we know that the latter is invariant
under any of the Poincar\'{e} symmetry operations.  For on energy-shell 
amplitudes ($S$-matrix elements), there is an exact equality, as was proved by
Ligterink and Bakker \cite{LB95b} and which is confirmed in our
results. Off energy-shell there is a breaking of Poincar\'e symmetry, 
which, however, is found to be surprisingly small in the case considered
in this chapter.

So why are we using these LF time-ordered diagrams in the first place,
when there is an equivalent covariant method available? We do, because
we want to determine the properties of the bound state using the
Hamiltonian form of dynamics. In this method, covariance can never be
fully maintained. However, one may try to apply it in such a way that
breaking of covariance is minimal.  In many applications in nuclear
physics, a one-meson exchange approximation is made for the interaction
and the scattering amplitude is computed by formally iterating this
interaction, leading to the Lippmann-Schwinger equation in the ladder
approximation. In this approximation one retains two- and
three-particle intermediate states and  neglects Fock states containing
four or more particles. These Fock sectors are needed to make the sum
of LF time-ordered diagrams equal to the covariant amplitude,
exhibiting the symmetries under all Poincar\'{e} transformations. If
these contributions are large, one can expect a significant breaking of
covariance, since the LF time-ordered diagrams are only invariant under
application of the kinematical symmetries.

For this reason we concentrate in this article on the determination of
the contributions of these higher Fock states. Our main concern shall be
the box diagram, defined on page~\ref{covbox}. 
Then we label the correction as ${\cal R}_{4^+}$. 
We shall calculate ${\cal R}_{4^+}$ explicitly for the box diagram with
scalar particles of different masses. The box diagram can be associated
with the two-meson exchange between two nucleons.  If spin were 
included, several well-known complications would arise, the most
important one being the occurrence of instantaneous propagators
\cite{LB95b,KS70}. We do not want these complications to interfere with
the main point of our investigation:  the connection between Fock-space
truncations and lack of covariance.  Therefore spin is omitted.  We
have not included crossed box diagrams, because they are not relevant for
a discussion on covariance, since both the crossed and noncrossed box
diagrams are covariant by themselves. 

However, it is well-known \cite{Gro82} that the use of ladder diagrams
alone in the Bethe-Salpeter equation does not lead to the proper
one-body limit when the mass of one of the nucleons goes to infinity.
Therefore, in order to use boson exchanges in bound state calculations
within the framework of LFD it is probably necessary to include
diagrams with crossed meson lines as well.

\subsection{Setup}
First, we explain the Lippmann-Schwinger formalism and the special
role of the box diagram. In \sec \ref{secbox} we describe how to
calculate both the covariant and the LF time-ordered amplitudes. After
this, we are ready for our numerical experiments.
In \sec \ref{secversus} the masses of the external particles are chosen
in such a way that on-shell singularities of the intermediate states
are avoided, and therefore it is easy to compare IF and LF Hamiltonian
dynamics. In that section it is shown that ${\cal R}_{4^+}$ is much
smaller in LFD than in IF dynamics (IFD), confirming the claim that in
LFD higher Fock states are more strongly suppressed. Moreover, it tells
us that covariance is more vulnerable in IFD than on the light-front!

After this exercise, we concentrate on the light-front, and in
\sec \ref{secabovet} we calculate the LF time-ordered diagrams
for the more interesting case in which we have particles of fixed masses $m$
(called nucleons) and $\mu$ (called mesons). As the process we are concerned 
with, scattering, is above threshold, we have to deal with on-shell
singularities. We show that the breaking of covariance is again small 
in the ladder approximation.

Although in \sec \ref{secundert}, where we discuss off-shell amplitudes below
threshold, no on-shell singularities are encountered,
matters become more complicated because the notion of the c.m. system (c.m.s)
becomes ambiguous, since the total momentum $P^z$ is dynamical and
found to be unequal to the combined momentum of the two particles, $p^z
+ q^z$. However, we are still able to relate the breaking of covariance and
Fock-space truncation.

The lack of covariance of the LF time-ordered amplitudes means that the
amplitude depends not only on the scalar products of the external
momenta, but on the angles between the quantization axis and the
external momenta as well.  Consequently, the amplitudes must have
singularities as a function of these angles in addition to the
familiar singularities as functions of the invariants.  The positions
of these singularities are found analytically in \sec \ref{seconshell},
in the framework of explicitly covariant LFD.  This gives a qualitative
understanding of the numerical results in \sex \ref{secversus}
and~\ref{secabovet}. In \sec \ref{secoffshell} explicitly covariant LFD
is applied to the off energy-shell results of \sec \ref{secundert}.

\section{The Lippmann-Schwinger formalism}
\label{seclippmann}
The Hamiltonian method aims at the determination of stationary states, i.e.,
eigenstates of the Hamiltonian. Here we take the Yukawa-type model with
scalar coupling
\begin{equation}
 {\cal L}_{\rm int} = g \Phi^2 \phi .
 \label{eq.lag}
\end{equation}
Two types of particles are considered:  ``nucleons'' (N, $\Phi$) with
mass $m$ and ``mesons'' (m, $\phi$) with mass $\mu$.  The Hamiltonian
$H\equiv P^-$ consists of a part $H_0$ which describes free particles
and a part $V$ which describes the interaction:
\begin{equation}
H = H_0 + V.
\end{equation}
We shall denote the second term on the right-hand side as the potential.
The problem of constructing the Hamiltonian from the underlying
Lagrangian has been recently reviewed by Brodsky {\em et al.} \cite{BPP97}.
Here we study two-nucleon states only. Moreover, we neglect self-energy 
diagrams.

We consider $H_0$, the kinematic part of the Hamiltonian in the two-nucleon 
(2N) sector. In the instant-form, quantization is carried out on planes of 
constant time ({equal-time planes}). 
Then we find for two particles of mass $m$ and momenta $p$
and $q$, respectively,
\begin{equation}
H_0^{\rm IF} = \sqrt{\vec{p}^{\;2} + m^2} + \sqrt{\vec{q}^{\;2} + m^2},
\end{equation}
which leads to both negative and positive energy solutions.  It is
well-known \cite{BKT79} that in this form the overall momenta and the
relative momenta are difficult to separate.

In LF quantization the square root, and therefore the negative energy
solutions are absent. 
The interaction-free part of the two-body Hamiltonian is
\begin{equation}
H_0^{\rm LF} = \frac{{p^\perp}^2 + m^2}{2 p^+} +
      \frac{{q^\perp}^2 + m^2}{2 q^+},
\end{equation}
which demonstrates that positive energies occur for positive plus-momenta.
Moreover, one can easily separate the motion of a many-particle system as a 
whole from the internal motion of its constituents in the LF case \cite{BKT79}.

We shall focus on light-front quantization of our model in which the 
interaction of the nucleons is due to meson exchange.  We write the potential
in the form
\begin{equation}
\label{Vexp}
V = V_1 + V_2 + V_3 + \dots
\end{equation}
where the subscript denotes the number of mesons simultaneously exchanged. 
The potentials only contain irreducible diagrams to prevent
double counting. $V_1$ contains one-meson exchanges only:
\begin{equation}
\label{obep}
V_1 = 
   \raisebox{-0.50cm}{\epsfxsize=2.55cm \epsffile{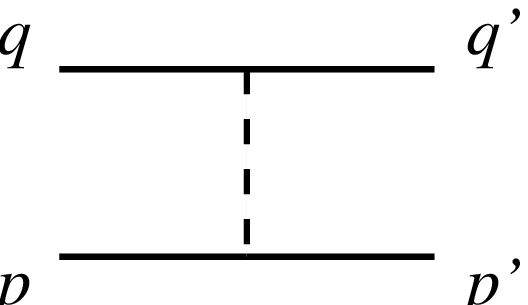}} 
 = \raisebox{-0.33cm}{\epsfxsize=2.0cm \epsffile{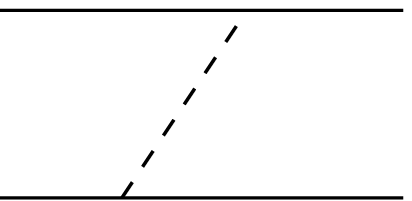}} +
   \raisebox{-0.33cm}{\epsfxsize=2.0cm \epsffile{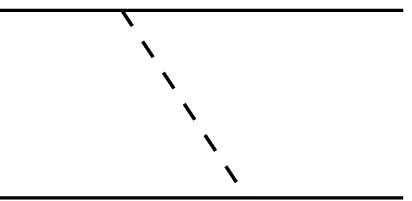}}\;.
\end{equation}
The irreducible diagrams contributing to $V_1$ are depicted in
Eq.~\r{obep}. 
In these diagrams time goes from left to right. The nucleons are denoted
by solid lines, and the mesons by dashed lines.
Irreducible diagrams contributing to $V_2$ are those diagrams 
of order $g^4$ that cannot be
separated into two pieces by cutting two nucleon lines or two nucleon lines and
one meson line only. In terms of Fock-space sectors this means that $V_1$
contains two-nucleon and one-meson intermediate states, and $V_2$ contains only
two-nucleon two-meson intermediate states.

The potential $V_1$ is a covariant object in case
the external lines are on shell.
The meaning of the equality sign in Eq.~\r{obep} is that the full
covariant amplitude can be written as a sum of two LF time-ordered diagrams.
Whereas the Feynman diagram contains the propagator $1/((q'-q)^2 - \mu^2)$, the
LF time-ordered diagrams contain the energy denominator $1/(P^- - H_0)$, 
$P^-$ being the parametric energy.  $H_0$ is the sum of the kinetic energies 
of the particles in the intermediate state: 
\begin{equation}
 H_0 = \sum_i \frac{{k^\perp_{i}}^2 + m^2_i}{2k^+_i} .
 \label{eq.ben1}
\end{equation}
The two diagrams contain $\theta$-functions of the plus-component of the 
momentum of the exchanged meson: one has the factor $\theta(p^+ - q^+)$, the
other $\theta(q^+ - p^+)$.

In a Feynman diagram the external lines are on mass-shell and the initial
and final states have the same energy, which coincides with the parametric
energy. Then the minus-component of the total four-momentum of a two-particle state
satisfies the relation
\begin{equation}
 P^- = p^- + q^- = p'^- + q'^- =
 \frac{{p^\perp}^{2} + m^2}{2 p^+} + \frac{{q^\perp}^{2} + m^2}{2 q^+} .
 \label{eq.ben2}
\end{equation}
As the minus-component of the total momentum is the only dynamical momentum
operator, the other three components are conserved in any LF time-ordered
diagram. For instance, $P^+ = p^+ + q^+ = p'^+ + q'^+$. This conservation law
is very important in LF quantization. It leads immediately to the
{spectrum condition}: in any intermediate state all massive particles have
plus-momenta greater than zero and the sum of the plus-components of
the momenta of the particles in that state is equal to the total plus-momentum.

The expansion in Fock space does not coincide with an expansion in powers of 
the coupling constant. This can easily be seen when one considers an approach
closely resembling the Lippmann-Schwinger method.
The eigenstates $|\psi>$ of the Hamiltonian
\begin{equation}
 H |\psi> = P^- |\psi>
\end{equation}
are also solutions of the Lippmann-Schwinger equation
\begin{equation}
 | \psi \rangle = | \phi \rangle + \frac{1}{P^- - H_0} V | \psi \rangle ,
 \label{eq.ben3}
\end{equation}
where $| \phi \rangle$ specifies the boundary conditions.
The  formal solution of this equation is
\begin{equation}
 |\psi> = \sum_{i=0}^\infty \left( \frac{1}{P^- - H_0} V \right)^i |\phi>.
\end{equation}
 
An equation similar to Eq.~(\ref{eq.ben3}) exists for the scattering amplitude:
\begin{equation}
\raisebox{-0.35cm}{\epsfysize=0.93cm \epsffile{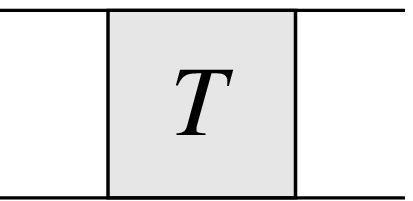}} =
\raisebox{-0.35cm}{\epsfysize=0.93cm \epsffile{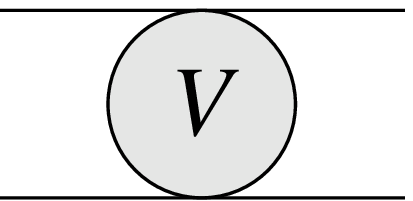}} +
\raisebox{-0.35cm}{\epsfysize=0.93cm \epsffile{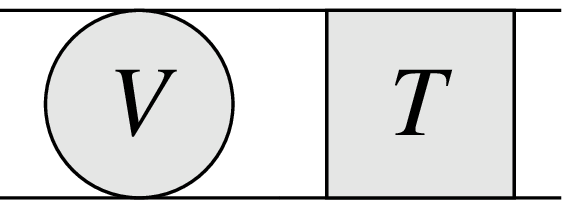}}\; .
\end{equation}
If one substitutes $V_1$ for $V$ in these equations, one obtains the {\em ladder
approximation}. This approximation does not generate all diagrams; so one needs
to add corrections. At order $g^4$ this correction is $V_2$:

\begin{eqnarray}
V_2 &=&\raisebox{-0.35cm}{\epsfxsize=2.01cm \epsffile{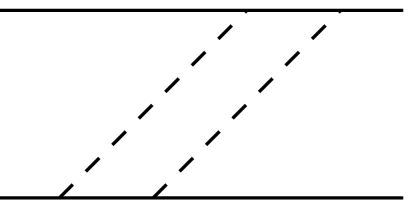}} +
       \raisebox{-0.35cm}{\epsfxsize=2.01cm \epsffile{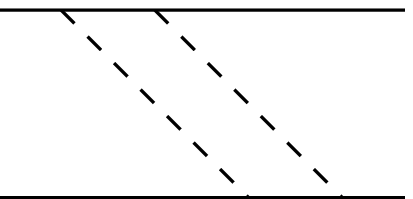}} \;.
\end{eqnarray}
If one takes into account all the contributions to $V$ from Eq.~\r{Vexp}, then
the full scattering amplitude is
\begin{eqnarray}
  \raisebox{-0.35cm}{\epsfysize=0.93cm \epsffile{T.eps}} =
  \raisebox{-0.35cm}{\epsfxsize=2.01cm \epsffile{oneba.eps}} &+&
  \raisebox{-0.35cm}{\epsfxsize=2.01cm \epsffile{onebb.eps}} \nonumber\\
+ \raisebox{-0.35cm}{\epsfxsize=2.01cm \epsffile{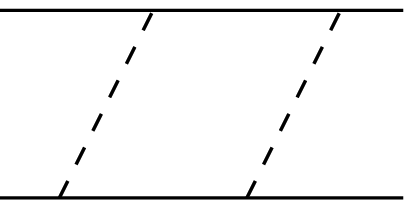}} +
  \raisebox{-0.35cm}{\epsfxsize=2.01cm \epsffile{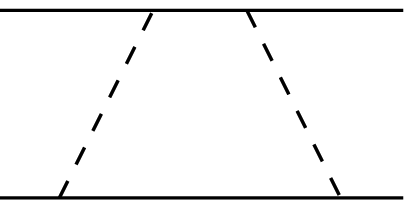}} &+&
  \raisebox{-0.35cm}{\epsfxsize=2.01cm \epsffile{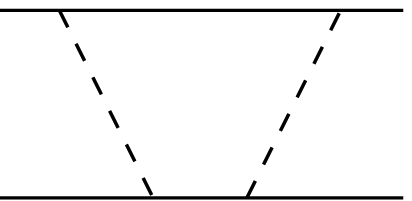}} +
  \raisebox{-0.35cm}{\epsfxsize=2.01cm \epsffile{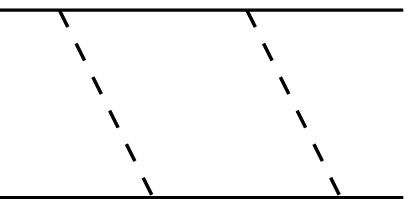}} \nonumber\\
+ \raisebox{-0.35cm}{\epsfxsize=2.01cm \epsffile{box0.eps}} + 
  \raisebox{-0.35cm}{\epsfxsize=2.01cm \epsffile{box6.eps}} &+&
  \quad {\cal O}(g^6).   \label{T1}
\end{eqnarray}
In the ladder approximation one only takes $V_1$ into account. 
Effectively, one then describes the full interaction between two nucleons by
\begin{eqnarray}
  \raisebox{-0.35cm}{\epsfysize=0.93cm\epsffile{T.eps}} =
  \raisebox{-0.35cm}{\epsfxsize=2.01cm \epsffile{oneba.eps}} &+&
  \raisebox{-0.35cm}{\epsfxsize=2.01cm \epsffile{onebb.eps}} \nonumber\\
+ \raisebox{-0.35cm}{\epsfxsize=2.01cm \epsffile{box2.eps}} +
  \raisebox{-0.35cm}{\epsfxsize=2.01cm \epsffile{box1.eps}} &+&
  \raisebox{-0.35cm}{\epsfxsize=2.01cm \epsffile{box3.eps}} +
  \raisebox{-0.35cm}{\epsfxsize=2.01cm \epsffile{box4.eps}} \nonumber\\
  &+& \quad {\cal O}(g^6). \label{T2}
\end{eqnarray}

In this approximation intermediate states containing more than three
particles do not occur.  This implies that time-ordered box diagrams
with four particles in the intermediate state are neglected, as we can
see if we compare the expansions in Eqs.~\r{T1} and ~\r{T2}.
As the individual diagrams contributing to $V_2$ are not covariant, the
sum of box diagrams produced by the ladder approximation is not
covariant.

Using equal-time quantization, 20 out of the 24 possible
time-orderings have intermediate states containing more than three
particles.  On the light-front, the spectrum condition destroys many of the
time-ordered diagrams.  There are six nonvanishing diagrams, of which
four only contain two- and three-particle intermediate states. One
concludes that the one-meson exchange kernel neglects the majority of the
contributing time-ordered box diagrams in equal-time quantization,
whereas on the light-front most of the nonvanishing diagrams are taken
into account.  This does not mean necessarily that in IF dynamics the
ladder approximation misses most of the amplitude, since the missing
diagrams have smaller sizes.  The contribution of the missing
diagrams needs to be investigated in order to see how much the higher
Fock sectors are suppressed.

There is one thing which seems to complicate matters on the
light-front.  The individual LF time-ordered diagrams are not
rotational invariant.  When a number of them is missing, the full
amplitude will also lack rotational invariance, as is mentioned often
in the literature.  This feature does not occur on the equal-time
plane, since there rotational invariance is a manifest symmetry.
However, in other types of Hamiltonian dynamics other symmetries are
nonmanifest.  In IF Hamiltonian dynamics, e.g., boost invariance is not
manifest.  Therefore we refer to breaking of covariance, which is a
general feature of any form of Hamiltonian dynamics, if one truncates
the Fock-space expansion.

We would like to estimate the contribution of the missing diagrams,
irrespective of the strength of the coupling.  It is not possible to do
this in a completely general way, so we perform our numerical
calculations for the box diagram only. We assume that our results will
be indicative for the higher orders too.

We define the fraction
\begin{equation}
{\cal{R}}^{\rm LF}_{4^+} = 
\frac{\;\;\raisebox{-.35cm}{\epsfxsize=2.01cm \epsffile{box0.eps}} +
      \raisebox{-.35cm}{\epsfxsize=2.01cm \epsffile{box6.eps}}\;\;}
     {\raisebox{-.7cm}{\epsfxsize=2.01cm \epsffile{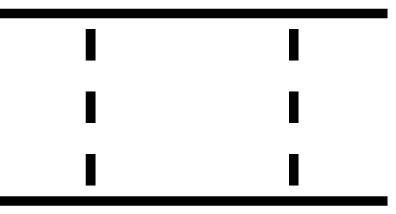}}^{\phantom{A}}}
\;.
\end{equation}
The subscript 4 indicates that this variable includes all diagrams
having at least four particles in some intermediate state. 
For ${\cal{R}}^{\rm IF}_{4^+}$ one would have to add the diagrams 
containing five- and six-particle intermediate states in the numerator,
as these give nonvanishing contributions in the instant-form.
The diagram in the denominator is the covariant diagram.

We shall show that the correction $V_2$ is indeed much less important
numerically in LFD than in IFD. 
So the 2N2m-state is
in LFD much less important than in IFD. We conjecture that this
property of LFD---that the Fock-state expansion converges much more rapidly
than in IFD---persists in higher orders in the coupling constant.

\section{The box diagram}
\label{secbox}
In the previous section we saw that the lowest level at which breaking
of covariance is to be expected is the two-meson exchange diagram, also
referred to as the box diagram. 
The discussion is limited to scalar particles.  Although a bound state
of scalar particles is not found in nature, we do not include spin
because we want to avoid in this investigation the complications due to
instantaneous terms.

\begin{figure}
\[
\epsfxsize=3.5cm \epsffile{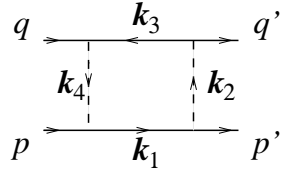}
\]
\vspace{-1cm}
\caption{\label{figkin}Kinematics for the box diagram. The arrows denote the momentum flow.}
\end{figure}

We look at the process of two nucleons with momenta $p$ and $q$,
respectively, coming in and exchanging two meson of mass $\mu$. The
outgoing nucleons have momenta  $p'$ and $q'$. 
The kinematics is given in Fig.~\ref{figkin}.
The internal momenta are
\begin{eqnarray}
\label{eqk1}
k_1 &=& k,        \\
k_2 &=& k - p',   \\
k_3 &=& k - p - q,\\
k_4 &=& k - p.
\label{eqk4}
\end{eqnarray}
We have to keep in mind that these relations only hold for those components
of the momenta that are conserved.

\subsection{The covariant box diagram}
The covariant box diagram is given by
\begin{eqnarray}
\label{covbox}
\raisebox{-.4cm}{\epsfxsize=2.01cm \epsffile{boxcov.eps}}
= \int_{\rm Min} \frac{-i\;{\rm d}^4k} 
{({k_1}^2 - m^2) ({k_2}^2 - \mu^2)
        ({k_3}^2 - m^2) ({k_4}^2 - \mu^2)},
\end{eqnarray}
where the imaginary parts $i \epsilon$ of the masses are not written
explicitly.  If the external states are on energy-shell, that is,
\begin{equation}
\label{onEcond}
P^- = p^- + q^- = p'^- + q'^-,
\end{equation}
then the time-ordered diagrams are the same as those derived
by integrating the covariant diagram over LF energy $k^-$.
In that case we have
\begin{eqnarray}
\raisebox{-.4cm}{\epsfxsize=2.01cm \epsffile{boxcov.eps}}
&=&
\raisebox{-.35cm}{\epsfxsize=2.01cm \epsffile{box1.eps}}
+
\raisebox{-.35cm}{\epsfxsize=2.01cm \epsffile{box3.eps}}
+
\raisebox{-.35cm}{\epsfxsize=2.01cm \epsffile{box2.eps}}\nonumber\\
&+&
\raisebox{-.35cm}{\epsfxsize=2.01cm \epsffile{box4.eps}}
+
\raisebox{-.35cm}{\epsfxsize=2.01cm \epsffile{box0.eps}}
+
\raisebox{-.35cm}{\epsfxsize=2.01cm \epsffile{box6.eps}}\;\;.
\label{equiv}
\end{eqnarray}
The example of the box diagram with scalar particles of equal masses
has been worked out before by Ligterink and Bakker~\cite{LB95b}.

\subsection{The LF time-ordered diagrams}

It is well-known~\cite{KS70,LB80} how to construct the LF time-ordered 
diagrams.  They are expressed in terms of integrals over energy denominators 
and phase-space factors. In the case of the box diagram, we need the ingredients
given below.  The phase space factor is
\begin{equation}
\label{phasespace}
\phi = 16 | k_1^+ k_2^+ k_3^+ k_4^+ |.
\end{equation}
Without loss of generality we consider the case $p^+ \geq p'^+$. The internal
particles are on mass-shell; however, the intermediate
states are off energy-shell.
A number of intermediate states occur.
We label the corresponding kinetic energies according to which of the 
internal particles, labeled
by $k_1$ \dots $k_4$ in Fig.~\ref{figkin}, are in this state:
\begin{eqnarray}
\label{cut1a}
H_{14} &=& q^- + \frac{{k_1^\perp}^2 +  m^2}{2 k_1^+}
               - \frac{{k_4^\perp}^2 +\mu^2}{2 k_4^+}, \\
\label{cut2a}
H_{13} &=&       \frac{{k_1^\perp}^2 +  m^2}{2 k_1^+}
               - \frac{{k_3^\perp}^2 +  m^2}{2 k_3^+}, \\
\label{cut3a}
H_{12} &=& q'^-+ \frac{{k_1^\perp}^2 +  m^2}{2 k_1^+}
               - \frac{{k_2^\perp}^2 +\mu^2}{2 k_2^+}, \\
\label{cut1b}
H_{34} &=& p^- - \frac{{k_3^\perp}^2 +  m^2}{2 k_3^+}
               + \frac{{k_4^\perp}^2 +\mu^2}{2 k_4^+}, \\
\label{cut2b}
H_{24} &=&q'^-+p^-+\frac{{k_2^\perp}^2 +\mu^2}{2 k_2^+}
                  -\frac{{k_4^\perp}^2 +\mu^2}{2 k_4^+}, \\
\label{cut3b}
H_{23} &=& p'^-+ \frac{{k_2^\perp}^2 +\mu^2}{2 k_2^+}
               - \frac{{k_3^\perp}^2 +  m^2}{2 k_3^+}.
\end{eqnarray}
A minus sign occurs if the particle goes in the direction opposite to
the direction defined in Fig.~\ref{figkin}.
All particles are on mass-shell, including the external ones:
\begin{eqnarray}
q^- = \frac{{q^\perp}^2 +  m^2}{2 q^+},\hspace{1cm}
q'^- &=& \frac{{q'^\perp}^2 +  m^2}{2 q'^+},\nonumber\\
p^- = \frac{{p^\perp}^2 +  m^2}{2 p^+},\hspace{1cm}
p'^- &=& \frac{{p'^\perp}^2 +  m^2}{2 p'^+}.
\end{eqnarray}

We can now construct the LF time-ordered diagrams. Diagrams
\r{box1} and \r{box3} will be later referred to as trapezium diagrams,
\r{box2} as the diamond, and \r{box0} as the stretched box.
\begin{eqnarray}
\label{box1}\raisebox{-0.35cm}{\epsfxsize=2.01cm \epsffile{box1.eps}} \!\!&=&\!\!
\int {\rm d}^2k^\perp \int_{  0}^{p'^+} \!\!
\frac{-2 \pi \;{\rm d}k^+}{\phi \; (P^-\!\!-H_{14}) \; 
                      (P^-\!\!-H_{13}) \;(P^-\!\!-H_{12})},\\
\label{box2}\raisebox{-0.35cm}{\epsfxsize=2.01cm \epsffile{box2.eps}} \!\!&=&\!\!
\int {\rm d}^2k^\perp \int_{p'^+}^{p^+} \!\!
\frac{-2 \pi \;{\rm d}k^+}{\phi \; (P^-\!\!-H_{14}) \; 
                      (P^-\!\!-H_{13}) \;(P^-\!\!-H_{23})},\\
\label{box3}\raisebox{-0.35cm}{\epsfxsize=2.01cm \epsffile{box3.eps}} \!\!&=&\!\!
\int {\rm d}^2k^\perp \int_{p^+}^{p^++q^+} \hspace{-.5cm} \!\!
\frac{-2 \pi \;{\rm d}k^+}{\phi \; (P^-\!\!-H_{34}) \; 
                      (P^-\!\!-H_{13}) \;(P^-\!\!-H_{23})},\\
\label{box0}\raisebox{-0.35cm}{\epsfxsize=2.01cm \epsffile{box0.eps}} \!\!&=&\!\!
\int {\rm d}^2k^\perp \int_{p'^+}^{p^+} \!\!
\frac{-2 \pi \;{\rm d}k^+}{\phi \; (P^-\!\!-H_{14}) \; 
                      (P^-\!\!-H_{24}) \;(P^-\!\!-H_{23})},\\
\raisebox{-0.35cm}{\epsfxsize=2.01cm \epsffile{box4.eps}} \!\!&=&\!\!
\raisebox{-0.35cm}{\epsfxsize=2.01cm \epsffile{box6.eps}}= 0. \label{boxo}
\end{eqnarray}
The factor $2 \pi$ matches the conventional factor $i$ in Eq.~\r{covbox}.
The last two diagrams are zero because we consider $p^+ \geq p'^+$ and
therefore these diagrams have an empty $k^+$-range. If we take
$p^+ \leq p'^+$, which case will also occur in forthcoming sections, 
diagrams \r{boxo} have nonvanishing contributions.

\section{A numerical experiment}
\label{numexp}
 
\begin{figure}
\[
\epsfxsize=10.8cm \epsffile{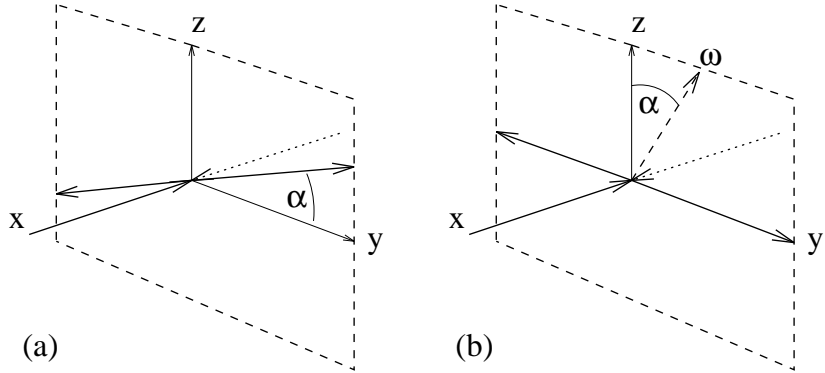}
\]
\vspace{-1cm}
\caption{(a) Two particles come in along the $x$-axis. They scatter
into the $y-z$ plane over an angle of $\pi/2$.  The azimuthal angle is
given by $\alpha$. (b) Another viewpoint.  The outgoing particles go
along the y-axis. The light-front vector $\omega$ makes an angle
$\alpha$ with respect to the $z$-axis.  \label{figscat}} 
\end{figure}

We look at the scattering of two particles over an angle of $\pi/2$.
In Fig.~\ref{figscat} the process is viewed in two different ways.

Fig.~\ref{figscat}a pictures the situation where the scattering plane is
rotated around the $x$-axis.
The viewpoint in Fig.~\ref{figscat}b concentrates on
the influence of the orientation of the quantization plane and is
connected to explicitly covariant LFD, as will be discussed in
\sec \ref{seconshell}.
Both viewpoints should render
identical results, since all angles between the five relevant directions
(the quantization axis and the four external particles) are the same.
We choose for the momenta
\begin{eqnarray}
\label{pmu}
p^\mu &=& ( v^0,+v^x ,\;\;\;\;\;   0,\;\;\;\;   0) ,\\
\label{qmu}
q^\mu &=& ( v^0,-v^x ,\;\;\;\;\;   0,\;\;\;\;   0),\\
\label{ppmu}
p'^\mu &=& ( v^0,\;\;\;\;\;  0 ,-v^y,-v^z),\\
\label{qpmu}
q'^\mu &=& ( v^0,\;\;\;\;\; 0 ,+v^y,+v^z).
\end{eqnarray}
indicating that we have chosen the fixed quantization plane $x^+=0$ 
(Fig.~\ref{figscat}a).
The incoming and outgoing particles have the same absolute values of
the momenta in the c.m.s. Therefore,
\begin{equation}
|\vec{v}|^2 = (v^x)^2 = (v^y)^2 + (v^z)^2 = |\vec{v'}|^2.
\end{equation}
The Mandelstam variables are
\begin{eqnarray}
\label{mandels}
s &=& (p+q)^2 = 4 (v^0)^2, \\
\label{mandelt}
t &=& (p-p')^2 =-2 |\vec{v}|^2,\\
\label{mandelu}
u &=& (p-q')^2 = -2 |\vec{v}|^2.
\end{eqnarray}
We are now ready to perform the numerical experiments for three cases, 
which are described in \sex \ref{secversus}-\ref{secundert}. In the
experiments two parameters are focused on. We shall vary the 
azimuthal angle $\alpha$ in the $y$-$z$-plane,
\begin{equation}
\alpha = \arctan \frac{v^z}{v^y},
\end{equation}
and the incoming c.m.s.-momentum
\begin{equation}
v = v^x.
\end{equation}
In the remainder we will omit the units for the masses, which are MeV$/c^2$.

\section{Light-front versus instant-form dynamics}
\label{secversus}
One of the claims of LFD is that higher Fock states are more strongly
suppressed than in IFD. We can investigate this claim
for the box diagram easily in the following case.

We take the external states on energy-shell, Eq.~\r{onEcond}, such that the
equality \r{equiv} holds. At the same time we avoid on-shell
singularities for the intermediate states by giving the external
particles a slightly smaller mass $m'$,
\begin{equation}
\label{mprime}
{m'}^2 = p^2 < m^2,
\end{equation}
such that we can still relate the amplitude to an $S$-matrix element. 

The process we look at is described in the previous section  and
has two scalars of mass $m'$ coming in along
the $x$-axis, interacting, and scattered over a scattering angle of
$\pi/2$.  Stretched boxes give maximal contributions (see next section)
if the quantization axis is in the scattering plane, which is the case
if the azimuthal angle $\alpha = \pi/2$.

\def \figdrie{
\begin{figure}
\[
\epsfxsize=7.5cm \epsffile{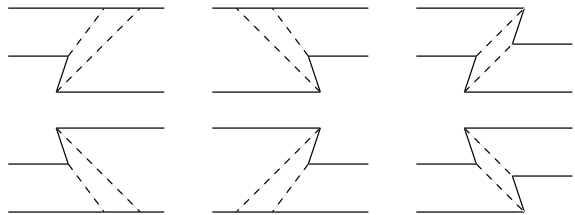}
\]
\vspace{-.5cm}
\caption{\label{figdiagsR5} Time-ordered diagrams that contribute to
${\cal{R}}_5$. The diagrams in the first column have five
particles in the first intermediate state. The diagrams in the second
column have five particles in the last intermediate state, and the
diagrams on the right have five-particle intermediate states for both
the first and third intermediate states.}
\end{figure}
}
\figdrie
${\cal{R}}^{\rm LF}_{4^+}$ is easily found by calculating the stretched
box.  ${\cal{R}}^{\rm IF}_{4^+} = {\cal{R}}^{\rm IF}_4 + 
{\cal{R}}^{\rm IF}_5 + {\cal{R}}^{\rm IF}_6$, however, has 20 nonzero
contributions. As an example, we show the six contributions to
${\cal{R}}_5$ in Fig.~\ref{figdiagsR5}. 

This illustrates why ${\cal{R}}^{\rm LF}_{5} = 0$. All
contributing diagrams contain vacuum creation or annihilation vertices,
which are forbidden by the spectrum condition.
There are 12 diagrams contributing to ${\cal{R}}_6$, and
all contain vacuum creation or annihilation vertices. Therefore,
${\cal{R}}^{\rm LF}_{6}$ vanishes.

We calculated
${\cal{R}}^{\rm IF}_{4^+}$ by subtracting the four diagrams only
containing three particle intermediate states from the full sum. This
sum can be obtained by doing the covariant calculation, or by adding
all LF time-ordered boxes.  Our results are given in
Fig.~\ref{figR4and5}. 
We also calculated ${\cal{R}}^{\rm IF}_{5^+}$.

\def \figvier{
\begin{figure}
\[
\epsfxsize=10.5cm \epsffile{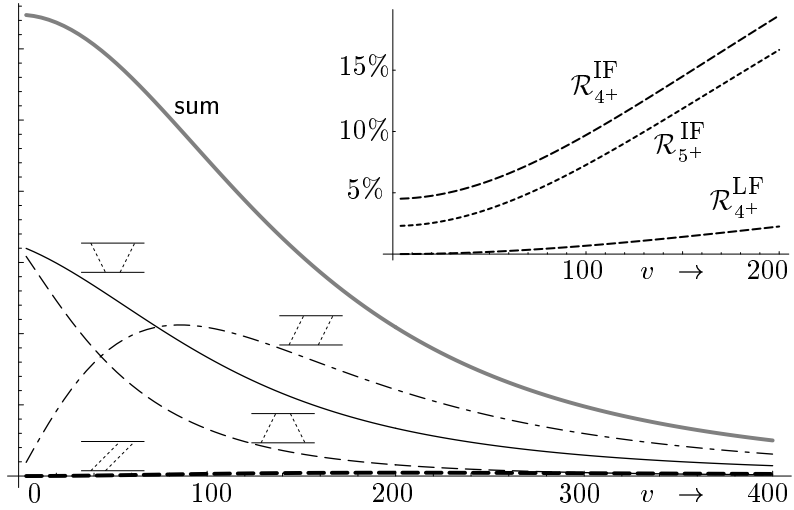}
\]
\caption{\label{figR4and5} LF time-ordered boxes for a scattering angle
of $\pi/2$ as a function of the incoming momentum $v$. We also  
give the ratios of boxes with at least four particles 
(${\cal{R}}^{\rm IF}_{4^+}$ and ${\cal{R}}^{\rm LF}_{4^+}$) or five particles 
(${\cal{R}}^{\rm IF}_{5^+}$, ${\cal{R}}^{\rm LF}_{5^+} =0$) in one
of the intermediate states.} \end{figure}}
\figvier

We conclude that on the light-front contributions of higher Fock states are
significantly smaller than in IFD. In the limit $v \rightarrow 0$, the
ratio ${\cal{R}}^{\rm LF}_{4^+}$ goes to zero, because the phase space
becomes empty. However, in IFD there is a finite contribution of
${\cal{R}}^{\rm IF}_{4^+} = 4.5 \%$ in this limit.  Even if one
includes five-particle intermediate states, the LF is the winner by far.

Note that $m'$, given by Eq.~\r{mprime}, varies as a function of $p^2$,
and therefore also as a function of $v$, but is independent of
$\alpha$. The deviation of $m'$ from $m$ is small: less than 2.3\% for
$v < 200$ and less than 9\% for $v < 400$.  As the deviation of the
mass $m'$ from $m$ is only small, we are convinced that these results are
indicative for calculations above threshold. However, we do not want to
do these calculations, because then one needs to subtract the on-shell
singularities of the equal-time ordered boxes.
\newpage
\section{Numerical results above threshold}
\label{secabovet}

As in \sec \ref{secversus}, we look at the scattering of two particles
over an angle of $\pi/2$. We focus on LFD, and therefore we simply
write ${\cal{R}}_4 = {\cal{R}}^{\rm LF}_{4^+}$.  We do not try to avoid
on-shell singularities by taking different masses for the internal and
external nucleons.  Two nucleons of mass $m= 940$  scatter via the
exchange of scalar mesons of mass $\mu = 140$.  Again, there is a
scalar coupling and no spin is included.

\subsection{Evaluation method}
Contrary to the case considered in \sec \ref{secversus}, 
now it is not straightforward
to evaluate the contributions of the LF time-ordered boxes, because
the nonstretched boxes contain on-shell singularities, thoroughly
analyzed in \sec \ref{seconshell}. Here we briefly sketch
how we deal numerically with the singularities. Using the analysis
of \sec \ref{seconshell}, we identify the singularity $I_{\rm sing}$ 
and rewrite the nonstretched boxes as
\begin{equation}
\label{singsub}
\int {\rm d}^3k \; I = \int {\rm d}^3k (I - I_{\rm sing})
+ \int {\rm d}^3k \; I_{\rm sing}.
\end{equation} 
The integrand $I_{\rm sing}$ has a simple algebraic form, such that
the integration in one dimension over the singularity can be done analytically,
and the remaining integral is regular. This integral is then done
numerically by {\sc mathematica}. 
The integral over $(I - I_{\rm sing})$ was implemented in {\sc fortran}. 
These two numbers are added to give the results presented is
\sex \ref{subsalpha} and \ref{subsvx}.

\def \figurefive{
\begin{figure*}
\[
\hspace{-.4cm}\epsfxsize=16.2cm \epsffile{fig05.eps}
\]
\caption{\label{figabove}Amplitudes above threshold 
from $\alpha = 0$ to $\alpha = 2 \pi$.
Here ${\cal{R}}_4$ is the maximal fraction 
of the stretched box to the absolute value of the sum.}
\end{figure*}
}
\figurefive
\subsection{\label{subsalpha}Results as a function of $\alpha$}
We shall now vary the direction of $\vec{v'}$, given by the azimuthal angle
$\alpha$, however not its length. Therefore the Mandelstam variables are
independent of $\alpha$, and we expect the full amplitude to be invariant.
We tested this numerically for a number of values of ${v}$.
In the region $0 \leq \alpha \leq \pi$, we used the formulas \r{box1} until
\r{box0}. In the region $\pi \leq \alpha \leq 2 \pi$ the diagrams
\r{box2} and \r{box0} vanish. However, then there are contributions
from the diagrams in \r{boxo}.
The results are shown in Fig.~\ref{figabove}.
The results are normalized to the value of the covariant amplitude.
The contributions from the different diagrams vary strongly with the
angle $\alpha$. Since the imaginary parts are always positive, they are
necessarily in the range $[0,1]$ when divided by the imaginary part of
the covariant amplitude. The real parts can behave much more eccentrically,
especially for higher values of the incoming c.m.s.-momentum $v$. An
analysis of the $\alpha$-dependence is given in \sec \ref{seconshell}.
Clearly, the LF time-ordered diagrams add up to the covariant amplitude;
so we see that in all cases we obtain covariant (in particular
rotationally invariant) results for both the real and imaginary parts.
\def \figuresix{
\begin{figure*}
\[
\hspace{-.4cm} \epsfxsize=16.2cm \epsffile{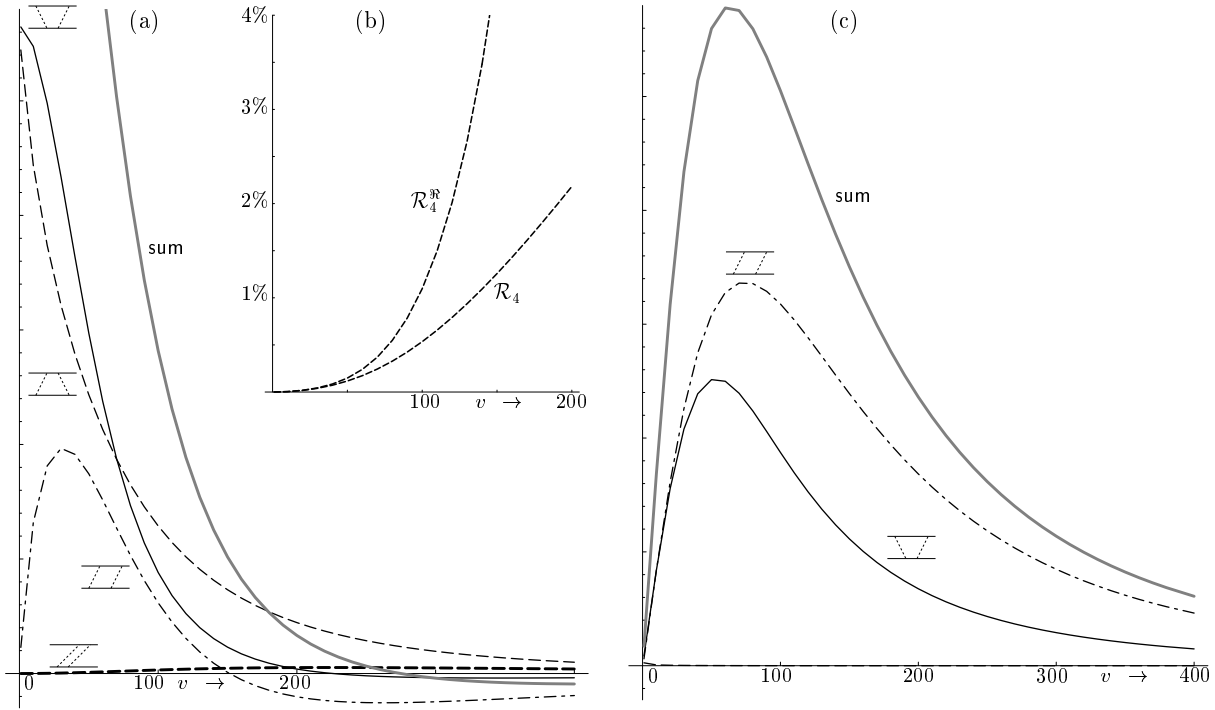} 
\]
\vspace{-1cm}
\caption{\label{figsix} 
Real (a) and imaginary (c) parts of the LF time-ordered boxes above threshold
for $\alpha = \pi/2$ as a function of the momentum
of the incoming particles $v$. The inset (b) shows the ratio of
the stretched box to the real part of the amplitude (${\cal R}_4^{\Re}$)
and to the absolute value (${\cal R}_4$).}
\end{figure*}
}

\newpage
\subsection{\label{subsvx}Results as a function~of~${v}$}
We look at scattering in the $x$-$z$-plane ($\alpha = \pi/2$),
because in that case the contributions from the
stretched boxes are maximized. The results are shown in Fig.~\ref{figsix}.

We depict the ratio of the stretched box, the diagram with two
simultaneously exchanged mesons, to both the real part and to the magnitude
of the total amplitude. Since the real part has a zero near $v= 280$,
the ratio ${\cal R}_4^{\Re}$ becomes infinite at that value of the
incoming momentum. Therefore ${\cal R}_4$ gives a better impression of
the contribution of the stretched box.  We conclude from our numerical
results that the stretched box is relatively small at low energies, but
becomes rather important at higher energies.

\figuresix

\section{Numerical results off energy-shell}
\label{secundert}

In the previous section, we tested covariance of the LF formalism for
amplitudes with on energy-shell external particles, 
by using the c.m.s., where $P^\perp = p'^\perp + q'^\perp = 0$
and $P^z = p'^z + q'^z = 0$.
 However, on the light-front the operator $P^z$ is  dynamical,
 and the last equality does not hold anymore off energy-shell,
as one can easily verify in the following way. 
Consider the case of a bound state with mass ${\cal M} < 2m$,
where ${\cal M}$ is related to the parametric LF energy $P^-$ by the mass-shell
relation
\begin{equation}
P^-=\frac{{\cal M}^2 + {P^\perp}^2}{2 P^+}.
\end{equation}
The bound state is off energy-shell and its mass $\cal{M}$ is smaller than
the sum of the constituent masses. Therefore we have
\begin{equation}
\label{eq1653}
p'^- + q'^- > P^-.
\end{equation} 
The plus and transverse momenta are kinematic; so
\begin{eqnarray}
\label{eq1253}
p'^+ + q'^+ &=& P^+, \\
\label{eq1654}
p'^\perp + q'^\perp &=& P^\perp.
\end{eqnarray} 
Adding Eqs.~\r{eq1653} and \r{eq1253} gives
\begin{equation}
\label{eq1655}
p'^z + q'^z > P^z.
\end{equation} 
If $P^z = 0$, then Eq.~\r{eq1655} implies that $p'^z + q'^z > 0$.
Therefore  the two outgoing particles cannot have exactly opposite momenta
as in Eqs.~\r{ppmu} and \r{qpmu}. In terms of the explicitly covariant
LFD, introduced in \sec \ref{seconshell}, this reflects the fact that the off  
energy-shell relation between $p'+q'$ and $P$ contains an extra four-momentum
like in Eq.~\r{eq29}, below.
What was the reason that
we chose opposite momenta in the previous sections in the first place? 
Our reason  was that we wanted to have a manifest symmetry of the
amplitude, because it is obvious that the Mandelstam variables $s$, $t$,
and $u$ given by Eqs.~\r{mandels}-\r{mandelu} remain the same under the
rotations we investigated.

In the present case where the states are taken off energy-shell, the 
full amplitude is not covariant. We can, however, study this breaking
of covariance by comparing amplitudes that satisfy the conditions
\r{eq1653}-\r{eq1655} and, at the same time, choosing the scattering angle
$\theta$, the plus-momentum $p'^+$ and the magnitude of $p'^\perp$ in such
a way that the Mandelstam variables $s$, $t$, and $u$ remain constant,
while the azimuthal angle $\alpha$ is varied. In the limiting case that
$P^-$ is equal to $p'^- + q'^-$, the amplitude becomes on energy-shell
and the dependence on $\alpha$ in the full amplitude drops.

The variation of the amplitude with $\alpha$ can be compared to the     
relative contribution of the stretched boxes. We shall do that in what
follows, but first we describe in detail the choice of momenta for the
particles.

\subsection{Determination of the momenta}

As in the previous sections, we shall fix the direction of the incoming
particles, as in Eqs.~\r{pmu} and ~\r{qmu}, and vary the direction of
the outgoing particles. 
For on energy-shell amplitudes, there are only two independent Mandelstam
variables. Off energy-shell, more independent Lorentz invariant objects are
found.  We construct the
momenta in such a way that all six inner products between them are constant.
We first look at $\inp{p}{q'}$ and $\inp{p}{p'}$, and later we verify if
the conditions found ensure the invariance of the four others:
\begin{eqnarray} \label{inpq}
 \inp{p}{q'} &=& p^+ q'^- + q'^+ p^-  -
 \inp{p^\perp}{q'^\perp} \hspace{30mm} \\
 &=&x_p P^+\frac{{q'^\perp}^2\!\!+ m^2}{2 (1\!-\!x_{p'}) P^+} +
 (1\!-\!x_{p'}) P^+ \frac{{p^\perp}^2\!\!+ m^2}{2 x_p P^+} - \inperp{p}{q'} ,
 \hspace{-1cm} \nonumber \\ 
 \label{inpp}
 \inp{p}{p'} &=& p^+ p'^- + p'^+ p^- - \inperp{p}{p'} \hspace{30mm} \\
 &=& x_p P^+ \frac{{q'^\perp}^2 + m^2}{2 x_{p'} P^+} + x_{p'} P^+
	  \frac{{p^\perp}^2 + m^2}{2 x_p P^+}  - \inperp{p}{p'}.
 \hspace{-1cm} \nonumber 
\end{eqnarray} 
We have introduced the fractions 
\begin{equation}
\label{xfrac}
x_p = p^+/P^+, \quad x_{p'}=p'^+/P^+.
\end{equation}
Since the perpendicular momenta are conserved, we have in the c.m.s.
$p'^\perp = - q'^\perp, $
so the inner products of the
perpendicular momenta can be written as 
\begin{eqnarray}
\label{inperppp} \inperp{p}{p'} = \phantom{-} |p^\perp| |p'^\perp| \cos \theta,  
\\ 
\label{inperppq} \inperp{p}{q'} = - |p^\perp| |p'^\perp| \cos \theta, 
\end{eqnarray} 
where $\theta$ is the scattering angle.  We can now solve
Eqs.~\r{inpq}-\r{inperppp} for $x_{p'}$, $|p'^\perp|$ and $\theta$.
There are many curves satisfying these conditions. For uniqueness, we
demand that the curve go through the point in which $x_{p'} = x_p =
1/2$, $|p'^\perp| = |p^\perp|$, and $\theta = \pi/2$.  We find that the
curve is then parametrized by

\begin{eqnarray} 
\label{condtheta} \theta &=& \pi/2 ,\\ 
\label{condperp}
\frac{{p'^\perp}^2 + m^2}{x_{p'} (1\!-\!x_{p'})} &=& 4 ({p^\perp}^2 + m^2).
\end{eqnarray}

Writing down the other four inner products between the four-vectors of
the external momenta, we checked that they are invariant
if the momenta satisfy Eqs.~\r{condtheta} and \r{condperp}.
Because the particles come in along the $x$-axis, the
above relations define an ellipse in the $y$-$z$-plane. In the case of IFD
these ellipses reduce to circles with their center at the origin and radius $v$.
Our procedure to obtain numerical values for the momenta was the following. 
We take $p'^x = 0$ and varied $p'^y = |p^\perp| \cos \alpha$. Now
Eq.~\r{condperp} gives us $x_{p'}$. All components of $p'$ and $q'$ are
then easily found.
\begin{figure} 
\[
\epsfxsize=10cm \epsffile{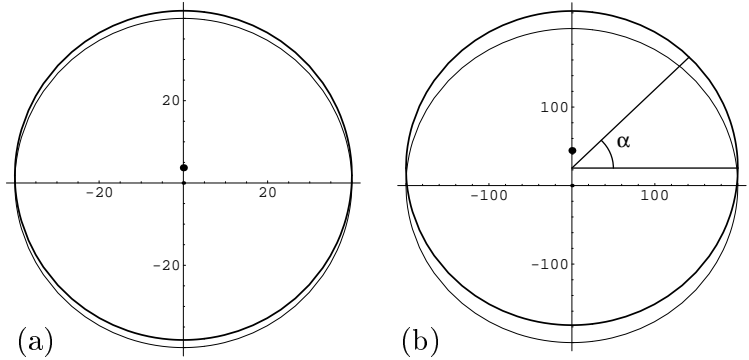}
\]
\vspace{-1cm}
\caption{\label{rotinv} Momenta $p'$ and $q'$ of the outgoing
particles  (thick line) in the scattering plane (horizontal $p'^y$ and $q'^y$;
vertical: $p'^z$ and $q'^z$) for two cases: (a) $v=40$
and (b) $v = 200$. The momentum $p'+ q'$ is indicated by the dot. As a
reference we have drawn the locus for on energy-shell external
particles (thin line): a circle centered at the origin. }
\end{figure} 
In Fig.~\ref{rotinv} we have indicated the $y$ and $z$-components of
the momenta of the external particles for the two cases we investigate.
The $z$-component, not being a LF variable, does not enter the computer
code. We determined it from the relation $p'^z = (p'^+ -
p'^-)/\sqrt{2}$ with the purpose of showing the effect of off-shellness in
this numerical experiment. We see that Eqs.~\r{eq1654} and \r{eq1655}
hold.  The off energy-shell momenta form an ellipse. However, the
deviation from a circle with radius $v$ is hardly visible.

\subsection{Calculation of the amplitude}
\begin{figure}
\[
\epsfxsize=10.5 cm \epsffile{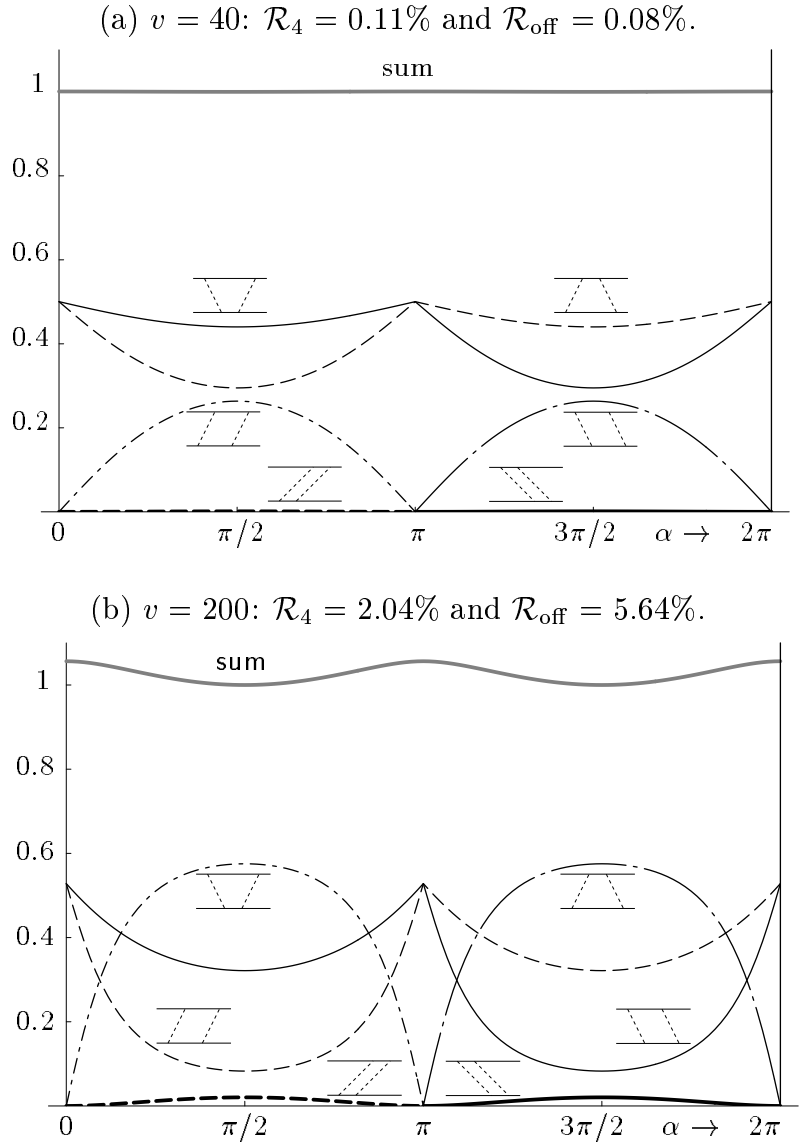}
\]
\vspace{-1cm}
\caption{\label{figoff} The LF time-ordered boxes as a function of
$\alpha$.}
\end{figure}

We did numerical experiments for particles that are weakly bound:
$2 m - {\cal M} = 2\; {\rm MeV}/c^2$.
In Fig.~\ref{figoff} we show the
contributions of the different boxes and their sum as we vary the angle
$\alpha$. The calculations are the same as in the previous section,
using the formulas \r{box1}-\r{boxo}, except that the momenta of the
outgoing particles have changed such that \r{condtheta} and
\r{condperp} are satisfied. As there does not exist a covariant amplitude in
the off energy-shell case, we normalized the curves shown by dividing the amplitudes
by their sum at $\alpha = \pi/2$.

\begin{figure}
\[
\epsfxsize=10.5cm \epsffile{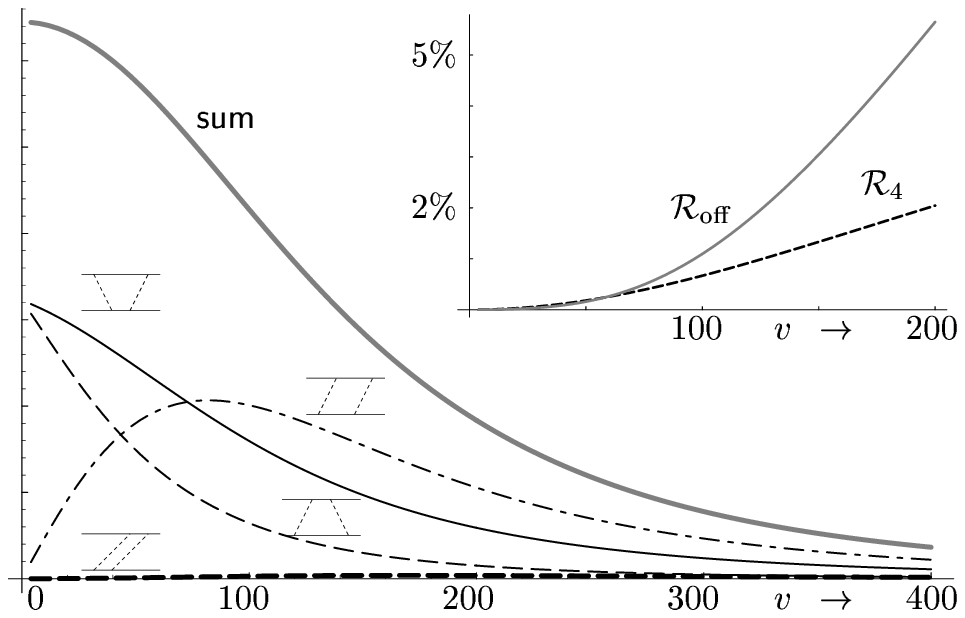}
\]
\vspace{-1cm}
\caption{\label{figmaxoff} LF time-ordered boxes as a function of
$v$ for $\alpha = \pi/2$. The inset shows the
maximum contributions of the stretched box and the maximal breaking
of covariance. }
\end{figure}

In Fig.~\ref{figoff} we see the consequences of the off energy-shell
initial and final states. Condition~\r{onEcond} is violated; therefore,
Eq.~\r{equiv} does not hold and breaking of covariance can be expected.  
We see that the contributions of the higher Fock states ${\cal R}_{4}$
are smaller than the effect of the off-shellness ${\cal R}_{\rm off}$,
defined as the largest difference between two full amplitudes at arbitrary
values of $\alpha$.
This is confirmed in Fig.~\ref{figmaxoff}, in which $\alpha$ is fixed 
and the incoming c.m.s-momentum $v$ is varied. 

From Fig.~\ref{figoff} we infer that the full amplitude is maximal at
$\alpha = 0$  and $\alpha= \pi$. The minimum is reached at 
$\alpha = \pi/2$ and $\alpha= 3 \pi/2$. Therefore the maximal breaking
of covariance of the amplitude can be calculated by taking the
difference of the total sum at the angles $\alpha = 0$ and
$\alpha = \pi/2$. We see that at typical values for incoming
momentum ($v \leq 50$) ${\cal R}_{\rm off}$ is small, even smaller
than ${\cal R}_{4}$. However, at higher momenta it 
dominates over the stretched box. In this region we see that 
the stretched box contributions remain small. 

A detailed explanation of the behavior of the off energy-shell
amplitudes is given in~\ref{secoffshell}.

\section{Analysis of the on energy-shell results}
\label{seconshell}
The angle dependence of the LF time-ordered amplitudes found numerically
can be understood analytically. 
The variation of the LF amplitudes with the 
angle $\alpha$ means that they have singularities in this variable, either
at finite values of $\alpha$ or at infinity. 
They should disappear when they are summed to give the covariant amplitude. 
These singularities can be 
most conveniently analyzed in the explicitly covariant version of LFD 
(see for a review \cite{CDKM98}). In this version the orientation of the 
light-front is given by the invariant equation 
$\omega\cd x=0$. The amplitudes are calculated by the rules of the graph 
technique explained in Ref.~\cite{CDKM98}. 
After a transformation of variables, these amplitudes
coincide with those given by ordinary LFD. However, they are 
parametrized in a different way. The dependence of the amplitudes on 
the angle $\alpha$ means, in the covariant version, that they depend on the 
four-vector $\omega$ determining the orientation of the LF plane:  
$M=M(p,q,p',q',\omega)$. Hence, besides the usual Mandelstam variables 
$s$ \r{mandels} and 
$t$ \r{mandelt} the amplitude $M$ depends on the scalar products of $\omega$ 
with the four-momenta.  Since $\omega$ determines the direction only 
(the theory is invariant relative to the substitution 
$\omega\rightarrow a\omega$), an amplitude should depend on the 
ratios of the scalar products of the four-momenta with $\omega$. 
Hence~\cite{Kar78}
\begin{equation}\label{eq1a}
M=M(s,t,x_p,x_{p'}),
\end{equation}
where
\begin{equation}\label{eq2a}
x_p=\frac{\omega\cd p}{\omega\cd (p+q)},\quad
x_{p'}=\frac{\omega\cd p'}{\omega\cd (p+q)}.
\end{equation}
Formulas \r{eq2a} coincide with the definitions \r{xfrac} if we use
the $z$-axis as the quantization axis.
The $\omega$-dependence is reduced to two scalar variables 
$x_p$ and $x_{p'}$, since the direction of $\vec{\omega}$ is determined by two angles.
Hence, this amplitude should have singularities in the variables $x_p$ and $x_{p'}$.
Their positions will be found below.  The amplitude corresponding to 
the sum of all time-ordered diagrams should not depend on 
$x_p$ and $x_{p'}$. 

\begin{figure}
\[
\epsfxsize=7cm \epsffile{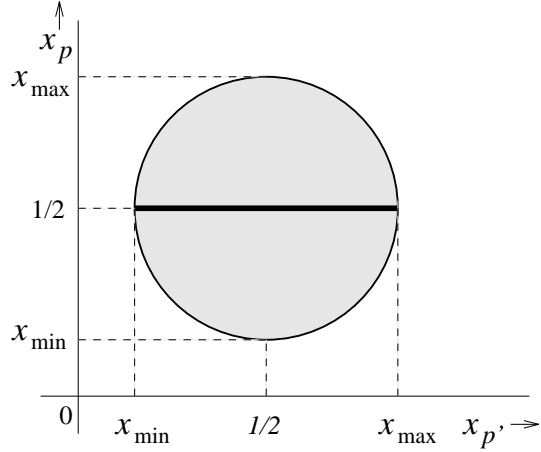}
\]
\vspace{-1cm}
\caption{\label{figphysreg}Physical region in the $x_p$-$x_{p'}$-plane 
for a scattering angle $\theta = \pi/2$. If the incoming particles are
in the $x$-$y$-plane the physical region reduces to the thick line at
$x_p = 1/2$.}
\end{figure}

Let us find the physical domain of the variables $x_p$ and $x_{p'}$, 
corresponding to all possible directions of $\vec{\omega}$ for fixed 
$s,t$. In the c.m.s., the variables Eqs.~(\ref{eq2a}) are represented 
as
\begin{equation}\label{eq3b}
x_p=\frac{1}{2}-\frac{v}{\sqrt{s}}
\inp{\hat{\omega}}{ \hat{p}}, \quad
x_{p'}=\frac{1}{2}-\frac{v}{\sqrt{s}}
\inp{\hat{\omega}}{ \hat{p}'},
\end{equation}
where, e.g., $\inp{\hat{\omega}}{ \hat{p}}$ is the scalar product of the unit
vectors $\hat{\omega} = \vec{\omega}/|\vec{\omega}|$ and $\hat{p} = \vec{p}/|\vec{p}|$
in three-dimensional Euclidian space, and $v=\sqrt{s/4-m^2}$ is the
momentum of the particle in the c.m.s.  The Eqs.~(\ref{eq3b})
determine an ellipse in the $x_p$-$x_{p'}$-plane.  Its boundary is
obtained when $\vec{\omega}$ is in the scattering plane, that is,
$\inp{\hat{n}}{ \hat{p}} = \cos \gamma$ or $\inp{\hat{n}}{ \hat{p}} 
= \cos (\gamma -
\theta)$, where $\gamma$ is the angle between  $\vec{p}$ and
$\vec{\omega}$ in coplanar kinematics and $\theta$ is the
scattering angle in the c.m.s. The case when $\vec{\omega}$ is out of
the scattering plane corresponds to the interior of the ellipse. For a
scattering angle $\theta=\pi/2$, the ellipse turns into a circle,
shown in Fig.~\ref{figphysreg}.

For the kinematics shown in Fig.~\ref{figscat} and Eqs.~\r{pmu}-\r{qpmu},
i.e., when $\vec{\omega}\perp \vec{p}$, it follows from Eqs.~(\ref{eq3b}) 
that the value $x_p$ is fixed: $ x_p=\frac{1}{2},  $
whereas for a given $\alpha$ we obtain
\begin{equation}\label{eq3d}
x_{p'}=\frac{1}{2}-\frac{v}{2v_{0}}\sin\alpha,
\end{equation}
with $v_{0}=\sqrt{m^{2}+v^{2}}$. So $x_{p'}$
varies along a straight line when $\hat{\omega}$ is rotated in the $y$-$z$-plane. 
The bounds of the physical region of $x_{p'}$ are
\begin{equation}\label{eq3c} 
 x_{\rm min} = \frac{1}{2}-\frac{v}{2v_{0}},\quad 
x_{\rm max} = \frac{1}{2}+\frac{v}{2v_{0}}. 
\end{equation}
When $0\leq \alpha \leq \pi/2$, $x_{p'}$ moves from 1/2 to $x_{\rm min}$.
When $ \pi/2 \leq \alpha \leq  \pi$, $x_{p'}$ moves in the 
opposite direction in the same interval.
This explains why all the curves in Figs.~\ref{figabove} 
and~\ref{figoff}  in the interval $0\leq \alpha \leq\pi$ are symmetric
relative to $\alpha=\pi/2$.

When $\pi \leq \alpha \leq 3\pi/2$, $x_{p'}$ moves from 1/2 to $x_{\rm
max}$ and, finally, when $3\pi/2 \leq \alpha \leq 2\pi$, it goes back
in the same interval. As in the previous paragraph, this explains why
all the curves in Figs.~\ref{figabove} and ~\ref{figoff} in the
interval $\pi\leq \alpha\leq 2\pi$ are symmetric relative to
$\alpha=3\pi/2$.
When $\alpha=\pi/2$ and $3\pi/2$, the values of $x_{p'}$ 
are on the boundary of the physical region.

Note also that the amplitudes for the trapezium (dashed and solid curves in 
Figs.~\ref{figabove} and~\ref{figoff}) 
are evidently obtained by the replacement 
$p\leftrightarrow  q, \quad p'\leftrightarrow  q'$, which, according to 
the definition in Eq.~(\ref{eq2a}), corresponds to $x_{p'}\rightarrow 
1-x_{p'}$. This is the same as the
replacement $\alpha\rightarrow 2\pi-\alpha$ in Eq.~(\ref{eq3d}). 
Therefore the curves for the other trapezium, when $\alpha$ goes 
from $2\pi$ to 0, are identical to the curves for the trapezium, when $\alpha$ 
increases from 0 to $2\pi$. The same is true for the other diagrams 
(diamonds and stretched boxes).

\subsection{Trapezium}\label{trap}
The method of finding the singularities of the LF diagrams was 
developed in~\cite{Kar78}. Here we restrict ourselves to the example of   
the diagram~\r{box1}. Its counterpart in the explicitly
covariant LFD is shown in Fig.~\ref{figtrap}.
\def \figelf{
\begin{figure}
\[
\epsfxsize=3.5cm \epsffile{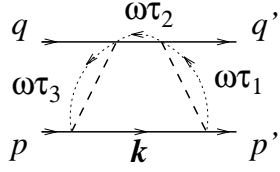}
\]
\vspace{-1cm}
\caption{\label{figtrap}Trapezium in explicitly covariant LFD.}
\end{figure}}
The dotted lines in this figure are associated with fictitious
particles (spurions), with four-momenta proportional to $\omega$. 
The four-momenta of the particles (the spurions not included) are not 
conserved in the vertices. Conservation is restored
by taking into account the spurion four-momentum. In the ordinary 
LF approach, this corresponds to nonconservation of the 
minus-components of the particles. The spurions make up for the difference.

According to the 
rules of the graph technique~\cite{CDKM98}, one should associate with 
a particle line with four-momentum $p$ and mass $m$  the factor 
$\theta(\omega\cd p)\delta(p^2-m^2)$ and associate with the spurion line with
four-momentum $\omega\tau_i$ the factor $1/(\tau_i-i\epsilon)$. 
Then one integrates, with measure d$^4k_i/(2\pi)^3$, over all the 
four-momenta $k_i$ not 
restricted by the conservation laws in the vertices and over all 
$\tau_i$.  The expression for the amplitude of Fig.~\ref{figtrap} is
\begin{eqnarray}
&&\hspace{-.5cm}\Mone=\int\theta(\omega\cd \nonumber
k)\delta(k^2-m^2) 
\theta(\omega\cd (p'-k))\ 
\delta((p'-k + \omega\tau_1)^2-\mu^2)\\ &&\hspace{.5cm}\times
\theta(\omega\cd (p+q-k))\ \delta((p+q-k+\omega\tau_2)^2-m^2) 
\theta(\omega\cd (p-k)) 
\delta((p-k+\omega\tau_3)^2-\mu^2) \nonumber \\ &&\hspace{.5cm}\times
\frac{{\rm d}\tau_1}{(\tau_1-i\epsilon)}\frac{{\rm d}
\tau_2}{(\tau_2-i\epsilon)} 
\frac{{\rm d}\tau_3}{(\tau_3-i\epsilon)}\frac{{\rm d}^4k}{(2\pi)^3}.  
\label{eq1} 
\end{eqnarray}                                                                 Like in Eq.~\r{covbox}, we omit the coupling constant.
Performing the integrals over $\tau_i$ and ${\rm d}k_{0}$ in Eq.~(\ref{eq1})
by means of the $\delta$-functions, we get
\begin{equation}\label{eq1p}
\Mone=\int\frac{\theta(\omega\cd (p'-k))}{\mu^2-(p'-k)^2}\ 
\frac{\theta(\omega\cd (p+q-k))}{m^2-(p+q-k)^2} 
\label{eq2} \frac{\theta(\omega\cd (p-k))}{\mu^2-(p-k)^2}\ 
\theta(\omega\cd k)\frac{{\rm d}^3k}{2\varepsilon_k(2\pi)^3}.
\end{equation}                                                                   
\figelf 

By transformation of variables (see for the details appendix~B of
Ref.~\cite{CDKM98}), expression~(\ref{eq2}) can be transformed
such that it exactly coincides with Eq.~(\ref{box1}).

For Feynman amplitudes the method to find their singularities was developed 
by Landau~\cite{Lan59}. A method very similar to that one can be applied to
time-ordered amplitudes.
If we would omit for a moment the $\theta$-functions in Eq.~(\ref{eq2}) 
and would not take into account that $k^2=m^2$, for finding the singularities
we should construct the function 
$\varphi_1=\alpha_1(\mu^2-(p'-k)^2)+\alpha_2(m^2-(p+q-k)^2) 
+\alpha_3(m^2-(p-k)^2)$ formed from the denominator of Eq.~(\ref{eq1p}).
The singularities of the trapezium are found by putting to zero the 
derivatives of $\varphi_1$ with respect to $\alpha_{1-3}$ and 
with respect to $k$.  However, the trapezium may have singularities 
corresponding to a coincidence of the singularities of its integrand 
with the boundary of the integration domain caused by the presence of the
$\theta$-functions. 
So we must find a conditional extremum. The restrictions 
can be taken into account using Lagrange multipliers \cite{Kar78}. 
Hence we should consider the function
\begin{eqnarray}
\varphi&=&\alpha_1(\mu^2-(p'-k)^2)+\alpha_2(m^2-(p+q-k)^2) \nonumber\\
\label{eq3}
&&+\alpha_3(\mu^2-(p-k)^2)+\alpha_4(k^2-m^2)
+\gamma_1 \omega\cd (k-p'),
\end{eqnarray}
where $\alpha_4$ and $\gamma_1$ are the Lagrange multipliers.
One should also consider the functions obtained from 
$\varphi$ at $\alpha_1=0$, subsequently at $\alpha_2=0$, at $\alpha_3=0$, 
at $\alpha_1=\alpha_2=0$, et cetera.  One should not consider the 
function obtained from Eq.~(\ref{eq3}) by $\alpha_4=0$, since the 
integral (\ref{eq2}) contains the three-dimensional integration volume
${\rm d}^3k$, and the condition $k^2=m^2$ cannot be removed. 
Therefore there is no need to introduce the term 
$\gamma_2\omega\cd k$, since the $k^2=m^2$ condition prevents
this term from being zero and, hence, does not impose any restrictions. 
The case $\gamma_1=0$ reproduces the singularities of the Feynman 
graph. Therefore below we shall consider the case 
$\gamma_1\neq 0$ resulting in the singularities in the 
variables $x_p$ and $x_{p'}$.
We suppose that  
$\omega\cd p \geq \omega \cd p'.$ 
This corresponds to the condition $p^+\geq p'^{+}$ of \sec \ref{secbox}.
In the kinematics shown in Fig.~\ref{figscat} this means that $x_{p'}\leq
1/2$ and $0\leq \alpha\leq \pi$. In this case, the second and third
$\theta$-functions in Eq.~(\ref{eq2}) do not give any restrictions and can be
omitted. Therefore we omit also the term $\gamma_3 \omega\cd (p-k) +
\gamma_4 \omega\cd (p+q-k)$.

The derivatives of $\varphi$ with respect to $k$, the $\alpha$'s, 
and $\gamma_1$ give
\begin{equation}
\partial \varphi/\partial k =\alpha_1 2(p'-k)+\alpha_2 2(p+q-k)
\label{eq19}   +\alpha_3 2(p-k)+\alpha_4 2k+\gamma_1 \omega=0,
\end{equation}
with
\begin{eqnarray}
(p'-k)^2 & = & \mu^2, \quad (p+q-k)^2=m^2, \nonumber\\
\label{eq18a}
(p-k)^2  & = & \mu^2, \quad k^2  =  m^2, \quad \omega\cd k=\omega\cd p'.
\end{eqnarray}
We multiply Eq.~(\ref{eq19}) in turn by $(p'-k)$, $(p+q-k)$, et cetera, 
and get
\begin{equation}\label{eq19a}
\begin{array}{lcl}
(\ref{eq19})\times p'-k&:&  \alpha_1 2\mu^2 +\alpha_2 \mu^2
 +\alpha_3(2\mu^2-t) -\alpha_4 \mu^2 =0,\\[0.3cm]
(\ref{eq19})\times p+q-k&:& \alpha_1\mu^2 +\alpha_2 2m^2  
+\alpha_3 \mu^2 +\alpha_4 (s-2 m^2) +\gamma_1 (1-x_{p'})=0,\\ [0.3cm]
(\ref{eq19})\times p-k&:& \alpha_1 (2\mu^2-t)+\alpha_2 \mu^2 
+\alpha_3 2\mu^2-\alpha_4 \mu^2 +\gamma_1 (x_p-x_{p'}) =0,\\[0.3cm]
(\ref{eq19})\times k&:&-\alpha_1\mu^2 +\alpha_2 (s-2m^2) 
-\alpha_3 \mu^2+2\alpha_4 m^2 + \gamma_1 x_{p'} =0, \\ [0.3cm]
(\ref{eq19})\times \omega&: & \alpha_2 (1-x_{p'}) +\alpha_3 
(x_p-x_{p'}) 
+\alpha_4 x_{p'}=0 .
\end{array} 
\end{equation}
These equations have a nontrivial solution if and only if
\begin{equation}\label{eq19b}
\left|
\begin{array}{lllll}
2\mu^2       & \mu^2   &(2\mu^2-t)   & -\mu^2   & 0 \\ [0.3cm]
\mu^2        & 2m^2    & \mu^2       &(s-2 m^2) &(1-x_{p'})\\ [0.3cm]
(2\mu^2-t)   & \mu^2   & 2\mu^2      &-\mu^2    &(x_p-x_{p'}) \\[0.3cm]
-\mu^2       & (s-2m^2)& -\mu^2      & 2m^2     &  x_{p'}  \\ [0.3cm]
0            & (1-x_{p'}) & (x_p-x_{p'})   & x_{p'}      & 0  
\end{array} 
\right| =0.
\end{equation}
Eq.~(\ref{eq19b}) is quadratic in $x_{p'}$. Its 
solution is simple but lengthy. We show it for the particular case of 
the kinematics of Fig.~\ref{figscat} supposing that the particles in the
c.m.s. have momenta~$v$. In this case 
$s$ and $t$ are given by Eqs.~(\ref{mandels}) and (\ref{mandelt}).
The solution of Eq.~(\ref{eq19b}) is
\begin{equation}\label{eq19d}
x_{p'}^0=\frac{1}{2}\pm \frac{v}{2v_{0}}
\frac{\sqrt{2\mu^4+8\mu^2v^2+4v^4}}
{\sqrt{\mu^4+8\mu^2v^2+4v^4}}.
\end{equation}
The position of the singularity in the variable $x_{p'}$ is denoted by $x^0_{p'}$.  
According to Landau~\cite{Lan59}  the behavior in the vicinity of $x_{p'}^0$ should be
either logarithmic, proportional to $ |x_{p'}-x_{p'}^0|^\beta$, or
proportional to $|x_{p'}-x_{p'}^0|^\beta \log(x_{p'}-x_{p'}^0)$, where
$\beta$ is a noninteger number.

When $\mu\ll v$ we find from Eq.~(\ref{eq19d})
\begin{equation}\label{20a} 
x_{p'}=\frac{1}{2}\pm \frac{v}{2v_{0}}
(1+\frac{\mu^4}{8v^4}).
\end{equation}
Comparing with Eq.~(\ref{eq3c}), we see that
at small $\mu$ or at large $v$ the singularities come 
closer to the physical region of $x_{p'}$.  We will see below that this 
will be a property of all the singularities depending on $\mu$ and 
$v$. This explains the numerical results, showing that with an increase 
of $v$ the graphs of the amplitudes versus $\alpha$ become more 
sharply peaked.
 
Now consider the case when one of the $\alpha$'s is zero. Let $\alpha_1=0$. 
Then Eq.~(\ref{eq3}) is reduced to
\begin{equation}
\varphi=\alpha_2(m^2-(p+q-k)^2) 
+\alpha_3(\mu^2-(p-k)^2)
\label{eq3e}
+\alpha_4(k^2-m^2)
+\gamma_1 \omega\cd (k-p').
\end{equation}
Similarly to the previous case, we get an equation for the
singularities, which can be obtained
from Eq.~(\ref{eq19b}) by deleting the first row and column:
\begin{equation}\label{eq13}
\left|
\begin{array}{lllll}
 2m^2    & \mu^2         &(s-2 m^2) &(1-x_{p'})\\ [0.3cm]
 \mu^2     & 2\mu^2        &-\mu^2      &(x_p-x_{p'}) \\[0.3cm]
 (s-2m^2)& -\mu^2        & 2m^2     &  x_{p'}  \\ [0.3cm]
 (1-x_{p'}) & (x_p-x_{p'})   & x_{p'}      & 0  
\end{array} 
\right| =0.
\end{equation}
Under the given kinematical conditions,
its solution with respect to $x_{p'}$ is
\begin{equation}\label{eq14}
x_{p'}^0=\frac{1}{2}\pm 
\frac{\mu}{4v}\sqrt{\frac{\mu^2+4v^2}{m^2+v^2}}.
\end{equation}
In the limit $\mu\rightarrow 0$ or $v\rightarrow \infty$, these singularities  
are again approaching the physical region.

Let $\alpha_2=0$. The singularity condition
 is obtained from Eq.~(\ref{eq19b}) by 
deleting the second row and column:
\begin{equation}\label{eq15}
\left|
\begin{array}{lllll}
2\mu^2       &(2\mu^2-t)   & -\mu^2   & 0 \\ [0.3cm]
(2\mu^2-t)   & 2\mu^2       &-\mu^2      &(x_p-x_{p'}) \\[0.3cm]
-\mu^2       & -\mu^2        & 2m^2     &  x_{p'}  \\ [0.3cm]
0            & (x_p-x_{p'})   & x_{p'}      & 0  
\end{array} 
\right| =0.
\end{equation}
Its solution reads
\begin{equation}\label{eq16}       
x_{p'}^0=\frac{x_p\mu(4m^2\mu-\mu^3- \mu t
\pm 2\sqrt{t}\sqrt{tm^2+\mu^4-4m^2\mu^2})}
{4m^2\mu^2-(t-\mu^2)^2}.
\end{equation}
In the limit $v\rightarrow \infty$, it is simplified:
\begin{equation}\label{eq16a}       
x_{p'}^0=-\frac{\mu^2}{4v^2}\pm \frac{\mu m}{2v^2}.
\end{equation}

Let $\alpha_3=0$.
The singularity condition
 is obtained from Eq.~(\ref{eq19b}) by deleting the third row
and column:
\begin{equation}\label{eq17}
\left|
\begin{array}{lllll}
2\mu^2       & \mu^2   & -\mu^2   & 0 \\ [0.3cm]
\mu^2        & 2m^2    &(s-2 m^2) &(1-x_{p'})\\ [0.3cm]
-\mu^2       & (s-2m^2)& 2m^2     &  x_{p'}  \\ [0.3cm]
0            & (1-x_{p'}) & x_{p'}      & 0  
\end{array} 
\right| =0.
\end{equation}
The determinant in Eq.~\r{eq17} can be evaluated:
\begin{equation}\label{eq17a}
4sx_{p'}^2-4sx_{p'}+4m^2-\mu^2=0.
\end{equation}
The solutions of Eq.~(\ref{eq17a}) are
\begin{equation}\label{eq18}
x_{p'}^0=\frac{1}{2} \pm \frac{\sqrt{v^{2}+\mu^2/4}}{2v_{0}}.
\end{equation}
For $\mu\rightarrow 0$ they also approach the boundary of the physical 
region of $x_{p'}$.

Now consider the cases when a number of coefficients are zero. 
Let $\alpha_1=\alpha_3=0$. 
The singularity condition
 can be obtained from Eq.~(\ref{eq19b}) by deleting the 
first and third rows and columns: 
\begin{equation}\label{eq6a}
\left|
\begin{array}{lllll}
 2m^2           &(s-2 m^2) &(1-x_{p'})\\ [0.3cm]
 (s-2m^2)       & 2m^2     &  x_{p'}  \\ [0.3cm]
 (1-x_{p'})       & x_{p'}      & 0  
\end{array} 
\right| =0.
\end{equation}
This equation reduces to
\begin{equation}\label{eq7}
x_{p'}^2-x_{p'} s+m^2=0.
\end{equation}
Its solutions are
\begin{equation}\label{eq8}
x_{p'}^0=\frac{1}{2} + \frac{v}{2v_{0}}=x_{{\rm max}}, \; \;
x_{p'}^0=\frac{1}{2} - \frac{v}{2v_{0}}=x_{{\rm min}}.
\end{equation}
Since we consider the interval $0\leq \alpha\leq \pi$ corresponding to 
$x_{\rm min} \leq x_{p'} \leq1/2$, the singularity 
at $x_{p'}^0=x_{\rm max}$ is beyond the 
physical region, whereas the singularity at 
$x_{p'}^0=x_{\rm min}$ is just on the boundary of 
the physical region. The amplitude in this point gets an 
imaginary part:
\begin{eqnarray}
\label{eq1001}
{\rm Im}\;\Mone \; \neq 0 {\;\;\rm at \;\;} x_{p'}> x_{\rm min},\\ 
\label{eq1002}
{\rm Im}\;\Mone \;     =0 {\;\;\rm at \;\;} x_{p'}= x_{\rm min}.
\end{eqnarray}
Eq.~\r{eq1002} corresponds to $\alpha= \pi/2$. This explains why all
the dashed curves of the imaginary parts in Fig.~\ref{figabove} go through zero
at the point $\alpha=\pi/2$.

Now put $\alpha_1=\alpha_2=0$. The corresponding 
singularity condition
 is obtained from Eq.~(\ref{eq19b}) by deleting the first and
second rows and columns:
\begin{equation}\label{eq9}
\left|
\begin{array}{lllll}
 2\mu^2        &-\mu^2      &(x_p-x_{p'}) \\[0.3cm]
 -\mu^2        & 2m^2     &  x_{p'}  \\ [0.3cm]
 (x_p-x_{p'})   & x_{p'}      & 0  
\end{array} 
\right| =0.
\end{equation}
Eq.~(\ref{eq9}) reads
\begin{equation}\label{eq10}       
x_{p'}^2m^2-x_{p'} x_p(2m^2-\mu^2)+x_p^2m^2=0.
\end{equation}
Its solution is
\begin{equation}\label{eq11}       
x_{p'}^0= \frac{x_{p}}{2m^2}\left(2m^2-\mu^2 
\pm i\mu\sqrt{4m^2-\mu^2}\right). 
\end{equation}
These two singularities are fixed points in the complex plane.
At $x_p=1/2$ and $\mu\ll m$, they are approaching the point 
$x_{p'}=1/2$ in the physical region, i.e., $\alpha=0$ and $\alpha=\pi$.

The case $\alpha_2=\alpha_3=0$ leads to the 
singularity condition
 obtained from Eq.~(\ref{eq19b}) by deleting the second and
third rows and columns:
\begin{equation}\label{eq12}
\left|
\begin{array}{lllll}
2\mu^2      & -\mu^2   & 0 \\ [0.3cm]
-\mu^2      & 2m^2     &  x_{p'}  \\ [0.3cm]
0           & x_{p'}   & 0  
\end{array} 
\right| =0.
\end{equation}
It gives $x_{p'}^{0}=0$. This is a fixed singularity in the nonphysical 
region.

Above, we have considered the region $ \omega\cd p \geq \omega\cd p'$.
In the region $ \omega\cd p \leq \omega\cd p' $ the integration domain
is restricted by the step function $\theta(\omega\cd (p-k))$ instead
of $\theta(\omega \cd (p'-k))$ in Eq.~(\ref{eq2}). 
The integrals defining these amplitudes define different analytic
functions depending on the region considered.
In the point $x_{p'}=1/2$, i.e., at $\alpha=0$ and $\alpha=\pi$, the
values of the functions coincide, but their analytic behavior is different.

This can indeed be seen in Fig.~\ref{figabove}. The slopes at $\alpha=0$ and 
$\alpha=\pi$ are different.

\subsection{Diamond}
\begin{figure}
\[
\epsfxsize=3.5cm \epsffile{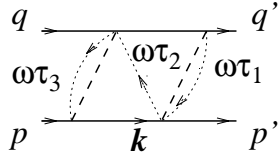}
\]
\vspace{-1cm}
\caption{\label{figdiamond}Diamond in explicitly covariant LFD.}
\end{figure}

The diamond corresponding to Eq.~\r{box2}
 is shown in Fig.~\ref{figdiamond}. 
The analytical expression is
\begin{eqnarray}\label{eq23} 
&&\hspace{-.5cm}\Mtwo =\int\theta(\omega\cd 
k)\delta(k^2-m^2) 
\theta(\omega\cd (k-p')) 
\delta((k-p'+\omega \tau_1 - \omega\tau_2)^2-\mu^2)
\nonumber \\&&\hspace{.5cm}\times
\theta(\omega\cd (p+q-k)) \delta((p+q-k+\omega\tau_2)^2-m^2) 
\theta(\omega\cd (p-k))                                       
 \delta((p-k+\omega\tau_3)^2-\mu^2) \nonumber\\ &&\hspace{.5cm}\times
\frac{{\rm d}\tau_1{\rm d}\tau_2{\rm d}\tau_3} 
 {(\tau_1-i\epsilon)(\tau_2-i\epsilon)(\tau_3-i\epsilon)} 
 \frac{{\rm d}^4k}{(2\pi)^3}. 
\end{eqnarray}                                                   
Performing the integrations in  Eq.~(\ref{eq23}) over $\tau_i$, we get
\begin{eqnarray}\label{eq24} 
\Mtwo\nonumber &=& 
\int\frac{\theta(\omega\cd (k-p'))}{\mu^2-(k-p')^2
+\frac{\displaystyle{\omega\cd (k-p')}}{\displaystyle{\omega\cd 
(p+q-k)}}[m^2-(p+q-k)^2]}
\nonumber\\
&&\times\frac{\theta(\omega\cd (p+q-k))}{m^2-(p+q-k)^2} 
\frac{\theta(\omega\cd (p-k))}{\mu^2-(p-k)^2}\ 
\theta(\omega\cd k)\frac{{\rm d}^3k}{2\varepsilon_k(2\pi)^3}.  
\end{eqnarray}

We still suppose that $\omega\cd p > \omega \cd p'$. However, now,
in contrast to the trapezium, $\omega \cd p' \leq \omega \cd k \leq 
\omega \cd p,$
and both restrictions have to be taken into account.

In order to find the singularities, one should consider the extremum of 
the function:
\begin{eqnarray}\label{eq25} 
\varphi&=&\alpha_1\left\{\mu^2-(k-p')^2
+\frac{\displaystyle{\omega\cd (k-p')}}{\displaystyle{\omega\cd 
(p+q-k)}}(m^2-(p+q-k)^2)\right\}
\nonumber\\
&&+\alpha_2\left\{m^2-(p+q-k)^2\right\} 
+\alpha_3\left\{\mu^2-(p-k)^2\right\}
+\alpha_4(k^2-m^2) 
\nonumber\\[0.1cm]
&&+\gamma_1\omega\cd (k-p')
+\gamma_2\omega\cd(k-p).
\end{eqnarray}  
At $\omega\cd p'=\omega\cd p$, i.e., at $\alpha=0$ and $\alpha=\pi$, the
integration domain vanishes and the diamond becomes zero, as shown in
Fig.~\ref{figabove}.  It remains zero in the interval $\pi \leq
\alpha\leq 2\pi$.


\subsection{Stretched box}
\begin{figure}
\[
\epsfxsize=3.5cm \epsffile{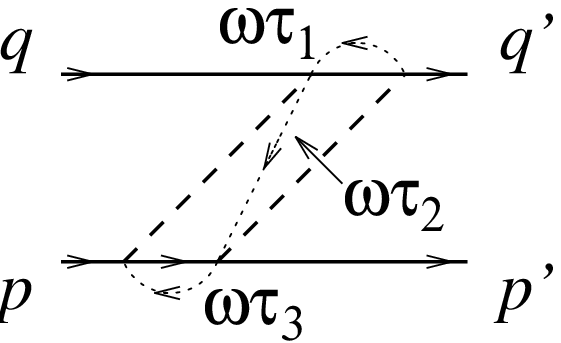}
\]
\vspace{-1cm}
\caption{\label{figsbox}Stretched box in explicitly covariant LFD.}
\end{figure}

The stretched box, corresponding to Eq.~\r{box0}, is shown in 
Fig.~\ref{figsbox}. 
The analytical expression is
\begin{eqnarray}
\Mthree&=&\int\theta(\omega\cd 
k)\delta(k^2-m^2) \nonumber 
\theta(\omega\cd (k-p')) 
\delta((k-p'+\omega\tau_2-\omega\tau_3))^2 -\mu^2) \nonumber\\
&\times&\theta(\omega\cd (p+q-k)) \delta((p+q-k +\omega\tau_1 + 
\omega\tau_3 -\omega \tau_2)^2 -m^2) \nonumber \\ &\times&
\theta(\omega\cd (p-k)) \delta((p-k+\omega \tau_3)^2 -\mu^2) 
\nonumber
\frac{{\rm d}\tau_1{\rm d}\tau_2{\rm d}\tau_3} 
 {(\tau_1-i\epsilon)(\tau_2-i\epsilon)(\tau_3-i\epsilon)} 
 \frac{{\rm d}^4k}{(2\pi)^3}\ .\\ \label{eq26}
\end{eqnarray} 
\newpage
Performing the integrations in  Eq.~(\ref{eq26}) over $\tau_i$, we get
\begin{eqnarray}\label{eq27} 
\Mthree&=&\int
\frac{\theta(\omega\cd (p+q-k))}{m^2-(p+q-k)^2
+\frac{\displaystyle{\omega\cd (p+q-k)}}{\displaystyle{\omega\cd 
(k-p')}}[\mu^2-(k-p')^2]}
\\
&&\hspace{-1cm}\times\frac{\theta(\omega\cd (k-p'))}{\mu^2-(k-p')^2+
\frac{\displaystyle{\omega\cd (k-p')}}{\displaystyle{\omega\cd 
(p-k)}}[\mu^2-(p-k)^2]} 
\frac{\theta(\omega\cd (p-k))}{\mu^2-(p-k)^2}\ 
\theta(\omega\cd 
k)\frac{{\rm d}^3k}{2\varepsilon_k(2\pi)^3}.    
\nonumber
\end{eqnarray}
In order to find the singularities, one must consider the extremum of 
the function:
\begin{eqnarray}\label{eq28} 
\varphi&=&\alpha_1\left\{m^2-(p+q-k)^2
+\frac{\displaystyle{\omega\cd (p+q-k)}}{\displaystyle{\omega\cd 
(k-p')}}(\mu^2-(k-p')^2)\right\}
\nonumber\\
&&+\alpha_2\left\{\mu^2-(k-p')^2+
\frac{\displaystyle{\omega\cd (k-p')}}{\displaystyle{\omega\cd 
(p-k)}}(\mu^2-(p-k)^2)\right\}
\\
&&+\alpha_3\{ \mu^2-(p-k)^2\} +\alpha_4(k^2-m^2) 
\nonumber
+\gamma_1\omega\cd (k-p')+\gamma_2\omega\cd(k-p).
\end{eqnarray}          

Calculating the derivative of Eqs.~(\ref{eq3e}), (\ref{eq25}), and
(\ref{eq28}), for example, with respect to $\alpha_{1-4}$, at
$\gamma_{1}=\gamma_{2}=0$, one finds identical equations determining
the singularities. These do not depend on $x_{p}$ and $x_{p'}$, and coincide
with the ones of the Feynman graph.  Similarly, one can see that any
singularity depending on $x_{p}$ and $x_{p'}$ cannot appear in a
separate diagram only. It appears at least in two amplitudes. These
singularities cancel each other in the sum of the amplitudes.


\section{Analysis of the off energy-shell results}
\label{secoffshell}
\begin{figure}
\hspace{-.2cm}
\[
\epsfxsize=8.5cm \epsffile{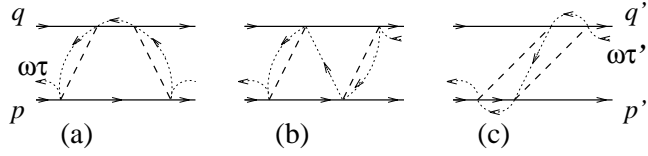}
\]
\vspace{-1cm}
\caption{\label{figkoff}Off energy-shell amplitudes in explicitly covariant
LFD: (a) The trapezium. (b) The diamond. (c) The stretched box. The external
momenta are the same for all diagrams.}
\end{figure}

The off energy-shell amplitude is shown graphically in Fig.~\ref{figkoff}.  
It contains incoming
and outgoing spurion lines with the momenta $\omega\tau$ and $\omega\tau'$,
respectively.  The conservation law has the form
\begin{equation}\label{eq29}
p+q-\omega\tau = p'+q'-\omega\tau' = P.
\end{equation}
From Eq.~\r{eq29} one can infer that, if $\vec{P}=0$, then $\vec{p'} + \vec{q'} \neq 0$, as
was indicated for the $z$-components in \sec \ref{secundert}.
To parametrize the off energy-shell amplitude, we introduce different
initial and final Mandelstam variables $s$
\begin{equation} s=(p+q)^2,\quad s'= (p'+q')^2, \end{equation}
and the total mass squared:
\begin{equation}  {\cal M}^2= (p+q-\omega\tau)^2= (p'+q'-\omega\tau')^2. \end{equation}
So, in general, the off energy-shell amplitude is parametrized as
\begin{equation}\label{0}
M=M(s,s',{\cal M}^2,t,x_p,x_{p'}).
\end{equation}
The on energy-shell amplitude Eq.~(\ref{eq1a}) is obtained from Eq.~(\ref{0})
by the substitution $s=s'={\cal M}^2$. One can also consider the half off energy-shell
amplitude with one incoming or outgoing spurion line. It is obtained
from Eq.~(\ref{0}) by the substitutions $s={\cal M}^2\neq s'$ or $s'={\cal M}^2\neq s$.

In the case of the trapezium, Fig.~\ref{figkoff}a, the external spurion
lines enter and exit from the diagram at the same points as the momenta
$p$ and $p'$. So they can be incorporated by the replacement
\begin{equation}
p \rightarrow p-\omega\tau,\quad p'\rightarrow p'-\omega\tau'.
\end{equation}
This
corresponds to new masses of the initial and final particles for the
bottom line:

\begin{eqnarray}
m_i^2=(p-\omega\tau)^2=m^2-x_p(s-{\cal M}^2), \nonumber\\
m_f^2=(p'-\omega\tau')^2=m^2-x_{p'}(s'-{\cal M}^2). 
\end{eqnarray}
With these new masses, one can repeat the calculations of \sec \ref{trap} 
and find the singularities of the off energy-shell amplitude
for the trapezium. The masses of the intermediate particles are not
changed.

For other diagrams, both for the diamond and the stretched box, in contrast to 
the 
trapezium,  the spurion line enters in the point where the momenta $q'$
go out from the graph. This means that the calculation has to be done with
the following external mass of this particle:
\begin{equation}m'^2=m^2\rightarrow (q'-\omega\tau)^2=m^2-(1-x_{p'})(s'-{\cal M}^2),\end{equation}
whereas the mass of the particle with momentum $p'$ is $m$.

As in the case when all masses are equal, the sum of all time-ordered
graphs with masses different from the internal ones, but 
the same in all the time-ordered graphs, would not depend on $\omega$.  
However,
now we take the sum of the graphs with different external masses in
different particular graphs. This sum cannot be obtained by the time-ordering 
of a given Feynman graph. In this case the $\omega$-dependence is
not eliminated in the sum of all the graphs, and the exact off
energy-shell amplitude in a given order still depends on $\omega$.
An example of this dependence is shown in Fig.~\ref{figoff}.

The off-shell amplitude is not a directly observable quantity. It may
enter as part of  a bigger diagram. Therefore, the off shell 
amplitude may depend on $\omega$. This $\omega$-dependence is not forbidden
by covariance and, hence, does not violate it. On energy-shell, this
dependence disappears.

\section{\label{secconc5} Conclusions}
 
If sufficient caution is exercised, invariance of $S$-matrix elements can be
maintained in Hamiltonian formulations of field theory. A necessary condition
to be fulfilled is that all Fock sectors included in the Feynman diagrams that
contribute to a perturbative approximation of the $S$-matrix be retained.
For the specific case of scalar field theory at fourth order in the coupling
constant, we have determined the 
magnitude of the breaking of covariance if only
the diagrams generated by the ladder approximation to Hamiltonian dynamics are
included. The remaining terms, the stretched boxes, were found to contribute a 
small fraction, less than 2\% for small to intermediate c.m.s. momenta, of the
total amplitude. This fraction is, however, increasing with energy.

It was found, in a calculation closely approximating the first one, that the
breaking of Lorentz invariance is much larger in IFD than in 
LFD, confirming quantitatively what has been claimed in the
literature.

In both  cases we determined quantitatively the dependence of the six LF
time-ordered diagrams on the orientation of the light-front. We verified that,
although the individual diagrams depend strongly on the orientation, their sum
does not, as it should not. This dependence of individual diagrams may be
interpreted as a breaking of rotational invariance.

Having established numerically that invariance of the $S$-matrix
elements is obtained only if all Fock sectors relevant to a certain
order in perturbation theory are included, we extended our
investigation to amplitudes that are off energy-shell. Such amplitudes
are not $S$-matrix elements, calculated between asymptotic states, from
$-\infty$ to $+\infty$ in time. They are elements of an $S$-matrix
calculated for finite light-front time, i.e., defined on a light-front
in the interaction region, not moved to $\pm\infty$ \cite{CDKM98}.
Therefore they depend on the orientation of this light-front. They
either occur as parts of larger diagrams that are invariant, or in the
calculation of LF wave functions. Not being invariant, the sum of the
six LF time-ordered diagrams corresponding to the box is expected to
depend on the orientation of the light-front. We found the variation of
the sum of these six diagrams to grow more strongly with increasing
relative momentum than the fraction carried by the stretched boxes.

All these results point to the conclusion that for low and intermediate
momenta, e.g., those relevant for the bulk of the deuteron wave
function, the higher Fock components are very small and are expected to
play a minor role in LFD. We conjecture that this conclusion
remains essentially valid for higher orders in perturbation theory.

Two remarks are in order here. First, if bound or scattering states at high
values of the relative momentum are to be calculated, the higher Fock states 
will become much more important. Second, in the present work we neglected
spin. It remains to be seen to what extent the special effects of spin, notably
instantaneous propagators, will influence our conclusions.

A final point concerns the dependence of the individual diagrams on the 
orientation of the light-front. By an analysis very close to the Landau method
for Feynman diagrams, we were able to explain all the peculiarities of the
angular dependence in terms of the occurrence and position of singularities of 
the time-ordered diagrams as a function of the angles and their locations. In
particular, the symmetries of the angular dependence and the cusps showing up at
specific orientations could be explained fully.

\chapter{\label{chap6}Summary and conclusions}
\begin{quote}
{\em
I often use the analogy of a chess game: one can learn all the
     rules of chess, but one doesn't know how to play well. The
     present situation in physics is as if we know chess, but we
     don't know one or two rules. But in this part of the board where
     things are in operation, those one or two rules are not operating
     much and we can get along pretty well without understanding
     those rules. That's the way it is, I would say, regarding the
     phenomena of life, consciousness and so forth.
}

\raggedleft Richard Feynman~\cite{SS}
\end{quote}

With this thesis we hope to bring the light-front formulation of Hamiltonian
dynamics closer to its final goal; the calculation of the spectrum of
bound states such as small nuclei and hadrons. In particular, we
hope to contribute to this advance by supplying a better
understanding of the basics of LFD.

In Chapter~\ref{chap1} we have explained some of the pros and cons of
the use of LFD. A Hamiltonian method is intuitively appealing. However,
a general feature of Hamiltonian methods is that manifest covariance is
lost. Our work deals with two topics in LFD, both concerning covariance:
equivalence of light-front and covariant perturbation theory, and
the entanglement of Fock-space expansion and covariance. 

\subsubsection{Equivalence of light-front and covariant perturbation theory}
The first part of this thesis is devoted to the proof of
equivalence between covariant perturbation theory and LFD in the Yukawa
model.  The Yukawa model is explained in Chapter~\ref{chap2}, where
also its longitudinal and transverse divergences are classified. The
proof takes places at the diagrammatic level; following Kogut and
Soper~\cite{KS70} a covariant diagram is integrated over the LF energy
component $k^-$. For convergent diagrams a rigorous proof of the
construction of all LF time-orderings is given by
Ligterink~\cite{LB95b,Lig95}. In the case of longitudinal divergences,
i.e., when the integration over $k^-$ is divergent, this proof needs to
be extended. This is done in Chapter~\ref{chap3} where also the concept
of minus regularization is introduced. In analogy to the well-known
BPHZ regularization method, minus regularization was developed by
Ligterink~\cite{LB95a,Lig95}. It is formulated in terms of LF
coordinates and we find that minus regularization exactly kills the
part of the amplitude that is ill-defined, the so-called forced
instantaneous loop (FIL). The FIL is a diagram with so many
instantaneous fermions that no energy denominators remain. Therefore it
is {\em not} considered as a proper LF time-ordered diagram.

Other advantages of minus regularization are that it deals with
longitudinal and transverse divergences in the same way, and that it
can be easily implemented numerically. We encounter these properties in
Chapter~\ref{chap4} where we deal with diagrams in the
Yukawa model that suffer from transverse divergences.  The formulas of
the one-boson exchange correction and the fermion triangle are too
complicated for us to be able to give an algebraic proof of
equivalence, and therefore we limit ourselves to a numerical
check of equivalence.

\begin{quote}
{\em 
We conclude that the combined use of $k^-$-integration and minus
regularization provides all the ingredients for the proof of equivalence
between covariant and LF perturbation theory in the Yukawa model.}
\end{quote}

Therefore, there is no need to add noncovariant counterterms to the
Lagrangian of the Yukawa model as claimed by Burkardt and
Langnau~\cite{BL91}. After our proof of equivalence for the Yukawa
model it it still  an open question whether our method is applicable
to models such as QCD and QED. The propagator for a spin-1 particle is
more complicated than those for spin-$0$ and spin-$1/2$ particles,
which makes the decomposition into propagating and instantaneous parts
more elaborate.  An additional complication is that in models with a
vector coupling two instantaneous particles can become neighbors,
leading to an effective five-point interaction. It is also unclear how
we should handle models with an effective coupling.  A similar problem
concerns the use of form factors in nuclear physics.  It would be
interesting to find out how they can be incorporated in a light-front
formulation.

\subsubsection{Entanglement of Fock-space expansion and covariance} 
Confident about the usefulness of the LF time-ordered diagrams as tools
in LFD, in Chapter~\ref{chap5} we make another step towards the
calculation of the bound state.  One has to solve the problem of
diagonalizing an infinitely dimensional Hamiltonian. An approximate
solution for this is provided by the use of an effective interaction.
This  method truncates the expansion of the interaction in higher Fock
states. The splitting of the covariant box diagram into LF time-ordered
diagrams with intermediate states containing at most three or four
particles (the latter are referred to as stretched boxes) shows that in
LFD these higher Fock states are heavily suppressed, not only because
the spectrum condition allows for few diagrams, but also because the
diagrams themselves are small.  In addition, we find that in LFD the
relative size of the diagrams containing higher Fock states is much
smaller than in the instant-form of Hamiltonian dynamics.

\begin{quote}
{\em We conclude that the 
     Fock-state expansion of LFD converges fast, and therefore the
     combined use of LFD and the ladder approximation offers good
     possibilities for an accurate calculation of the spectrum of
     bound states.}
\end{quote}
Based on these results, we conjecture that the inclusion of stretched boxes in
the kernel will not lead to a great change of the spectrum. This 
explains why bound-state calculations done in the past in LFD using the
ladder approximation have given accurate results~\cite{Kar98}.
Note that the stretched boxes are not covariant by
themselves, as they should not be. Therefore their omission leads to a
breaking of covariance. Recently, Ji {\em et al.}~\cite{JKM98} made an
attempt to solve the problem by averaging over all directions of the
orientation of the light-front. However, the stretched boxes have the
same sign for all angles, and averaging over them cannot give a
vanishing contribution. Therefore, upon angular averaging 
a small part of the amplitude will still be missing.

We have to remind the reader that our calculations were done for scalar
particles.  To analyze the influence of spin one  can apply the same
method.  It may be very interesting to also investigate the contributions of
crossed boxes, because it was shown by Gross~\cite{Gro82} that in the
heavy-mass limit they are necessary to retain the proper one-body
equation.  However, we would like to stress that for a discussion on the
entanglement of covariance and the Fock-space expansion one can look at the
properties of box and crossed box separately, as both are covariant
objects. For a discussion on the validity of the ladder approximation
one should both discuss stretched and crossed boxes.
\paragraph{}
We are confident that our investigations have contributed to an improved
understanding of LF Hamiltonian field theory, fifty years after it was
given birth to by Dirac.  Its usefulness to solve the bound state can
be considered millenniumproof.

\appendix
\renewcommand{\chaptermark}[1]{\markboth{\footnotesize \sf Appendix
              \thechapter\ \hspace{.3cm} #1}{#1} \markright{\footnotesize
               \sf Appendix \thechapter\ \hspace{.3cm} #1}}
\chapter{Relations between Euclidian integrals}
\label{appeuclint}
The two basic formulas that are used in Chapter~\ref{chap3} to compare
LF time-ordered and covariant diagrams in $d$ space-time dimensions are
\begin{eqnarray}
\label{bas1}
\int {\rm d}^dk f(k^2) &=& \frac{2 \pi^{d/2}}{\Gamma(d/2)} \int_0^\infty
{\rm d}k k^{d-1} f(k^2) , \\
\label{bas2}
\int_0^\infty {\rm d}k \frac{k^{d-1} }{(k^2 + C^2)^m} &=&
\frac{\Gamma(d/2) \Gamma(m \!-\! d/2)}{2 \Gamma(m)} (C^2)^{d/2 - m} ,
\end{eqnarray}
with $d \geq 1$ and $m > 0$.
If we take $d \geq 2$ and $m > 1$ the following manipulations are valid.
Formulas \r{bas1} and \r{bas2} can be combined to give
\begin{eqnarray}
&&\int {\rm d}^dk \frac{1}{(k^2 + C^2)^m} =
\pi^{d/2} \frac{\Gamma(m\!-\!d/2)}{\Gamma(m)} (C^2)^{d/2 - m},\\
&&\int {\rm d}^dk \frac{A + B k^2}{(k^2 + C^2)^m} =
\pi^{d/2} \frac{\Gamma(m\!-\!1-\!d/2)}{\Gamma(m)} (C^2)^{d/2 - m}
\nonumber\\ &&\hspace{1cm} \times
\left( (m\!-1\!-\!d/2) A + d B C^2/2 \right).
\end{eqnarray}
We can formulate the same equation for $d-2$ dimensions and $m-1$ as the
power
in de denominator. We find that the right-hand sides differ only slightly
\begin{eqnarray}
&&\int {\rm d}^{d-2}k \frac{A + B k^2}{(k^2 + C^2)^{m-1}} =
 \frac{\pi^{d/2}}{\pi} \frac{\Gamma(m\!-\!1\!-\!d/2)}{\Gamma(m\!-\!1)} (C^2)^{d/2 - m}
\nonumber\\
&&\hspace{1cm} \times\left( (m\!-\!1\!-\!d/2) A + (d\!-\!2) B C^2/2 \right) .
\end{eqnarray}
A comparison of these formulas gives
\begin{equation}
\label{rel1}
\int {\rm d}^dk \frac{A + B k^2}{(k^2 + C^2)^m} =
\frac{\pi}{m-1} \int {\rm d}^{d-2}k
\frac{A + B \frac{d}{d-2} k^2}{(k^2 + C^2)^{m-1}} ,
\end{equation}
provided we have $d>2$ and $m>1$. 

\chapter{The fermion self-energy in closed form}
\label{exactfse}
Here we give the results for the integral \r{fsepropr} in closed form.
We write for the renormalized self-energy
 
\vspace{-.5cm}
\begin{equation}
\fseprop \put(-27,10){r}\hspace{-.4cm}=\slash{q} \; F_1(q^2) + m \; F_2(q^2) .
 \label{eqB1}
\end{equation}

\noindent
Then the two functions $F_{1,2}$ are found to be
\begin{equation}
\frac{F_1(q^2)}{\pi^2 i} = - \int^1_0 {\rm d} x \; x \log \left( 1 - \frac{x(1-x)q^2}
 {(1-x)m^2 + x \mu^2} \right),
\label{eqB2}
\end{equation}
and
\begin{equation}
\frac{F_2(q^2)}{\pi^2 i}= - \int^1_0 {\rm d} x \log \left( 1 - \frac{x(1-x)q^2}
 {(1-x)m^2 + x \mu^2} \right).
\label{eqB2b}
\end{equation}
For $\mu = 0$ we find the result to be in agreement with the formula given
by Ligterink and Bakker~\cite{LB95a} and by Bjorken and Drell \cite{BD64}. 
They use the vector coupling appropriate for the photon and therefore
overall numerical factors are different.  
\begin{equation}
\frac{F_1(q^2)}{\pi^2 i} =  \frac{1}{4} + \frac{m^2}{2 q^2} -
\left( \frac{1}{2} - \frac{m^4}{2 q^4} \right) 
\log \frac{m^2 - q^2}{m^2} ,
 \label{eqB3}
\end{equation}
\begin{equation}
\frac{F_2(q^2)}{\pi^2 i} = 1 - \left( 1 - \frac{m^2}{q^2} \right)
 \log\frac{m^2 - q^2}{m^2} .
 \label{eqB4}
\end{equation}
For $\mu > 0$ we have
\[
\frac{F_1(q^2)}{\pi^2 i} =  \frac{1}{4} + \frac{(\mu^2 - m^2)^2 - \mu^2 q^2}
{2 (m^2 - \mu^2) q^2} 
%
+ \left( \frac{(m^2 - \mu^2 + q^2)^2 - 2 m^2 q^2}{4 q^4} - 
  \frac{m^4}{2 (m^2 - \mu^2)^2} \right) \log \frac{\mu^2}{m^2} 
\]
\begin{equation}
+\left( \log \frac{D^\frac{1}{2} + m^2 - \mu^2 - q^2}
                 {D^\frac{1}{2} - m^2 + \mu^2 + q^2}
     - \log \frac{D^\frac{1}{2} + m^2 - \mu^2 + q^2}
                 {D^\frac{1}{2} - m^2 + \mu^2 - q^2}
\right) 
\label{eqB5}
\frac{D^\frac{1}{2} (m^2 - \mu^2 + q^2)}{4 q^4} ,
\end{equation}
and
\[
\nonumber
\frac{F_2(q^2)}{\pi^2 i} =  1 + 
\left( \frac{m^2}{\mu^2 - m^2} + \frac{m^2 - \mu^2 + q^2}{2q^2} \right)
\nonumber
\log \frac{\mu^2}{m^2} 
\]
\begin{equation}
+ \frac{D^\frac{1}{2}}{2q^2}
\left( \log \frac{D^\frac{1}{2} + m^2 - \mu^2 - q^2}
                 {D^\frac{1}{2} - m^2 + \mu^2 + q^2}
     - \log \frac{D^\frac{1}{2} + m^2 - \mu^2 + q^2}
                 {D^\frac{1}{2} - m^2 + \mu^2 - q^2}
\right) ,
\label{eqB6}
\end{equation}
where the variable $D$ contains the threshold behavior
\begin{equation}
D = \left(q^2 - (m + \mu)^2 \right) \left(q^2 - (m - \mu)^2 \right) .
\label{eqB7}
\end{equation}
We checked that the limit $\mu \to 0$ of Eqs.~\r{eqB5} and \r{eqB6}
exists and is equal to Eqs.~\r{eqB3} and \r{eqB4} respectively.

\chapter{Internal and external variables}
\label{app1}
We get more insight into the properties of the structure functions used
in Chapter~\ref{chap4} if
we rewrite them in terms of internal and external variables.
This can be done by defining
\begin{eqnarray}
\label{a1}
x' &=& \frac{k^+}{q_1^+} = (x-1) \chi, \\
\label{a2}
x &=& \frac{k^+ + q_2^+}{q_2^+} = \frac{x' + \chi}{\chi}.
\end{eqnarray}
Or, equivalently,
\begin{eqnarray}
k^+ &=& x' q_1^+ = (x-1) q_2^+,\\
k_1^+ &=& (x'-1) q_1^+,\\
k_2^+ &=& x q_2^+.
\end{eqnarray}
In the numerator of the integrals defining
LF structure functions we encounter 
on-shell spin projections. They can be rewritten in terms of
internal variables using
\begin{eqnarray}
k^-_{1\rm on}&=& \frac{{k_1^\perp}^2 + m^2}{2 (x'-1) q_1^+},\\
k^-_{2\rm on}&=& \frac{{k_2^\perp}^2 + m^2}{2 x q_2^+}.
\end{eqnarray}
The energy denominators can also be written 
in terms of internal and external variables.
The poles are given by Eqs.~\r{pole2}, \r{pole3} and \r{pole1b}:
\begin{eqnarray}
2 q_1^+(H_1^-\!-\!H_2^-)&=&
  2 q_1^+\left(p^- + \frac{{k_1^\perp}^2 + m^2}{2k_1^+}
       - \frac{{k_2^\perp}^2+ m^2}{2k_2^+}\right)
\hspace{-.5cm}\nonumber\\
&=& (p^2\!+{p^\perp}^2) \frac{1+\chi}{\chi}
  - \frac{{k_1^\perp}^2 + m^2}{1-x'}
  - \frac{{k^\perp_2}^2+ m^2}{x \chi},
\\
2 q_1^+(H_1^-\!-\!H^-)&=&
  2 q_1^+\left(q_1^-\!- \frac{{k^\perp}^2+\mu^2}{2k^+}
       + \frac{{k_1^\perp}^2 + m^2}{2k_1^+}\right)
\hspace{-.5cm}\nonumber\\
&=& q_1^2 + {q_1^\perp}^2 - \frac{{k^\perp}^2+\mu^2}{x'}
       - \frac{{k_1^\perp}^2 + m^2}{1-x'},\\
2 q_2^+(H^-\!-H_2^-)&=&
  2 q_2^+\left(q_2^-+ \frac{{k^\perp}^2+\mu^2}{2k^+}
       - \frac{{k_2^\perp}^2 + m^2}{2k_2^+}\right)
\hspace{-.5cm}\nonumber\\
&=& q_2^2 + {q_2^\perp}^2 - \frac{{k^\perp}^2+\mu^2}{1-x}
       - \frac{{k_2^\perp}^2 + m^2}{x}.
\end{eqnarray}
The integration measures can be rewritten as follows:
\begin{eqnarray}
2 \pi i \int_0^{q_1^+} \frac{{\rm d}k^+ 4 q_1^+ q_2^+}{8 k_1^+ k_2^+ k^+},
&=&- \pi i \int_0^1 \frac{{\rm d}x'}{(1-x')x x'},\\
-2 \pi i \int_{-q_2^+}^0 \frac{{\rm d}k^+ 4 q_1^+ q_2^+}{8 k_1^+ k_2^+ k^+}
&=&- \pi i \int_0^1 \frac{{\rm d}x}{(1-x')x (1-x)}.
\end{eqnarray}
We conclude that it is possible to write the structure functions
in terms of the external variables
$q_1^-,\; q_2^-,\; q_1^\perp,\; q_2^\perp$ and $\chi$
and integrals over the internal variables $x$ or $x'$ and $k^\perp$.
The divergent parts of the structure
functions $F^2_i$ can now be written as
\begin{eqnarray}
\hspace{-1.8cm}&&f^{2-}_1 =- \pi i \int_0^1 \frac{{\rm d}x'}{(1-x')x x'} \;
\frac{m}{(x'-1) q_1^+}\; \frac{q^+_1}{q^+_2} 
\left( \frac{1}{1-x'} + \frac{1}{x \chi}\right)^{-1}
\left( \frac{1}{x'} + \frac{1}{1-x'}\right)^{-1},\\
\hspace{-1.8cm}&&f^{2-}_2 = - \pi i \int_0^1 \frac{{\rm d}x}{(1-x')x (1-x)} \;
\frac{m}{x q_2^+} 
\left( \frac{1}{1-x'} + \frac{1}{x \chi}\right)^{-1}
\left( \frac{1}{x} + \frac{1}{1-x}\right)^{-1}.
\end{eqnarray}
Upon cancelling common factors, and using Eq.~\r{a2}, we can evaluate
the integrals and obtain
\begin{equation}
f^{2-}_1 = - f^{2-}_2 = \pi i \; \frac{\chi}{1 + \chi} \; \frac{m}{q_2^+}
= \pi i \; \frac{m}{p^+}.
\end{equation}
Therefore condition~\r{cancel} is verified.

\chapter*{Samenvatting}
\addcontentsline{toc}{chapter}{Samenvatting}
\markboth{\footnotesize\sf Samenvatting}{\footnotesize \sf Samenvatting}
\hyphenation{re-a-lis-tisch}
\hyphenation{tijd-ge-or-den-de}
\hyphenation{ver-wij-derd}
\hyphenation{re-gu-la-ri-sa-tie}
\hyphenation{be-schrij-ving}
\hyphenation{be-schrij-ven}
\hyphenation{nood-zake-lijker-wijze}
\hyphenation{natuur-kundige}
\hyphenation{be-re-ke-ningen}

\subsubsection{Lichtfront Hamiltoniaanse veldentheorie\\
Naar een relativistische beschrijving van gebonden toestanden }

\vspace{.4cm}
Ik zal eerst de plaats van mijn onderzoek binnen de natuurkunde
aangeven, en daarna op het onderzoek zelf ingaan.
Het veelomvattende terrein waarop het onderzoek zich bevindt heet
hoge-energiefysica, vanwege het feit dat de bij de botsingsexperimenten 
betrokken deeltjes dermate hoge energie\"{e}n hebben dat
ze bijna met de lichtsnelheid voortbewegen. Het doel van dit onderzoek is
om de structuur van de materie, met name op de allerkleinste schaal,
te begrijpen.

Het atoom, waar we op de middelbare school vertrouwd mee zijn geraakt,
bestaat uit een kern die is omgeven door een wolk van elektronen. De kern,
op haar beurt, bestaat uit protonen en neutronen. Al geruime tijd is
bekend dat deze {\em  hadronen} uit twee of drie {\em quarks} bestaan, en deze
blijven bij elkaar door de uitwisseling van lijmdeeltjes, veelal
{\em gluonen} genoemd. De massa van het hadron is groter
dan de optelsom van de massa's van de quarks. E{\'{e}}n van de vragen in
de hoge-energiefysica is of
wij de massa van het hadron kunnen bepalen als we 
de wijze waarop quarks gluonen uitwisselen kennen.

Hadronen zijn voorbeelden van {\em gebonden toestanden}. In dit
proefschrift beschrijven we een wiskundige formulering waarmee wordt
geprobeerd een relativistische beschrijving te geven van gebonden
toestanden, die goed aansluit bij de {\em Hamiltoniaanse} technieken
die bekend zijn uit de quantummechanica. Zij heet {\em lichtfront
Hamiltoniaanse veldentheorie}, een formulering waarbij als het ware
met het licht wordt 'meegereisd', en daarom kan men zich voorstellen dat het
geschikt is om hoog-energetische deeltjes en hun gebonden toestanden te
beschrijven. Bedenk wel dat reizen met de lichtsnelheid niet mogelijk
is, en dat deze beschrijving dus abstract is.

De eerste die deze aanpak voorstelde was Dirac~\cite{Dir49} in 1949.
Aangezien mijn promotie bijna vijftig jaar na het verschijnen van dit
baanbrekende werk plaatsvindt, is dit een toepasselijke gelegenheid om
hier bij stil te staan. Helaas duurde het tot de jaren zeventig voordat
wetenschappers die ge\"{\i}nteresseerd zijn in gebonden toestanden het
belang van Diracs werk inzagen. Het moet dan ook wel gezegd worden dat
er aan deze lichtfrontdynamica (LFD) enige moeilijkheden zijn
verbonden. Zo is rotatiesymmetrie niet vanzelfsprekend in deze theorie,
wat ongeveer betekent dat de berekening van een meetbare grootheid,
zoals de massa van een samengesteld deeltje, niet noodzakelijkerwijze
tot hetzelfde antwoord leidt als je het van een andere kant beschouwt.
Alle symmetrie\"{e}n die samenhangen met Einsteins relativiteitstheorie,
waaronder rotatiesymmetrie,
worden in de natuurkunde wel samengevat onder
de noemer {\em Lorentz-covariantie}.  Natuurkundige theorie\"{e}n die
deze Lorentz-covariantie schenden worden, met recht, ernstig
gewantrouwd.                                            

Het doel van dit proefschrift is tweeledig. Ten eerste willen we laten zien
dat de bouwstenen van LFD, de {\em lichtfront-tijdgeordende diagrammen}
in principe tot Lorentz-covariante resultaten leiden. 
Vervolgens willen we laten zien dat de LFD, meer dan andere methodes,
geschikt is het voor het doen van zo nauwkeurig mogelijke berekeningen
aan gebonden toestanden.  Hieronder zullen we per hoofdstuk op de
details ingaan.

\subsubsection{Hoofdstuk \ref{chap1} \hspace{.5cm}
Introductie van lichtfront Hamiltoniaanse dynamica}

We beginnen met uit te leggen hoe een Hamiltoniaanse theorie moet
worden opgezet en vergelijken de twee belangrijkste
mogelijkheden: {\em quantisatie} op een vlak met gelijke tijd (de wijze
die bekend is uit de niet-relativistische quantummechanica)
en de methode die wij adverteren:  quantisatie op een vlak dat raakt
aan de {\em lichtkegel}: het lichtfront.  Om een aantal problemen die
bij quantisatie optreden te omzeilen introduceren we een methode die
vanuit covariante veldentheorie tot diagrammatische regels voor LFD
leidt: de $k^-$-integratie, dat wil zeggen de integratie over de
energiecomponent van de interne impuls.  Deze methode zullen we 
veelvuldig toepassen in de volgende hoofdstukken.

\subsubsection{Hoofdstuk \ref{chap2} \hspace{.5cm}
Het Yukawa-model}

Wij gebruiken het Yukawa-model om twee
redenen. Het is het meest eenvoudige model dat een beschrijving geeft
van de interactie tussen fermionen met spin $1/2$ via bosonen (in ons
geval zonder spin).
Bovendien werd onze aandacht getrokken door een artikel van Burkardt
en Langnau~\cite{BL91} die beweerden dat in het Yukawa-model 
Lorentz-covariantie gebroken wordt. Wij zullen laten zien dat als
de {\em divergenties} op de juiste wijze behandeld worden
Lorentz-covariantie niet in gevaar komt.

\subsubsection{Hoofdstuk \ref{chap3} \hspace{.5cm}
Longitudinale divergenties in het Yukawa-model}

Het eerste gedeelte van het bewijs van 
Lorentz-covariantie van LFD is reeds
geleverd door Ligterink in zijn proefschrift~\cite{Lig95}. Hij liet
door een ingewikkeld wiskundig hercomberingsschema zien hoe uit
een covariant diagram alle bijbehorende {lichtfront}-{tijd\-geordende}
diagrammen volgen door integratie over $k^-$.  Dit bewijs is niet geldig wanneer de integraal
over $k^-$ divergent is.  Dit type divergenties worden longitudinale
divergenties genoemd.  We laten zien dat met de {\em
minus-regularisatiemethode} de divergenties kunnen worden verwijderd en
dat Lorentz-covariantie gewaarborgd is. Minusregularisatie is door
Ligterink opgezet en door ons uitgebreid. Het is een in lichtfrontco\"ordinaten
geformuleerde regularisatiemethode analoog aan de BPHZ-methode die bekend
is in covariante veldentheorie.

\subsubsection{Hoofdstuk \ref{chap4} \hspace{.5cm}
Transversale divergenties in het Yukawa-model}

Een tweede complicatie is de aanwezigheid in diagrammen
van divergenties in de transversale componenten $k^\perp$ van de impuls.
We laten zien dat deze transversale divergenties ook met minusregularisatie
kunnen worden aangepakt. Omdat de diagrammen waar deze divergenties
optreden ingewikkelder zijn dan in hoofdstuk~\ref{chap3} 
is het niet mogelijk in alle
gevallen een wiskundig bewijs te geven. Wij laten echter in een
numerieke berekening zien dat ook in deze gevallen Lorentz-covariantie
gehandhaafd is. Opnieuw is het toverwoord hierbij minusregularisatie.

\subsubsection{Hoofdstuk \ref{chap5} \hspace{.5cm}
Verstrengeling van de Fock-ruimteontwikkeling en covariantie}

De volgende stap die men neemt op weg naar de berekening van de
gebonden toestand is het oplossen van de {\em Hamiltoniaanse
eigenwaardevergelijking}. Dit exact doen is schier onmogelijk, omdat de
Hamiltoniaan een oneindig grote matrix is. Om toch het massaspectrum te
kunnen uitrekenen maakt men meestal de {\em ladderbenadering}.  Deze
gaat er vanuit dat bij de berekening van de gebonden toestand van twee
deeltjes, die via de uitwisseling van bosonen verbonden zijn, nooit
twee of meer bosonen op hetzelfde moment worden uitgewisseld. Dat houdt
in dat {\em Fock-toestanden} met vier of meer deeltjes worden
genegeerd.

Echter, als men covariante storingsrekening
tot in de vierde orde in de koppelingsconstante uitvoert, dan
zullen er zowel diagrammen met maximaal drie deeltjes als diagrammen
met maximaal vier deeltjes in een intermediaire Fock-toestand
voorkomen. De laatste categorie wordt echter genegeerd in de
ladderbenadering.
Hierdoor gaat een gedeelte van de amplitude verloren. Bovendien
wordt Lorentz-covariantie gebroken, omdat de diagrammen die
weggelaten zijn op zich niet covariant zijn in LFD, een eigenschap die
wordt gedeeld met elke andere Hamiltoniaanse theorie.

Wij laten zien in een numerieke berekening dat de ladderbenadering toch
redelijk goed is, omdat in gevallen met waarden van massa's en
snelheden die typisch zijn voor realistische gebonden toestanden (in
ons geval het deuteron) uitwisseling van meerdere deeltjes
tegelijkertijd maximaal tot 2\% bijdraagt tot het resultaat.  In andere
typen Hamiltoniaanse theorie\"{e}n zijn deze bijdragen veel groter, en
wordt de ladderbenadering dus veel slechter.  Dit verklaart ook waarom
berekeningen die in het verleden zijn gedaan met de ladderbenadering in
LFD goede resultaten hebben gegeven.

\subsubsection{Hoofdstuk \ref{chap6} \hspace{.5cm}
Samenvatting en conclusies}

In de voetsporen tredend van Feynman (zie citaat op bladzijde~\pageref{chap6})
wil ik besluiten met een vergelijking tussen
de natuurkunde en het schaakspel. Het doel van mijn onderzoek  is niet zo
zeer om de regels waarmee de schaakstukken (de deeltjes, zoals {het}
pion) zich bewegen te ontdekken, maar om de wijze waarop zij
een eenheid vormen beter te begrijpen~\cite{sp}. 

\bibliographystyle{utphys}
\bibliography{articles,books}

\markright
{\sf \footnotesize obrigado, spasiba, remerciement, tack, takk, te\c{s}ekk\"urler, thanks} 
\markboth
{\sf \footnotesize dank, Danke, gracias, grazie, kiit\=a\=a, k\"osz\"on\"om, mamnoon, multumire,}
{\sf \footnotesize obrigado, remerciement, spasiba, tack, takk, te\c{s}ekk\"urler, thanks}

\newpage

\vspace{5cm}
\renewcommand{\thepage}{$\infty$}


\hspace{-1.5cm}\epsfxsize=3cm \raisebox{-1.65cm}{\epsffile{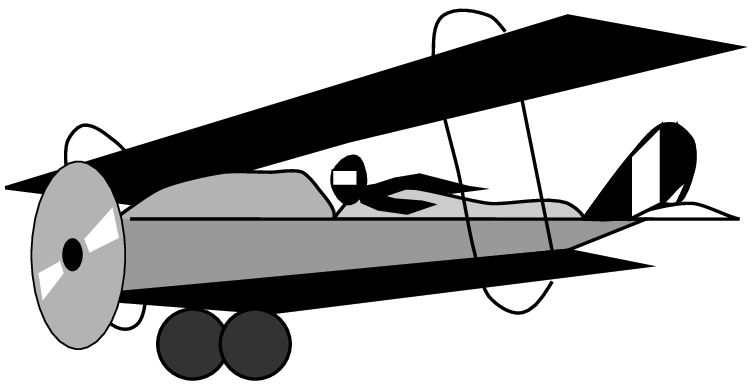}} 
\hspace{-.7cm}
\raisebox{-1.95cm}{\epsfxsize=16cm \epsffile{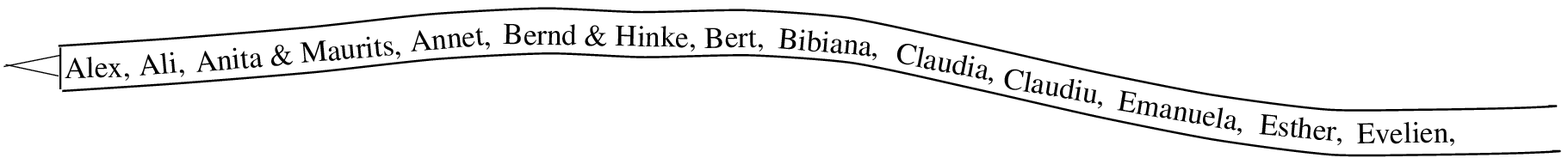}}

\vspace{6.4cm}
{\epsfxsize=18cm \epsffile{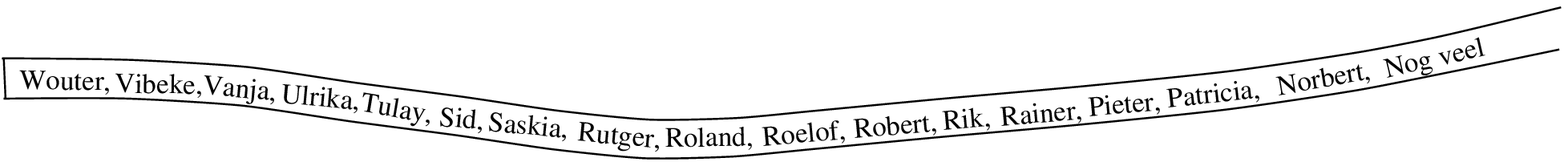}}

\newpage

\vspace{17cm}
\begin{picture}(200,400)(0,0)
\put(-50,-250){\epsfxsize=18cm \epsffile{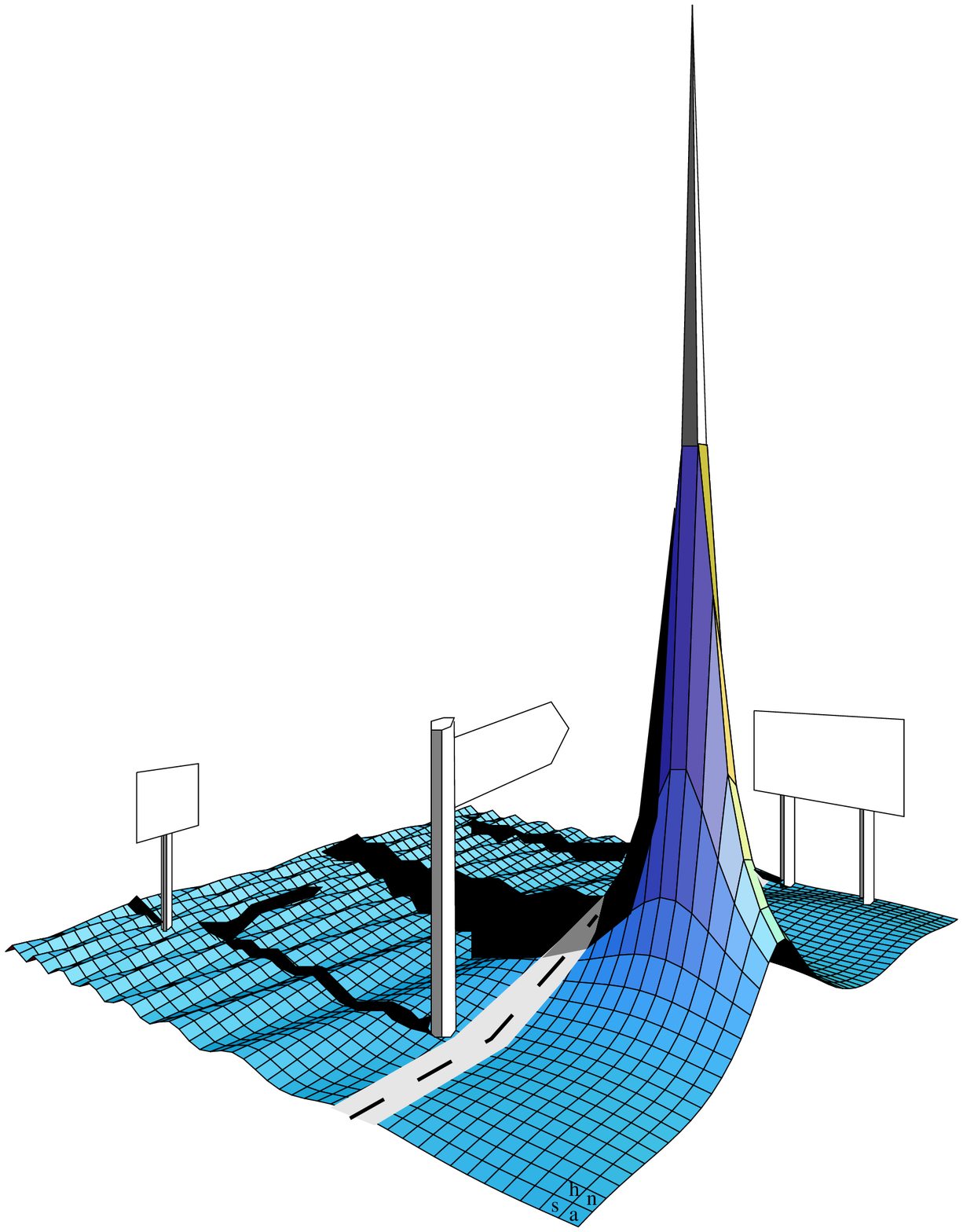}}
\put(-65,130){\epsfxsize=13.5cm \epsffile{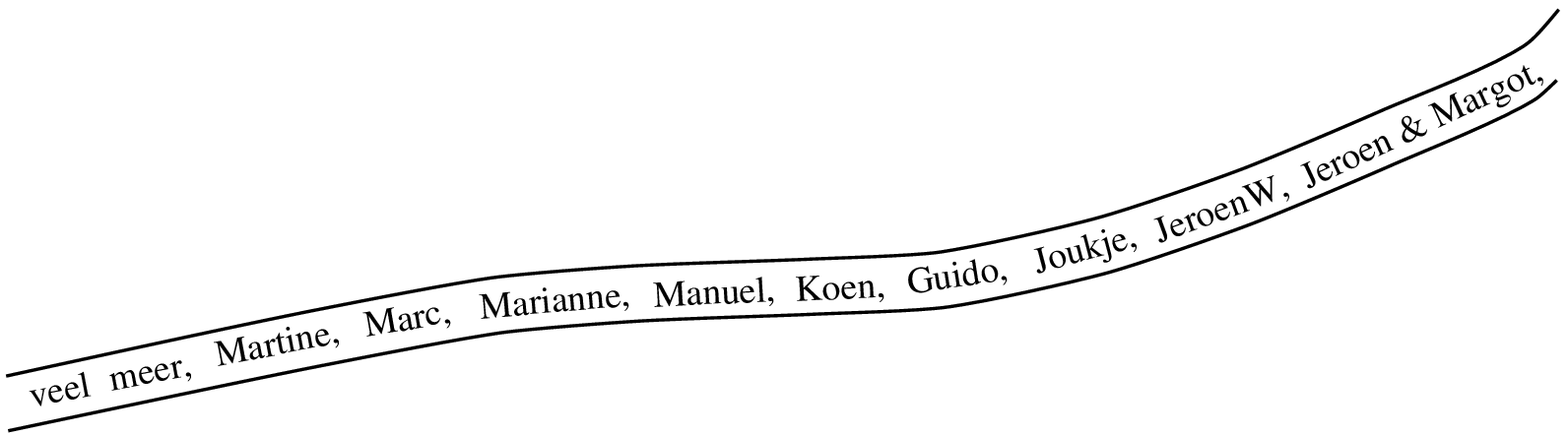}}
\put(-50,250){\epsfxsize=13cm \epsffile{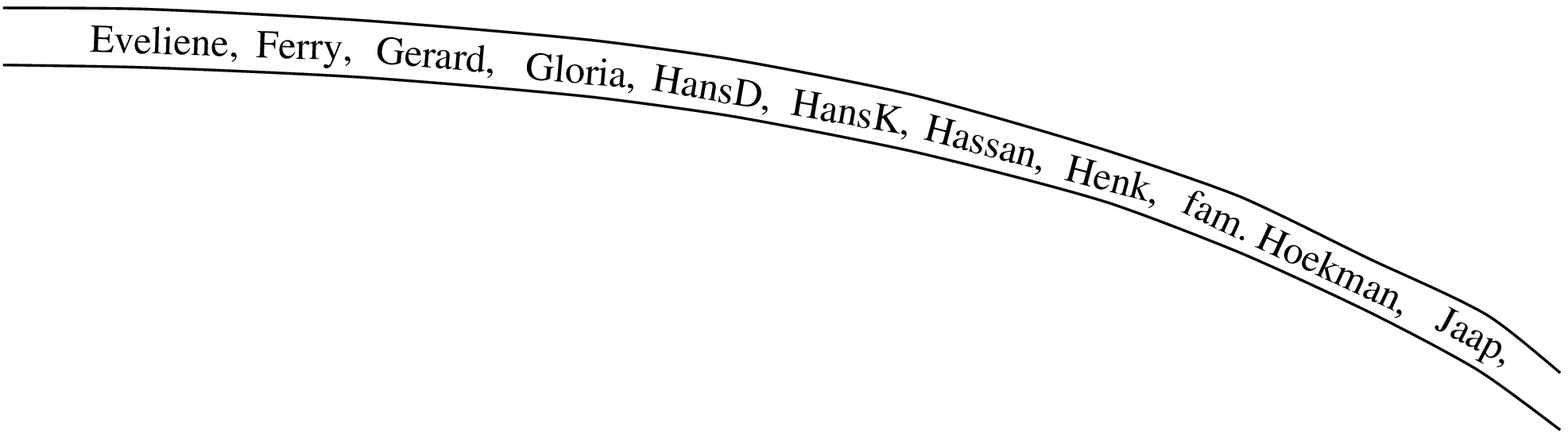}}
\put(50,35){\Huge $i$}
\put(204,54){\epsfxsize=1.8cm \epsffile{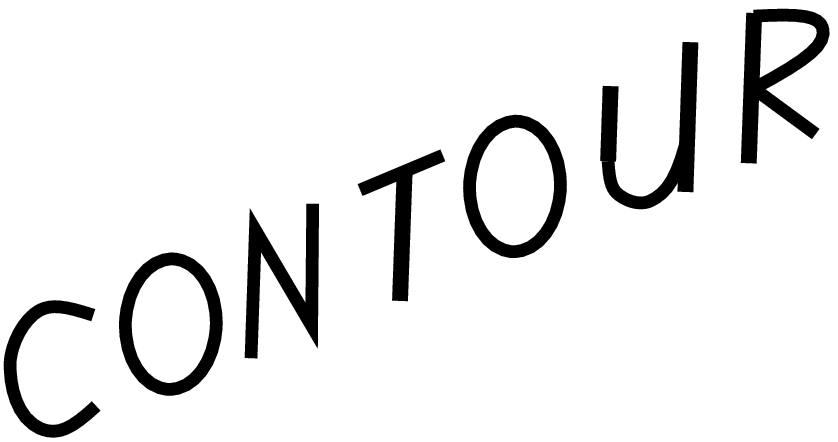}}
\put(355,45){\epsfxsize=2.5cm \epsffile{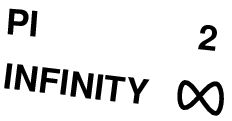}}
\end{picture}
\newpage
\epsfxsize=.5cm \epsffile{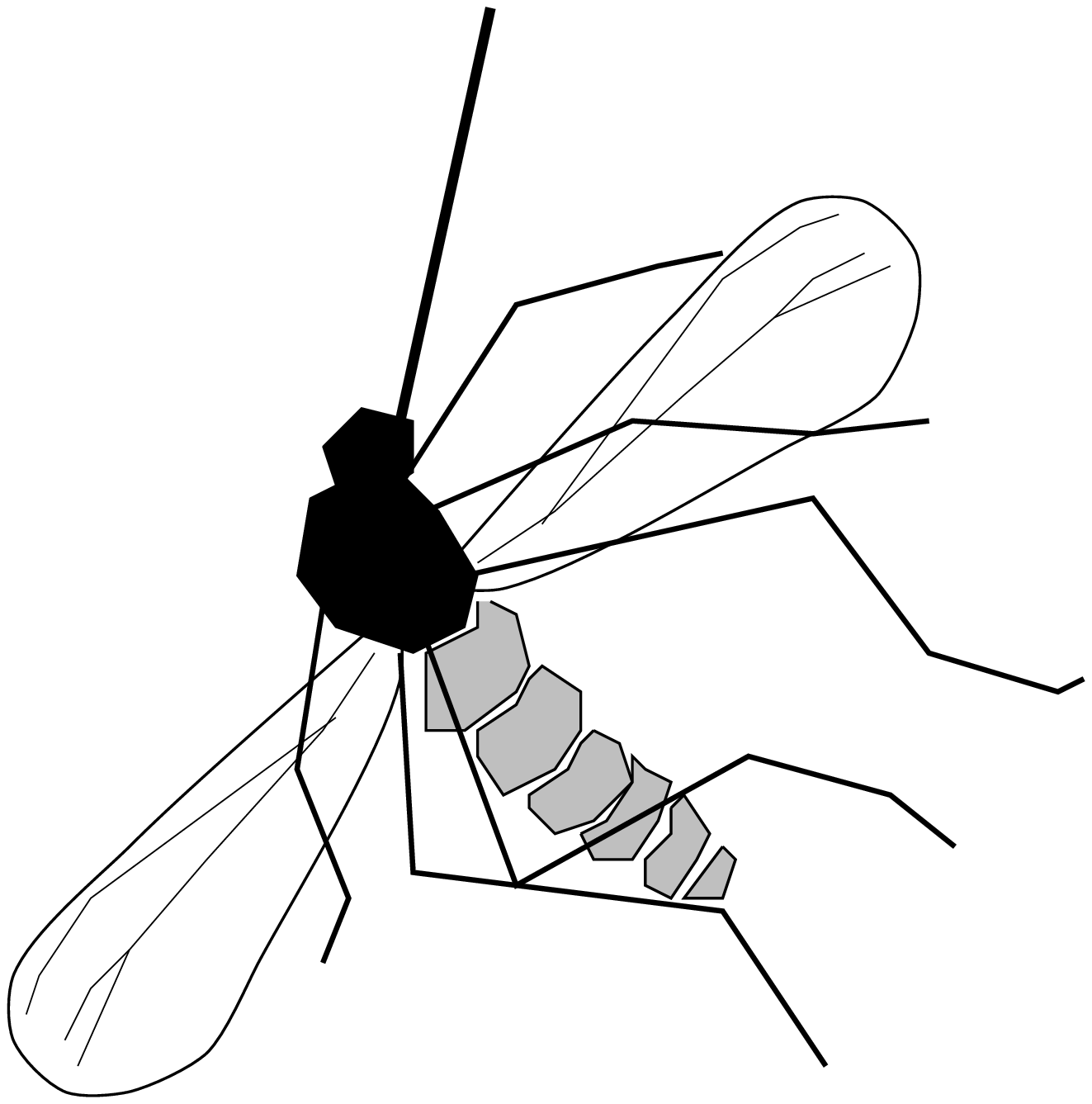}

\pagestyle{empty}

$\phantom{Bedankt voor het lezen, tot ziens!}$

\end{document}